%% file: main.tex
\documentclass[conference]{IEEEtran}
\IEEEoverridecommandlockouts
\usepackage{cite}
\usepackage{amsmath,amssymb,amsfonts}
\usepackage{algorithmic}
\usepackage{graphicx}
\usepackage{textcomp}
\usepackage{xcolor}
\usepackage[hyphens]{url}
\usepackage{fancyhdr}
\usepackage{hyperref}
\usepackage{xspace}
\usepackage{amsmath}
\usepackage{algorithmic}
\usepackage{graphicx}
\usepackage{textcomp}
\usepackage{xcolor}
\usepackage{xspace}
\usepackage[table,xcdraw]{xcolor}
\usepackage{pifont}
\usepackage{tabularx}
\usepackage{array}
\usepackage{diagbox}
\usepackage{tablefootnote}
\usepackage{multirow}
\usepackage{amsmath}
\usepackage{makecell}
\usepackage{colortbl} % Add this in your preamble for table coloring
\usepackage{booktabs}
\usepackage{soul}
\usepackage{adjustbox} % in preamble
\usepackage{enumitem} 
\usepackage{tabularx}
\definecolor{darkgreen}{rgb}{0.0, 0.5, 0.0}
\usepackage{tikz}
\usepackage{tcolorbox}

\def\BibTeX{{\rm B\kern-.05em{\sc i\kern-.025em b}\kern-.08em
    T\kern-.1667em\lower.7ex\hbox{E}\kern-.125emX}}
\begin{document}

% Ensure letter paper
\pdfpagewidth=8.5in
\pdfpageheight=11in

\pagenumbering{arabic}

%%%%%%%%%%%%%%%%%%%%%%%%%%%%%%%%%%%%%%%%
%%%%%%%%%%%%%% -- UPDATE -- %%%%%%%%%%%%%%%
\newcommand{\iscasubmissionnumber}{875}
\newcommand{\sys}[0]{\textsc{STAGE}\xspace}

% autoref section names
\renewcommand*\sectionautorefname{\Snospace}
\def\sectionautorefname{Sec.}
\def\subsectionautorefname{Sec.}
\def\subsubsectionautorefname{Sec.}
\def\figureautorefname{Fig.}
\def\algorithmautorefname{Alg.}

\newcommand{\TK}[1]{\textcolor{orange}{TK: #1}} % Tushar
\newcommand{\ju}[1]{{\color{blue}{JU: #1}}}
\newcommand{\ch}[1]{\textcolor{red}{CH: #1}} % Changhai
\newcommand{\hw}[1]{\textcolor{magenta}{HW: #1}} % Hanjiang
\newcommand{\hx}[1]{\textcolor{darkgreen}{HX: #1}} % Huan
\newcommand{\sr}[1]{\textcolor{purple}{SR: #1}} % Srinivas
\newcommand{\jy}[1]{\textcolor{red}{JY: #1}}

\newcommand{\revise}[1]{{\color{red}#1}}
\newcommand{\shephered}[1]{{\color{blue}#1}}

\newcommand{\squishlist}{
 \begin{list}{$\bullet$}
  { \setlength{\itemsep}{0pt}
     \setlength{\parsep}{3pt}
     \setlength{\topsep}{3pt}
     \setlength{\partopsep}{0pt}
     \setlength{\leftmargin}{1.5em}
     \setlength{\labelwidth}{1em}
     \setlength{\labelsep}{0.5em} } }

\newcommand{\squishlisttwo}{
 \begin{list}{$\bullet$}
  { \setlength{\itemsep}{0pt}
     \setlength{\parsep}{0pt}
    \setlength{\topsep}{0pt}
    \setlength{\partopsep}{0pt}
    \setlength{\leftmargin}{2em}
    \setlength{\labelwidth}{1.5em}
    \setlength{\labelsep}{0.5em} } }

\newcommand{\squishend}{
  \end{list}  }

%\title{Scalable Synthesis of distributed LLM Benchmarks through Symbolic Tensor Graphs}
\title{Scalable Synthesis of distributed LLM workloads through Symbolic Tensor Graphs}
%%%%%%%%%%%%%%%%%%%%%%%%%%%%%%%%%%%%%%%%

%%%%%%%%%%%---SETME-----%%%%%%%%%%%%%
% \author{\normalsize{ISCA 2026 Submission
    % \textbf{\#\iscasubmissionnumber} -- Confidential Draft -- Do NOT Distribute!!}}
\vspace{-1em}
\author{
Changhai Man$^{1}$,
Joongun Park$^{1}$,
Hanjiang Wu$^{1}$,
Huan Xu$^{1}$,
Srinivas Sridharan$^{2}$,
Tushar Krishna$^{1}$\\
$^{1}$Georgia Institute of Technology\quad
$^{2}$NVIDIA Inc.\\
{\{cman8,jpark3234,hwu419,hxu398\}@gatech.edu}
\quad
{srisridharan@nvidia.com}
\quad
{tushar@ece.gatech.edu}
\vspace{-1em}
}
%%%%%%%%%%%%%%%%%%%%%%%%%%%%%%%%%%%%

% \begin{titlepage}
% \twocolumn
% \thispagestyle{empty}
% \input{sections/cover-page}
% \end{titlepage}

\maketitle
\thispagestyle{plain}
\pagestyle{plain}

%%%%%% -- PAPER CONTENT STARTS-- %%%%%%%%

\input{sections/abstract}

\input{sections/intro}
\input{sections/background}
\input{sections/motivation}

\input{sections/design}
\input{sections/validation}
\input{sections/evaluation}

\input{sections/discussion}
\input{sections/relatedwork}
\input{sections/conclusion}

\vspace{-0.5em}
\section*{Acknowledgments}
We thank Jinsun Yoo for helpful discussions and feedback. We also thank the anonymous reviewers for their constructive comments, which helped strengthen this paper. We thank Matthieu Bloch and Aaron Jezghani for helping us use the College of Engineering AI Makerspace (RRID:SCR\_028058) at Georgia Tech, provided by PACE (RRID:SCR\_027619), to collect validation traces for this work. This work was supported in part by the ACE Center for Evolvable Computing, an SRC JUMP 2.0 Center.

%%%%%%% -- PAPER CONTENT ENDS -- %%%%%%%%

%%%%%%%%% -- BIB STYLE AND FILE -- %%%%%%%%
\bibliographystyle{IEEEtranS}
\bibliography{reference}
%%%%%%%%%%%%%%%%%%%%%%%%%%%%%%%%%%%%

\end{document}

%% file: sections/abstract.tex
\begin{abstract}
% Optimizing the performance of large language models (LLMs) on large-scale AI training/inference systems necessitates a nimble mechanism to represent distributed workload execution - to enable pre-deployment analysis of system-level optimizations (such as parallelism) on current platforms and enable design-space exploration of next-generation acceleration platforms.
% Recent efforts have introduced techniques to collect execution traces from real AI platforms. Unfortunately, access to large-scale AI systems is limited to a few hyperscalars today. Moreover, the collected traces cannot be easily extended to study larger scale systems.
% In this work, we introduce \sys\footnote{\textbf{S}ymbolic \textbf{T}ensor gr\textbf{A}ph \textbf{GE}nerator}, a framework that synthesizes execution traces to accurately capture LLM workload characteristics. \sys supports a variety of parallelization strategies, allowing users to explore different scaling and optimization scenarios. The generated workloads have been validated against results from latest ML frameworks and optimizations, demonstrating their effectiveness in optimizing distributed ML systems. 
% This project is publicly available at \url{https://github.com/astra-sim/symbolic_tensor_graph}.  
% By sharing \sys, we aim to promote progress in distributed machine learning across both industrial and research domains, making LLM training optimizations more accessible and effective.
Optimizing the performance of large language models (LLMs) on large-scale AI training and inference systems requires a scalable and expressive mechanism to model distributed workload execution. Such modeling is essential for pre-deployment system-level optimizations (e.g., parallelization strategies) and hardware design-space explorations.
While recent efforts have proposed collecting execution traces from real systems, access to large-scale infrastructure remains limited to major cloud providers. Moreover, traces capturing execution on a specific platform cannot be easily adapted to study alternate
software and/or hardware configurations, especially at scale.
%larger-scale system configurations.
We introduce \sys\footnote{\textbf{S}ymbolic \textbf{T}ensor gr\textbf{A}ph \textbf{GE}nerator}, a framework that synthesizes high-fidelity execution graphs to accurately model distributed AI workloads (including LLMs and MoEs). 
\sys supports a comprehensive set of parallelization strategies, allowing users to systematically explore a wide spectrum of model architectures and system configurations.
\sys demonstrates its scalability by synthesizing high-fidelity LLM traces spanning over 128K GPUs, while preserving tensor-level accuracy in compute, memory, and communication.
\sys is publicy available at \url{
https://github.com/astra-sim/stage}
%\sys will be publicly available to facilitate further research in distributed machine learning systems.
\end{abstract}

%% file: sections/intro.tex
\section{Introduction}

%\ju{Very intro}

The rapid growth of machine learning models, especially Large Language Models (LLMs), including GPT \cite{gpt-3}, Llama \cite{llama}, DeepSeek \cite{deepseek-moe}, and Mistral \cite{mistral}, has revolutionized the field of machine learning, driving massive advancements in natural language processing and generative AI.
%, machine learning, and beyond.
However, the scale and complexity of LLMs have introduced unprecedented computational challenges. These models often require massive amounts of computation and memory~\cite{wang_hoti2024,llm_memory},
not only during training but also for inference, necessitating distributed AI systems. 
Several such systems exist in practice today, including NVIDIA HGX \cite{nvidiaHGX}, Google TPU \cite{TPUv4Cluster}, Amazon Trainium \cite{AWSTrainium}, Cerebras CS-3 \cite{cerebrasCS3}, and others. Optimizing compute, memory and communication resources optimally in these systems is crucial for performance~\cite{astrasim2,themis}.
The need for scalable and efficient distributed training is only growing, as evidenced by the recently released Llama 4 model that leverages a Mixture-of-Experts (MoE) architecture with up to 2 trillion parameters~\cite{meta2025llama4}, pushing the limits of current AI system infrastructure.

%, necessitating distributed AI systems where compute and communication overheads dominate performance~\cite{themis}.
%Llama 4, for example, leverages a Mixture-of-Experts (MoE) architecture with up to 2 trillion parameters, pushing the limits of current infrastructure and further amplifying the need for scalable and efficient distributed training.

%\ju{Why we need traces? What can you do with Traces?}

Standardized benchmarks play a crucial role in our community, serving two key purposes: optimizing the performance of current AI systems and guiding the design choices for next-generation systems. Efforts like MLPerf~\cite{mlperf} have been leading the way in identifying representative benchmarks in the domain of AI. 
Unfortunately, deploying the full software stack of distributed AI benchmarks for the sole purpose of running optimization and design-space exploration (DSE) studies is prohibitive in practice, as they require extensive framework (PyTorch/JAX/TensorFlow) expertise and continued access to large-scale systems. Furthermore, it is extremely difficult to isolate hardware versus software bottlenecks, and compute versus memory versus network behaviors. 

%due to several reasons:
%\squishlist
%\item Requires extensive framework (PyTorch/JAX/TF) expertise
%\item Need continued access to expensive large-scale systems.
%\item Difficult to isolate hardware versus software bottlenecks
%\item Difficult to isolate compute, memory and network behavior
%\item Standardized benchmark suites find it hard to keep up with the pace of AI innovation.
%\squishend

% as it requires continued access to a large-scale systems which is extremely expensive. 
% Moreover, the extremely fast pace of AI innovation demands a more light-weight mechanism for reproduction of workload behavior in current production systems and extension to future AI system co-design.
%It is also hard to keep up with the pace of AI models, with several AI labs (OpenAI, Anthropic, Google, Meta) releasing new models at a rapid pace.
%Moreover, deploying full benchmarks does not help answer ``what-if" to design next-generation systems.

Acknowledging the aforementioned challenges, 
%To this end, 
recent efforts~\cite{sridharan2023chakra,madmax} have proposed the idea of execution traces (ET) as a mechanism to capture the \textit{coarse-grain (i.e., operator-level) compute and communication dependence behavior} during AI training.
%Precise workload modeling is critical for optimizing these systems, guiding decisions that balance resource utilization, computation efficiency, and cost. 
%Execution traces (ET) of distributed model execution have been proposed by recent works~\cite{sridharan2023chakra,madmax} as a mechanism to capture the coarse-grain (i.e., operator-level) compute and communication dependence behavior during AI training.
In particular,  MLCommons Chakra~\cite{sridharan2023chakra} has introduced specific support within PyTorch to trace the dependence graph (with timing) of distributed AI workloads \textit{post-execution} from real systems.
Selective replay of the ETs~\cite{mystique}, and analysis of the captured metadata (type, size and data volume) can help expose computation, memory, and communication bottlenecks, in turn guiding optimization tools.

\begin{table}[!t]
    \centering
    \caption{Number of Operations within Single Epoch per GPU. (Batch Size: 128 for DeepSeek, 32 for others)}
    \vspace{-1em}
    \resizebox{0.5\textwidth}{!}{
    \begin{tabular}{|p{2cm}|>{\raggedleft\arraybackslash}p{1.8cm}|>{\raggedleft\arraybackslash}p{1.5cm}|>{\raggedleft\arraybackslash}p{1.8cm}|>{\raggedleft\arraybackslash}p{1.8cm}|}
        \hline
        \textbf{Model} & \textbf{\# of Param.} & \textbf{\# of GPU} & \textbf{\# of Comp.} & \textbf{\# of Comm.} \\
        \hline
        \multirow{1}{*}{GPT-3} 
        % & 5B & 8 & 31,492 & 1,267 \\
        % \cline{2-5}
        & 175B & 32 & 156,317 & 30,978 \\
        \hline
        \multirow{1}{*}{LLaMA-3} 
        % & 8B & 8 & 42,144 & 4,227 \\
        % \cline{2-5}
        & 70B & 16 & 164,099 &  38,434 \\
        \hline
        \multirow{1}{*}{Mixtral} 
        % & 8x7B & 8 &  57,846 & 7,335 \\
        % \cline{2-5}
        & 8x22B & 32 &  24,102 &  3,180 \\
        \hline
        DeepSeek-MoE & 16B & 8 & 76,111 & 1,867 \\
        \hline
    \end{tabular}
    }
    \label{tab:numops}
    \vspace{-0.8cm}
\end{table}

While ETs are expected to play a crucial role in AI system design, we believe that ETs alone are insufficient for guiding optimization and DSE for the following reasons:

%\begin{itemize}
\squishlist
  \item \textbf{High cost and limited accessibility:} Generating ETs requires large-scale infrastructure—often hundreds or thousands of GPUs—accessible only to a few hyperscalers.
  %\item \textbf{Data sharing restrictions:} 
  Further, even when ETs are collected, privacy and proprietary constraints may prevent them from being shared broadly with the research community.
%  \item \textbf{Lack of generality and scalability:} 
 \item \textbf{Tied to AI platform:}
  ETs from real-systems are inherently tied to the system they were collected on, with platform-specific software optimizations and hardware bindings baked in. This limits scalability and generality to study larger and diverse systems. As ~\autoref{tab:numops} shows, even a single training epoch of a mid-sized LLM involves tens of thousands of operations per GPU, making trace analysis and scaling a nontrivial task. Efforts to scale ETs~\cite{madmax,chollmservingsim} have focused on mimicking pre-existing system and model behaviors rather than enabling exploration of diverse configurations or novel parallelization strategies.
  \item \textbf{Tied to AI Model.} In the arms race of AI models, there continues to be rapid evolution of LLM architectures—driven by innovations such as MoEs \cite{mixtral, deepseek-moe}, attention mechanism variants \cite{vaswani2023attention, deepseekv3, MQA}, and state space models \cite{mamba}, aimed at improving model accuracy and training efficiency. This can render ETs from real-systems obsolete in a matter of months.
 \squishend

\begin{figure}[t]
    \centering
    \includegraphics[width=8.5cm]{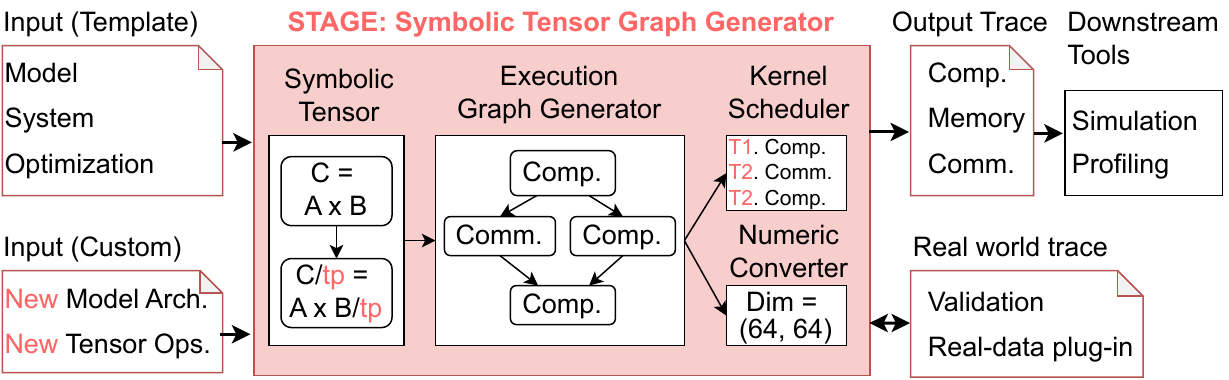}
    \vspace{-8pt}
    \caption{Overview of \sys}
    \label{fig:overview}
    \vspace{-0.75cm}
\end{figure}

These challenges point to a growing need for a more agile framework for distributed AI workload generation that can flexibly adapt to emerging AI model structures and support fast iteration across diverse hardware platform architectures. 
To this end, we present \sys, a novel framework for generating high-fidelity, scalable, and configurable execution graphs (EG) for distributed LLM workloads\footnote{For terminology purposes, we define execution \textit{graphs} to refer to the structure (nodes and dependencies) of the distributed workload, while execution \textit{traces} capture an EG with timing after execution on a real system.}. ~\autoref{fig:overview} shows the overall flow of \sys.
At the front-end, \sys accepts user-defined input workloads in tensor format and supports both predefined model templates and customized inputs for future extensibility.
A key innovation in \sys is the use of a symbolic tensor representation to generate a graph representation that compactly captures distributed ML workloads, enabling scalability by describing their shared computational structure while flexibly incorporating variations in tensor dimensions.
%It then uses symbolic tensor representations to compactly model distributed ML workloads that share a common computational structure but differ in tensor dimensions. 
Our abstraction enables flexible tensor partitioning and systematic support for all major parallelization strategies, as well as their arbitrary combinations—including hypothetical configurations beyond those seen in existing systems.
Once the distributed execution graph is constructed, \sys converts it into a schema that can be integrated with either a downstream simulator or augment a collection of real-system ETs for system optimization/analysis.
%to enable end-to-end simulation.
% \sys further automates the insertion of collective communication operations based on tensor distribution mismatches, allowing simulation of diverse system configurations without reliance on fixed templates or hand-tuned patterns.

% \sys leverages a symbolic tensor representation for distributed ML workloads that allows us to efficiently manage variations across workloads that share the same computational graph structure but differ in specific tensor dimensions.
% \sys enables arbitrary tensor partitioning and supports a wide variety of parallelization strategies, providing researchers with the tools needed to explore the full design space of distributed machine learning systems. 
% \sys also includes an algorithm to identify and encode the specific collective communication operation depending on the tensor distribution.
% Unlike existing methods, \sys is not constrained by predefined system configurations; instead, it offers a flexible, scalable approach to simulate and optimize systems across diverse settings.

The key contributions of this paper are as follows:
\begin{itemize}

    \item \textbf{Symbolic Representation for Diverse AI Model Architectures:} 
    \sys uses symbolic operations to abstract and generalize LLMs, enabling graph-based workload generation across a wide range of model architectures including dense (e.g., LLaMA, GPT), MoE (e.g., DeepSeek, Mixtral), and state-space-style (e.g., Mamba).

    \item \textbf{Comprehensive Parallelism Modeling:} 
    \sys systematically supports all viable combinations of parallelism with a novel producer-consumer-based communication matcher. It enables exhaustive exploration of parallelization configurations for diverse systems. 

    \item \textbf{Compute, Memory, and Network Modeling:}
    \sys accurately models computation, memory, and communication at tensor granularity by analyzing tensor dimensions, lifetimes, and synchronization behavior. This fine-grained modeling enables deeper insights into bottlenecks and resource utilization.

    \item \textbf{Validation with Real-World Traces:}  
    \sys generates execution graphs that model computation, communication, and memory behavior, and we validate their fidelity using real ETs collected from a single GPU to production-scale 128-GPU H100/H200 HGX clusters executing large-scale LLM training workloads.
    %\footnote{Upon acceptance of this paper, we will release these traces as publicly available complementary resources.} 
    %\hw{Should we add this footnote?}. 

    \item \textbf{Scalable and Open Framework:}  
    \sys can synthesize training traces for models on 32K GPUs in less than 30 minutes without compromising accuracy. This enables fast and scalable system analysis. The framework is publicly released to support the research community.

\end{itemize}

%% file: sections/background.tex
\section{Backgrounds}
% \vspace{-2mm}

% \hw{Background is updated but can be more interesting to show some more variants of LLMs. Will update this later}
% \ch{Mayber it has some overlap with motivation 3.2?}

\subsection{Large Language Models (LLMs)}
Large Language Models (LLMs) have scaled to unprecedented levels due to the effectiveness proven by the scaling law \cite{scaling-law}. These models, trained on various datasets, span billions or even trillions of parameters \cite{ZeRO}.
%The size and complexity of LLMs stimulate substantial computational and memory demands and necessitate advanced parallelization techniques and corresponding collective communications to make training feasible and efficient.
For the current LLMs, the decoder-only transformer is adopted by many popular models such as LLaMA \cite{llama} and GPT \cite{gpt-3}. 
The general architecture of the decoder-only transformer consists of repeated blocks, each of which includes a series of main operations such as LayerNorm, Multihead-Attention, MLP.
% as shown in Figure \ref{fig:transformer_with_parallel}.  \ju{Fig 2?}
Within Multihead Attention and MLP layers, the computation is further decomposed into finer-grained operations, such as matrix multiplications (e.g., Linear and MatMul), activation functions (e.g., Softmax and GeLU), and regularization components like Dropout and LayerNorm.

% \TK{This section can be shortened in the interest of space}

% \TK{shouldn't Figure 2 get referred to from here??}

% \vspace{-1.5mm}
%\subsection{Distributed Training with 
\subsection{Multi-dimensional Parallelization Strategies}
\label{sec:parallelization-strategies}
% \vspace{-1.5mm}

% \ju{I think this background section talks too much about parallelization. we may reduce it later.}
% \begin{table}[h!]
%     \label{tab:communication}
%     \caption{Collective Communication Operations for Various Parallelization Strategies}
%     \centering
%     \resizebox{\linewidth}{!}{
%     \begin{tabular}{|l|c|c|}
%         \hline
%         \textbf{Parallelization Strategy} & \textbf{Forward} & \textbf{Backward} \\
%         \hline
%         Data Parallel (DDP) & - & AllReduce \\
%         Fully-Sharded Data Parallel (FSDP) & AllGather & ReduceScatter, AllReduce \\
%         Tensor Parallel (TP) & AllGather & ReduceScatter \\
%         Expert Parallel (MoE) & All-to-All & All-to-All \\
%         Pipeline Parallel (PP) & Send/Receive & Send/Receive \\
%         Sequence Parallel (SP) & ReduceScatter & AllGather \\
%         Context Parallel (CP) & AllGather & ReduceScatter \\
%         \hline
%     \end{tabular}
%     }
% \end{table}

To support large-scale LLM training, sufficient memory is required to store both model weights and input activations, along with adequate computational resources to complete training within a reasonable timeframe. Consequently, the following parallelization strategies are used in practice. 
\squishlist
    \item \textbf{Data Parallelism (DP)}: Splits input data across devices with replicated weights; synchronizes gradients after backward pass~\cite{ZeRO}
    
    \item \textbf{Fully Sharded Data Parallelism (FSDP)}: Shards both input batches and model parameters across devices, reducing memory use but adding communication to gather parameters during training~\cite{zhao2023pytorchfsdp}.

    \item \textbf{Tensor Parallelism (TP)}: Shards model weights across devices while replicating input data; requires AllReduce to exchange activations after each layer.
    
    \item \textbf{Sequence Parallelism (SP)}: Splits input sequences into tokens; complements TP by replacing AllReduce with more efficient AllGather and ReduceScatter.
    
    \item \textbf{Pipeline Parallelism (PP)}: Divides model into stages and pipelines microbatches for concurrent execution across devices.

    \item \textbf{Expert Parallelism (EP)}: For MoE models, uses AllToAll to route tokens to specialized experts after attention layers.
    
    % \item \textbf{Expert Parallelism (EP) and Mixture of Experts (MOE)}: EP, used in Mixture of Experts (MOE) models, partitions specific components of the model into "experts" that specialize in different aspects of the computation. In an MOE setup, only a subset of experts is activated per input, allowing for significant memory and compute savings. EP distributes these experts across devices, while gating mechanisms dynamically determine which experts to activate. This strategy is particularly efficient for LLMs, as it scales well without a proportional increase in memory or compute resources.
    
\squishend

Each strategy introduces unique patterns for computation, memory access and network communications for a large-scale system \cite{ZeRO,megatron-lm}. To maximize efficiency and scalability, LLM frameworks today combine multiple parallelism strategies for training different workloads, such as DP, TP, SP, PP and EP within a single model. 
More details are covered in \autoref{sec:motivation}.

% \ju{Check} Figure~\ref{fig:transformer_with_parallel} shows an brief illustration for 3D parallelization with DP, TP and PP. 
% For DP Region, the collective communication takes place after the backward pass among the partitioned data partitioned group for weight gradients aggregation. For TP, the AllReduce is needed to converge the processed activations tensors before passing to the ensuing stages. For PP, point-to-point communication is needed to send the activations to the next stage. For more details about communication patterns, they will be covered in section \ref{coll_match}.
\vspace{-1mm}
\subsection{Execution Traces}
\vspace{-1mm}
%\TK{TODO - pls add explanation + ideally a small figure showing an ET}
%\ju{Updated}
% \begin{figure}[t]
%     \centering
%     \includegraphics[width=8.5cm]{figures/STAGE_SC_Dependency.pdf}
%     \vspace{-1pt}
%     \caption{An Example of Execution Trace} \label{fig:dependency}
%     \vspace{-15pt}
% \end{figure}

\begin{figure}[t]
    \centering
    \includegraphics[width=0.9\columnwidth]{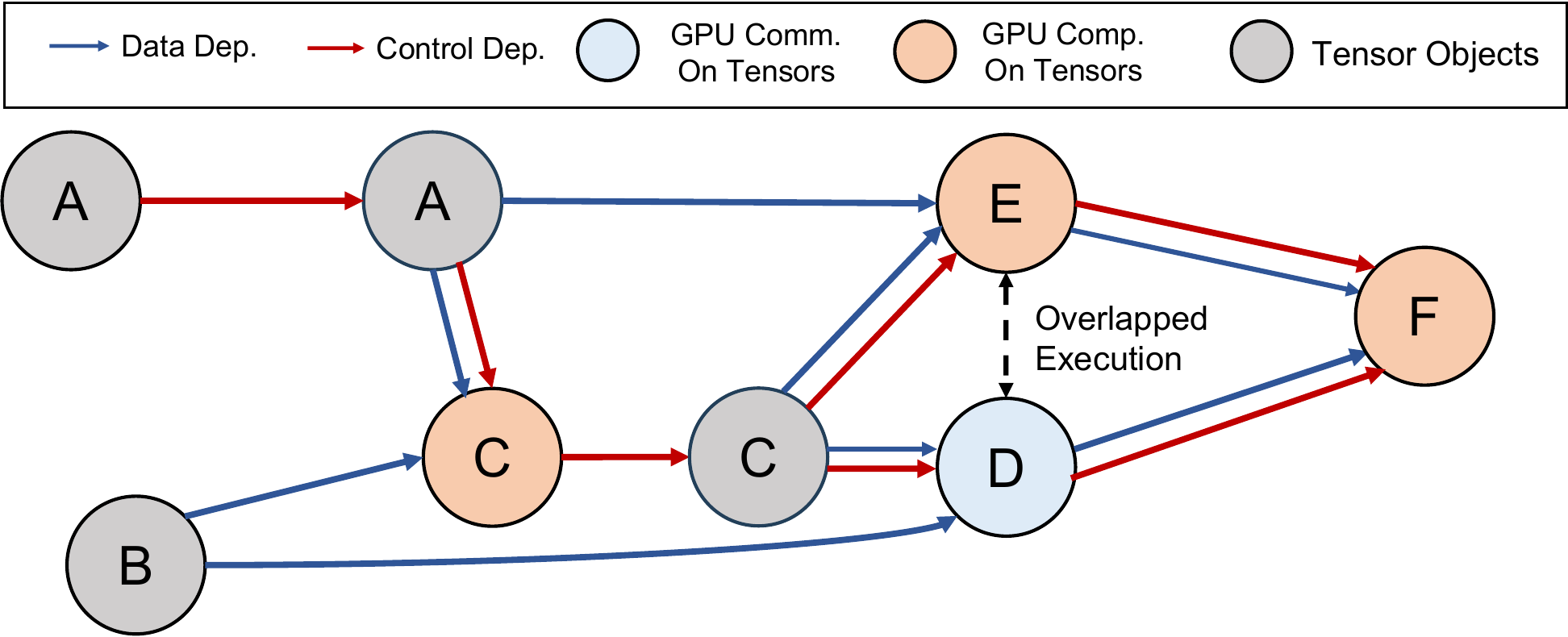}
    \vspace{-8pt}
    \caption{An Example of Execution Graph for GPU Operations.}
    \label{fig:dependency}
    \vspace{-16pt}
\end{figure}

Modern large-scale machine learning systems consist of interleaved compute and communication operations, often with complex execution order and data flow.
%Execution traces (ETs) provide a structured view of these operations by capturing actual runtime behavior. They record the sequence of operations along with metadata such as device type, execution time, and memory usage. Tools like PyTorch profiler~\cite{pytorch-profiler}, Kineto~\cite{kineto}, PARAM~\cite{param}, and Chakra~\cite{chakra-schema} collect these traces at various abstraction levels.
Graph-based formats are particularly useful to represent this behavior. They represent compute and communication operations as nodes, and encode data and control dependencies as edges. This structure enables analysis of execution order, critical paths, operator overlap, and performance bottlenecks. \autoref{fig:dependency} illustrates a simplified execution graph 
%(omitting tensor sizes and operation names) 
depicting GPU computations, communications, and their inter‑operation dependencies. 
Operation specific parameters, such as the tensor size of a GeMM operation, are encoded as attributes within each node.
Following the control and data dependencies from left to right, it reveals which operations must run sequentially and which can execute in parallel across different tensor objects. Labeled tensor sizes are updated following each computation or communication operation between dependent tensors. 
%\jy{Operation specific parameters, such as the tensor size of a GeMM operation, are encoded as attributes within each node.}

Execution traces (ETs) provide a structured view of these operations by capturing actual runtime behavior. They record the 
%sequence of operations 
execution graph, along with additional metadata such as device type, execution time, and memory usage. Tools like PyTorch profiler~\cite{pytorch-profiler}, Kineto~\cite{kineto}, PARAM~\cite{param}, and Chakra~\cite{chakra-schema} collect these traces at various abstraction levels.

%% file: sections/motivation.tex
\vspace{-6mm}
\section{Motivation}
% \vspace{-2mm}
\label{sec:motivation}

%In this section, 
In this section, we will identify a few challenges of the current approach, then summarize the \sys design principle.
\subsection{Challenges}

\textbf{Challenge 1: Limitations of Real-System ETs.}
%\vspace{-1mm}
Access to high-fidelity workloads is crucial for optimization and DSE efforts. However, obtaining real ETs is extremely prohibitive in practice due to the high computational and financial cost of running LLMs over large clusters.
Moreover, data-sharing limitations prevent organizations with resources from making internal ETs publicly available due to security concerns. Furthermore, even if system designers/optimizers have access to real ETs, their properties are inherently tied to the model architecture, parallelism strategy, and underlying hardware platform (e.g., fused operators depend on compiler support within the platform, compute and communication volumes are tied to the system size, and so on). This makes it challenging to extend the ETs or perform DSE for hypothetical future platforms.

\begin{table}[t]
    \centering
    \caption{Categorization of Modern LLM Components and Parallelism Strategies with Native Support in \sys}
    \vspace{-8pt}
    \resizebox{0.5\textwidth}{!}{%
    \begin{tabular}{@{}lll@{}}
        \toprule
        \textbf{Category} & \textbf{Component} & \textbf{Origin Source} \\
        \midrule
        \multirow{4}{*}{\textbf{Attention Mechanism}}   
            & Multi-head & Transformer \cite{vaswani2023attention} \\
            & Group Query Attention & LLaMA \cite{llama}   \\
            & Multi-latent & DeepSeek-V2\cite{deepseek-v2} \\
            & State Space Model & Mamba \cite{mamba}\\
        \midrule
        \multirow{2}{*}{\textbf{Feedforward Network}}   
            & Up-down FFN & GPT \cite{gpt-3} \\
            & Gate-up-down FFN & LLaMA \cite{llama}\\
        \midrule
        \multirow{2}{*}{\textbf{Normalization}}   
            & RMSNorm & LLaMA \cite{llama}    \\
            & Elem-wise Norm & BERT \cite{bert}  \\
        % \toprule
        % \textbf{Category} & \textbf{Component} & \textbf{Originality} \\
        \midrule
        \multirow{2}{*}{\textbf{Mixture-of-Experts}}    
            & MoE & Gshard \cite{lepikhin2020gshard}, Switch Transformer \cite{switch-transformer}  \\
            & MoE with Shared Experts & DeepSeek-MoE \cite{deepseek-moe} \\
        \midrule
        \multirow{5}{*}{\textbf{Sharding Strategy}}
    & Data Parallelism & PyTorch DDP \cite{li2020pytorchdistributed}  \\
    & Tensor Parallelism & Megatron-LM \cite{megatron-lm} \\
    & Pipeline Parallelism & GPipe \cite{huang2019gpipe}, PipeDream \cite{harlap2018pipedream} \\
    & FSDP (ZeRO-3) & DeepSpeed \cite{ZeRO}, PyTorch-FSDP\cite{zhao2023pytorchfsdp} \\
    & Expert Parallelism & Switch Transformer \cite{switch-transformer} \\
        \bottomrule
    \end{tabular}%
    }
    \label{tab:llm_components_categorized}
    \vspace{-0.5cm}
\end{table}

% \subsection{Need a easy way to modeling varies distributed LLM workload}
%\subsection{Challenge 2: Modeling the Architectural and Parallelism Diversity}

%\vspace{-3mm}
\textbf{Challenge 2: Limitations of Graph Capture from ML Frameworks.}
Today ML frameworks enable the capture of pre-execution graphs, such as PyTorch's FX graph representation. Unfortunately, relying purely on ML frameworks to obtain workload representations is also quite restrictive. First, capturing a distributed workload's pre-execution graph still requires access to a real cluster. This also limits the degree of parallelization to the cluster size. Second, and more importantly, the dependency on frameworks limits the generated representations to the set of AI models and parallelization strategies that are supported by the frameworks. This significantly restricts co-design opportunities to existing concepts that already made it to mainstream software stacks\footnote{As an anecdotal example, while FSDP made sense conceptually when it was developed, it could not be evaluated until support was added in PyTorch.}.

\textbf{Challenge 3: Limitations of Manually Describing AI Workloads.}
The wide range of LLM architectures and parallelization strategies makes synthetic modeling of distributed ML workloads particularly challenging. ~\autoref{tab:llm_components_categorized} summarizes commonly used model components and parallel strategies optimized for specific training or inference objectives. 
Previous efforts that have performed synthetic workload generation (such as Calculon~\cite{calculon}, MadMax~\cite{madmax}, SimAI~\cite{simai}) rely on customized templates or analytical first-order equations to describe AI workloads. 
While this approach has demonstrated promising performance analysis of distributed AI systems for realistic workloads, which was their goal, enabling arbitrary AI workload modeling was not.
As a result, their templates are over-optimized for specific target workloads / operators and require deep understanding with the codebase for extensions. 
Moreover, their analytical nature limits the ability to capture realistic system and hardware behaviors, such as compute–communication dependencies.

\textit{Challenge 3.1: Diverse and Rapidly Evolving Model Architectures.} Modern LLM architectures exhibit substantial diversity. %significantly increasing complexity in workload modeling.
For instance, LLaMA incorporates Group Query Attention (GQA) in its attention mechanism alongside a unique three-layer feed-forward network, differing substantially from traditional GPT architectures. Recent models, such as DeepSeek-R1~\cite{deepseekr1}, further increase complexity by employing MoE layers with shared experts and Multi-head Latent Attention (MLA). Additionally, non-transformer architectures, exemplified by Mamba~\cite{mamba}, replace conventional attention with selective state-space models.
Emerging hybrid architectures combining transformer and state-space models, such as Zamba~\cite{zamba} and Jamba~\cite{jamba}, further compound the complexity.

\textit{Challenge 3.2: Complexity and Variability in Parallelization Strategies.} Practical deployments of LLMs often employ a hybrid mix of parallelization strategies (\autoref{sec:parallelization-strategies}) to optimize system performance and resource utilization.

Furthermore, in practice, LLM developers rarely rely on a single model architecture or parallelization strategy. Instead, they often combine multiple design components, resulting in compositional and complex workloads that existing templated / analytical workload generators struggle to systematically represent and evaluate, highlighting a critical gap in current distributed AI workload modeling capabilities.

\subsection{\sys Design Principles}
\label{sec:design_principle1}
\noindent\textbf{Design Principles 1: Decoupling Workload Generation from Simulation.}
Existing synthetic workload generators, such as Calculon, MADMAX, and SimAI, couple workload construction with performance modeling and simulation assumptions. As a result, workload structure and communication behavior are often derived from analytical models that implicitly encode system characteristics.

In contrast, \sys treats the workload as a first-class, standalone artifact. Generation of execution graphs is independent of any specific simulators, performance model, or system topology. Simulation and profiling tools operate strictly downstream of the generated workload representation. By decoupling them, we enable: 1) reuse of the same workload across different simulators, 2) isolation of workload modeling from performance modeling assumptions, and 3) extensibility to future simulation backends, with no redundant work for each simulator backend. In the evaluation, we show that \sys can be adapted to multiple simulators, including AstraSim~\cite{astrasim2}, Genie~\cite{genie}, ScaleSim~\cite{scalesim}, and SimAI~\cite{simai}.

\noindent\textbf{Design Principle 2: Decoupling Workload Semantics from System Realization.}
\label{sec:design_principle2}
Execution traces collected from real systems inevitably embed system-specific realizations, including hardware characteristics, topology constraints, communication library implementations, compiler transformations, and framework-level optimizations. As a result, such traces are tightly bound to a particular system instantiation. Any change in topology, hardware generation, software stack, or parallel runtime typically requires re-collecting traces, making them unsuitable for systematic cross-system exploration.

\sys instead separates workload semantics from system realization. We model distributed workloads at the level of symbolic tensor operations and parallelization semantics, independent of any specific hardware platform, topology, or runtime stack. 

The underlying system is abstracted as a collection of devices connected via a configurable multi-dimensional topology model. This abstraction captures communication structure without embedding hardware-specific execution artifacts. While \sys supports optional modeling of system-specific optimizations, these are treated as extensible components rather than baked-in assumptions.

By maintaining a system-agnostic workload representation, \sys enables a fair comparison across heterogeneous systems, while providing portability to future hardware platforms. Furthermore, it makes a separation between algorithmic operators and hardware-specific implementations, like kernel implementation or collective scheduling, which introduce transferability between systems when running simulations.

%% file: sections/design.tex
\vspace{-2mm}
\section{\sys: Symbolic Tensor Graph Generator} \label{sec:stg_design}
% \vspace{-2mm}

\sys addresses the challenges discussed in \autoref{sec:motivation} by representing LLM workloads based on symbolic abstractions. %\sys enables users to simulate distributed training, explore parallelization strategies, and perform trace-driven optimization—without requiring large-scale infrastructure.
%\sys models workloads and parallelism using symbolic shared tensors.
Users simply specify high-level parameters such as model size and parallelism degrees, and \sys automatically generates execution graphs capturing computation, communication, and memory behavior.
By simplifying workload modeling while preserving key execution characteristics, \sys bridges the gap between synthetic and real-world traces, supporting scalable and systematic design exploration.

\begin{figure}[t]
    \centering
    \includegraphics[width=1.00\linewidth]{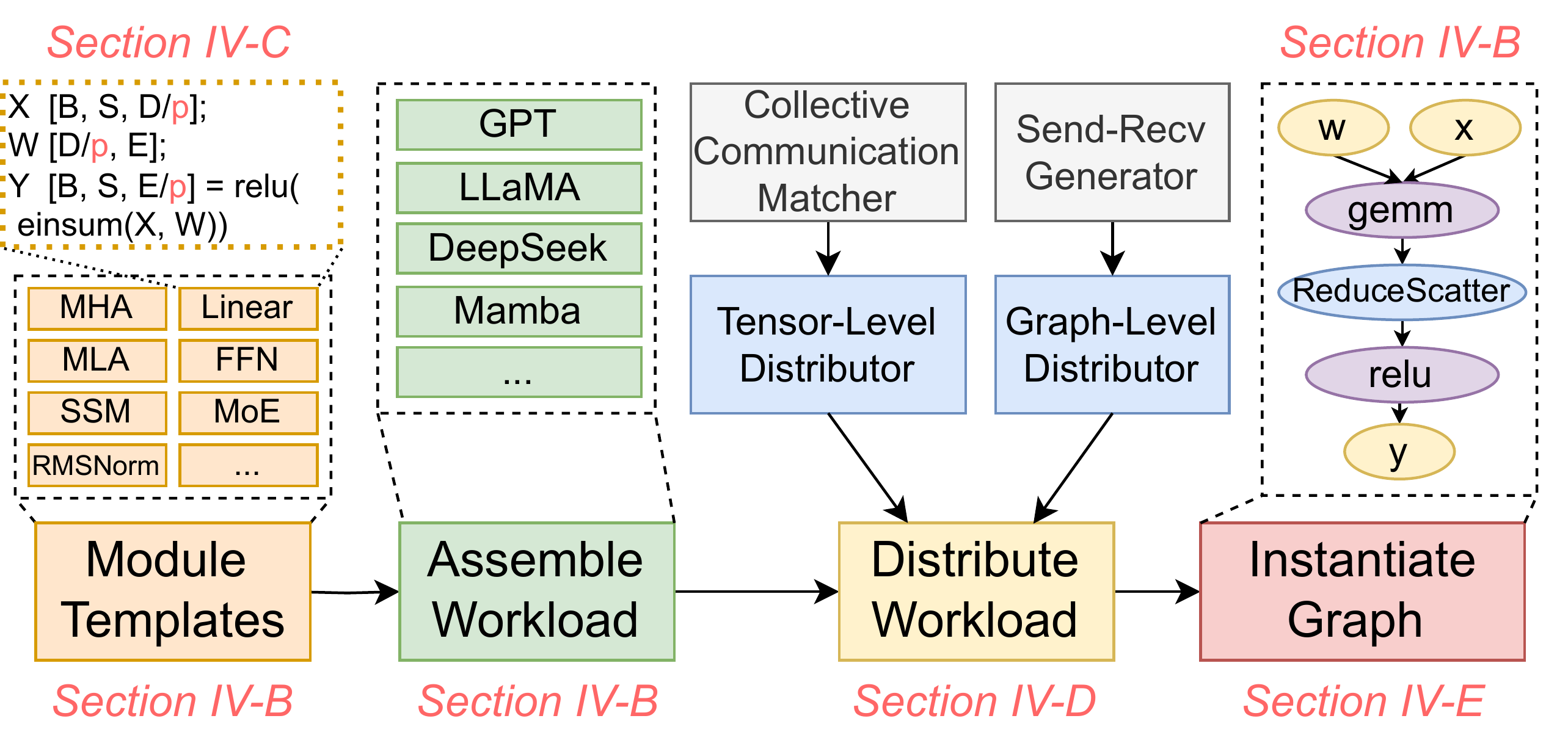}
    \vspace{-20pt}
    \caption{\sys Generation Flow Overview.}
    \label{fig:stg_generation_overview}
    % \vspace{-8mm}
\end{figure}

% \ju{NEED overview figure here}
\vspace{-1mm}
\subsection{\sys Overview} 
\vspace{-1mm}
\autoref{fig:stg_generation_overview} provides a high-level overview of \sys, illustrating its workflow from model specification to workload simulation.
\textcircled{1} Symbolic Tensor Graphs (STG): The flow starts from a set of templates of commonly used modules in LLMs. These modules are integrated in the \sys framework in the format of Symbolic Tensor Graph Intermediate Representation (STG IR).
\textcircled{2} \sys then assembles these modules into the whole model by repeating and connecting each module into a large STG for the whole model.
\textcircled{3} With the assembled model, \sys distributes the workload from a single piece to multiple accelerators by doing tensor-level distribution and graph-level distribution. \sys analyzes the communication required by each parallelization strategy based on the tensor/graph shardings.
\textcircled{4} 
%\jy{\st{Trace Generation: }(All other bullets don't have a 'title:' format in front.)}
Finally, \sys interprets the STG IR and generates a directed acyclic graph (DAG) with explicit operation dependencies for downstream tasks.
%The resulting graph is formatted as a directed acyclic graph (DAG) with explicit operation dependencies.
\vspace{-1mm}
\subsection{Workload definition}
\vspace{-1mm}
\sys is designed to be both simple to use and highly flexible, providing a systematic pipeline from model specification to workload generation.
\subsubsection{\textbf{Input - Model and Module Templates}}
To ensure ease of use, \sys requires only two user inputs: the target model (e.g., GPT, LLaMA) and a selection of module templates (e.g., MHA, FFN, MoE) in STG IR format. This design allows users to generate symbolic tensor graphs without manually specifying the entire model structure. In addition, \sys supports user-defined operations beyond the built-in templates and models, enabling researchers to extend the framework with custom computations. This flexibility is essential for supporting future system-level optimizations and accommodating emerging model architectures.

\subsubsection{\textbf{Output - Execution Graph}}
By default, \sys leverages the \textit{Chakra} schema since it is being standardized by MLCommons~\cite{chakra-wg}.
This schema captures the dependencies between compute and communication tasks, essential for identifying bottlenecks, critical paths, and opportunities for computation-communication overlap during distributed training. Using execution graphs to explicitly model task dependencies is widely adopted in both workload benchmarking \cite{sridharan2023chakra, li2020pytorchdistributed} and workload modeling \cite{duan2023proteus, liang2025lumos}. While Chakra is the default, \sys can be flexibly adapted to other output formats, by introducing suitable translation modules.
\vspace{-2mm}
\subsection{Symbolic Tensor Representation} 
\vspace{-1mm}
\label{subsec:symbolic_representation}
% \ch{Module Templates and Assemble Workload are all done with STG IR and the only difference is subgraph and the whole graph. Mismatched with figure 3, maybe need to better section structure or figure?}
% \ju{Please check: Two subsection has been Merged}

\sys introduces the \textit{Symbolic Tensor Graph (STG)} as an intermediate representation (IR) to model ML workloads. STG abstracts tensor shapes, operations, and distribution strategies symbolically, enabling efficient reuse across workloads that share the same graph structure but differ in dimensions.

\noindent
\textbf{Symbolic Tensor Format: }
Tensors are represented as:
{\abovedisplayskip=1pt
 \belowdisplayskip=1pt
\[
\texttt{Tensor[Shape @ Hidden]}
\]
}
Here, \texttt{Shape} includes symbolic dimensions such as \texttt{Batch} (\texttt{B}) and \texttt{Sequence} (\texttt{S}) and may also contain \textit{partition symbols}, such as data parallelism (\texttt{dp}), tensor parallelism (\texttt{tp}) or sequence parallelism (\texttt{sp}). The optional \textit{Hidden} (\texttt{H}) field denotes partial sums across devices. In the ML context, the hidden dimension typically corresponds to the model’s embedding size or other feature dimensions.
For instance, a tensor \texttt{x} with \texttt{dp} is represented as \texttt{x[B/dp, H]} with the batch dimension sharded across devices.

% \TK{it would be nice to have some example figure showing a STG for say a transformer partitioned across two nodes. Or maybe that *is* what Figure 2 is, in which case I suggest move it here}
% \TK{pls add text in this subsection pointing to this Fig 4 }

% \begin{figure}[t]
%     \centering
%     \includegraphics[width=8.5cm]{figures/Tensor_MHA.pdf}
%     \vspace{-5pt}
%     \caption{Using Symbolic Tensor Representation to Annotate MultiHead Attention with Sequence and Tensor Parallelism.}\label{fig:tensor_mha}
%     \vspace{-5mm}
% \end{figure}

\begin{figure}[t]
    \centering
    \includegraphics[width=8.5cm]{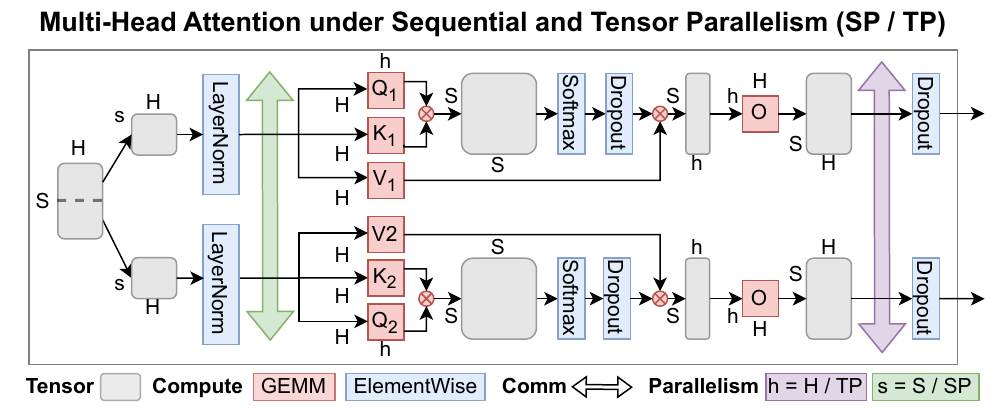}
    \vspace{-5pt}
    \caption{Using Symbolic Tensor Representation to Annotate MultiHead Attention with Sequence and Tensor Parallelism}\label{fig:tensor_mha}
    \vspace{-1em}
    % \vspace{-5mm}
\end{figure}

~\autoref{fig:tensor_mha} illustrates how tensor representations are used to model multihead attention sp and tp. For clarity, the input tensor assumes a batch size of 1, and key intermediate tensors undergoing shape transformations are highlighted in grey.

% \begin{figure}[t]
%     \centering
%     \includegraphics[width=1\linewidth]{figures/stg_graph_ffn1.pdf}
%     \vspace{-20pt}
%     \caption{\ju{Need to remove or update}Example flow of symbolic graph on the first linear and relu step for Feed Forward Network. The first linear operator is represented with op=Einsum[bsd,de->bse] according to the predefined tensor operation term and ReLU is an elementwise operation which has the representation of op=ElementWise1[1](ReLU).}
%     \label{fig:symbolic_tensor_graph}
% \end{figure}

\noindent
\textbf{Tensor-Level Distribution Types: }
\sys defines three symbolic distribution semantics:
\begin{itemize}
    \item \textit{Duplicated}: Full copy on all devices.
    \item \textit{Partition}: Tensor is disjointly sharded across devices along a specific dimension.
    \item \textit{PartialSum}: Each device holds a partial result; reduction is required.
\end{itemize}

% \ju{Matched the color with table 3.}
These distribution types can be composed to represent complex parallelization strategies. For example, the following notation combines \textcolor{blue}{\textit{dp}}, \textcolor{orange}{\textit{sp}}, and \textcolor{red} {\textit{tp}}:
{\abovedisplayskip=2pt
 \belowdisplayskip=2pt
\[
\texttt{X[B/\textcolor{blue}{dp}, S/\textcolor{orange}{sp}, H] @ 1/\textcolor{red}{tp}}
\]
}
Here, \textcolor{blue}{dp} applies to the batch dimension, \textcolor{orange}{sp} to the sequence dimension, and \textcolor{red}{tp} at the end indicates that the tensor is in \textit{PartialSum} form across the hidden dimension.

\noindent
\textbf{Symbolic Operations: }
Operators are expressed using a concise format:
{\abovedisplayskip=2pt
 \belowdisplayskip=2pt
\[
\texttt{output = op[op\_attr](input1, input2, ...)}
\]
}
For example, matrix multiplication is written symbolically as:
{\abovedisplayskip=2pt
 \belowdisplayskip=2pt
\[
\texttt{y = einsum[bm,mn$\rightarrow$bn](x, w)}
\]
}
Here, \texttt{x} has shape \([b, m]\), \texttt{w}  has shape \([m, n]\), and the output \texttt{y}  has shape \([b, n]\). \sys adopts \textit{einsum} to express all tensor multiplications, allowing representation of preserved, reduced, and shared dimensions. By encoding partitioning strategies directly into symbolic tensor shapes, \sys offers a unified abstraction that captures both computation and parallel execution, serving as the foundation for STG construction and downstream simulation.

% In \sys, we model the tensor-level partition with partition symbols by adding these partition symbols into the shape of tensors in the workload. The partition symbols are also symbols, for example, $dp$, $tp$, etc, and we can specify parallel degree by assigning different values to these partition symbols, just like model dimension symbols. 

% Here we show an example of a tensor-level partition with linear layer and data-parallel strategies. In this example, the computation is following einsum notation:
% \[
% \texttt{einsum("bm,mn->bn", x, w)},
% \]
% where $x$, $w$, and $y$ are input tensors, weights, and output tensors, respectively. Table~\ref{tab:tensor_partition_examples} shows how tensor shapes are modified for different partition strategies in a linear layer.

% To apply data-parallelism, the \texttt{Batch} dimension is divided using a partition symbol, $dp$, such that each device handles a portion of the batch. Similarly, tensor-parallelism divides the \texttt{In\_features} dimension using a partition symbol, $tp$. These partition symbols are appended to tensor shapes to represent the modified dimensions. For example, in a tensor-parallel setup, partial sums along the \texttt{In\_features} dimension are distributed across devices. To represent this, a "hidden" field is introduced to track the partial sums, updating the tensor representation from \texttt{tensor[shape]} to \texttt{tensor[shape@hidden]}.

\begin{table}[t]
    \centering
    \caption{Tensor‑Level Distribution in a linear layer. We use [H, 4H] to denote the weight matrix for the up‑projection. }
    \vspace{-5pt}
    \label{tab:tensor_partition_examples}
    {\renewcommand{\arraystretch}{1.2} % More row spacing
    \setlength{\tabcolsep}{4pt} % Adjust column padding
    % \Large % Increase font size here
    \resizebox{7cm}{!}{ % Keep this if you want to enforce 8cm width
    \begin{tabular}{|c|c|}
        \hline
        \textbf{Parallel Strategy} & \textbf{Symbolic Tensor Representation} \\
        \hline
        No Parallel & 
        \begin{tabular}{@{}c@{}}
            \texttt{x[B, H]} \\ 
            \texttt{w[H, 4H]} \\ 
            \texttt{y[B, 4H]}
        \end{tabular} \\
        \hline
        Data-Parallel (\textit{\color{blue}dp}) & 
        \begin{tabular}{@{}c@{}}
            \texttt{x[B/{\color{blue}dp}, H]} \\ 
            \texttt{w[H, 4H]} \\ 
            \texttt{y[B/{\color{blue}dp}, 4H]}
        \end{tabular} \\
        \hline
        Tensor-Parallel (Row) (\textit{\color{red}tp}) & 
        \begin{tabular}{@{}c@{}}
            \texttt{x[B, H/{\color{red}tp} @ 1]} \\ 
            \texttt{w[H/{\color{red}tp}, 4H @ 1]} \\ 
            \texttt{y[B, 4H @ 1/{\color{red}tp}]}
        \end{tabular} \\
        \hline
        Tensor-Parallel (Column) (\textit{\color{red}tp}) & 
        \begin{tabular}{@{}c@{}}
            \texttt{x[B, H]} \\ 
            \texttt{w[H, 4H/{\color{red}tp}]} \\ 
            \texttt{y[B, 4H/{\color{red}tp}]}
        \end{tabular} \\
        \hline
        Fully Sharded Data Parallel (\textit{\color{darkgreen}fsdp}) & 
        \begin{tabular}{@{}c@{}}
            \texttt{x[B/{\color{darkgreen}fsdp}, H]} \\ 
            \texttt{w[H/{\color{darkgreen}fsdp}, 4H]} \\ 
            \texttt{y[B/{\color{darkgreen}fsdp}, 4H]}
        \end{tabular} \\
        \hline
        \begin{tabular}{@{}c@{}}\textbf{Hybrid-Parallel} (\textit{\color{purple}hp})\\
        {(Column Tensor Parallel} \\ 
        {w/ Activation Sharded)}\end{tabular} & 
        \begin{tabular}{@{}c@{}}
            \texttt{x[B/{\color{purple}hp}, H]} \\ 
            \texttt{w[H, 4H/{\color{purple}hp}]} \\ 
            \texttt{y[B, 4H/{\color{purple}hp}]}
        \end{tabular} \\
        \hline
    \end{tabular}
    }
    } % Close resizebox
    \vspace{-5mm}
\end{table}
~\autoref{tab:tensor_partition_examples} enumerates some of the common distributing techniques used for a linear layer, and also a hybrid one to show the flexibility. 
% In the example, dp and tp have independent generation of the output tensors labeled in y. 
% Especially for the case of row tp, we show when partial sum is generated through tp being applied to the In\_features dimension of the tensor w and the number of partial sums matches the way we defined in section \ref{symbloc_tensor_def}.  
With this systematic design, STAGE reduces the need for user intervention while retaining flexibility for defining custom distribution strategies.
\autoref{sec:discussion} discusses how conventional parallel strategies can be defined with \sys. 

\vspace{-1em}
\subsection{Workload Distributor}
\vspace{-1mm}

%\ch{Originally comm matcher, merge coll matcher with graph-level distributor 1. Introduce tensor/graph level distribution. 2 For tensor one, show coll-natcher. 3 For graph one, make it brief(something about inserting send-recv pairs)}
\label{coll_match}
\label{subsec:communication_matcher}
% \subsubsection{Challenges and Necessity of Modeling Collective Communication}

% \hw{Updated this section with more clarity based on Jinsun's feedback, pls take a look at the section with labelled "updated"}

% \hw{updated}
In \sys, distributed workloads are handled using two approaches: (1) \textit{Tensor‑level distribution} where each machine holds shards of a tensor and collaborates to execute a single operator and (2) \textit{Graph‑level distribution} where each machine is responsible for a portion of the computation graph and exchanges data via send–receive pairs when information flows across graph partitions. Depending on the deployed parallelization strategy, \sys employs components to implement either the tensor‑level or graph‑level distribution approach. 

\subsubsection{\textbf{Tensor-level distributor}}

Tensor-level distribution transforms the initial tensor representations with corresponding parallel dimensions, enabling efficient workload distribution across multiple devices. However, these strategies inherently introduce the need for \textit{collective communication}, which is essential for maintaining consistency and data alignment between devices during computation. Accurately modeling these communications is crucial to reflect the real-world behavior of parallel workloads. \sys encodes tensor shardings in the assembled models. Then the tensor-level distributor will apply the corresponding parallel strategies, analyze and generate the collective communications required by the parallel strategies using Collective Communication Matcher.

\begin{figure}[t]
    \centering
    \includegraphics[width=0.95\linewidth]{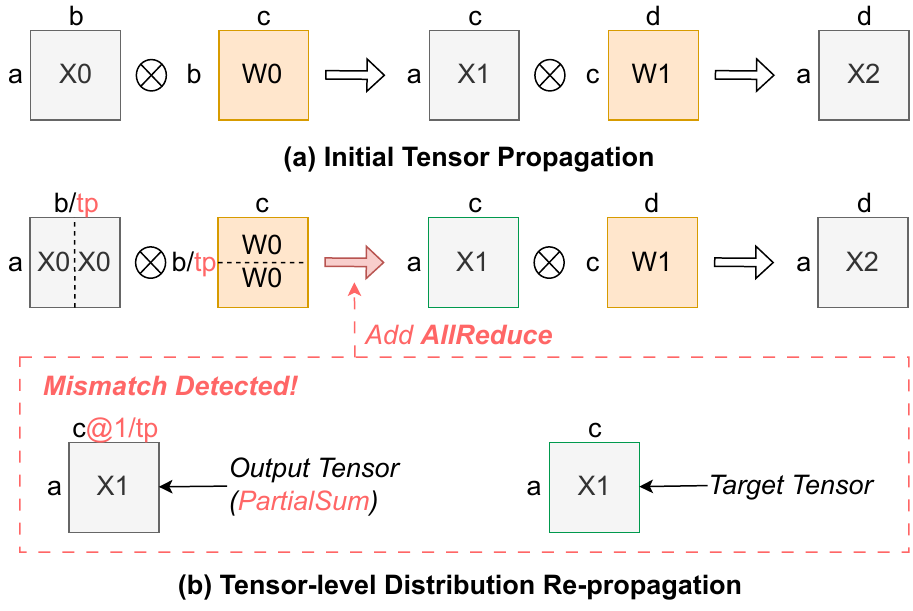}
    \vspace{-4mm}
    \caption{Tensor Distribution Mismatch: After applying tensor-level distribution}
    \vspace{-2em}
    \label{fig:mismatch_tensor_level_distribution}
\end{figure}

In \autoref{fig:mismatch_tensor_level_distribution}, we illustrate how propagating the initial parallelization across an undistributed compute graph—to avoid manually defining every sharded tensor—can create tensor distribution mismatches, where the producer and consumer of a tensor expect different sharding layouts. In \sys, before applying tensor‑level distribution, we first propagate the compute graph to infer each tensor’s shape.
Then we apply the tensor distribution separately for each operator, and repropagate the shape. 
This reveals a distribution mismatch for tensor \texttt{x1} when viewed from the producer and consumer side. From the producer side, \texttt{x1=einsum(x0,w0)} produces an output with layout \texttt{[a, c @ 1/\textcolor{red}{tp}]}. From the consumer side, \texttt{x2=einsum(\texttt{x1}, w1)} expects \texttt{x1} to have layout \texttt{[a, c]}. This mismatch necessitates an AllReduce operation to aggregate the partial sums across distributed tensors.

% This reveals a distribution mismatch of tensor \texttt{x1} from different views. From the producer view of \texttt{\texttt{x1}=einsum(x0, w0)}, it has output of \texttt{[a, c @ 1/\textcolor{red}{tp}]}. Meanwhile from the consumer view of \texttt{x2=einsum(\texttt{x1}, w1)}, it needs of \texttt{[a, c]}. The resulting tensor mismatch necessitates a collective AllReduce operation to aggregate the partial sums across distributed tensors.

\subsubsection{\textbf{Collective Communication Matcher}}
\label{sec:coll_comm_matcher_matching_algo}

\begin{figure}[t]
    \centering
    \includegraphics[width=1\linewidth]{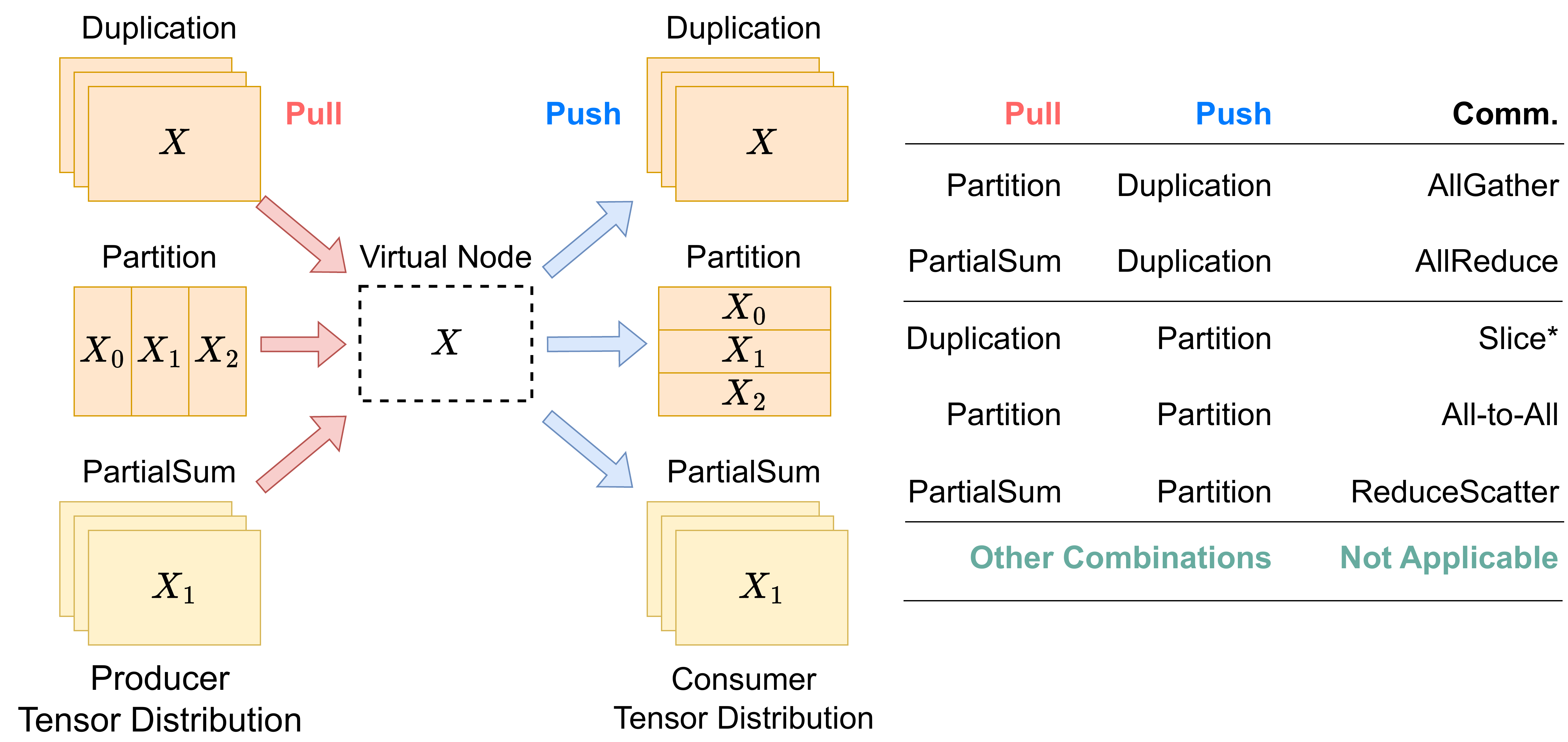}
    \vspace{-6mm}
    \caption{Collective Communication can be divided into two steps: Pull + Push. Note that Slice* is a special case on a single machine.}
    % \vspace{-10pt}
    % \vspace{-6mm}
    \vspace{-2em}
    \label{fig:coll_comm_pull_push}
\end{figure}

To handle the diverse communication requirements arising from various parallelization strategies, \sys uses a communication matcher to systematically identify and encode the required communications. The matcher operates by analyzing the distribution patterns of tensors across devices and matches the appropriate collective communication operations based on the relationship between the distribution from the \textit{producer} that produces the tensor, and the distribution from the \textit{consumer} that consumes this tensor. This producer-consumer model is divided into two conceptual steps \textit{Pull} and \textit{Push} as \autoref{fig:coll_comm_pull_push} shows.

During \textit{Pull}, data is gathered from all devices in the producer distribution to assemble a complete tensor. During \textit{Push}, this tensor is distributed to devices according to the consumer distribution. To bridge these two steps, we introduce a virtual node that serves as an intermediate conceptual connector, enabling a flexible mix‑and‑match between \textit{Pull} and \textit{Push}.

For \textit{Pull}, the process of reconstructing the complete tensor from different distributions is defined as follows:

\begin{itemize}
    \item \textit{Duplicated}: each device already holds a complete copy of the tensor. As a result, the head node does not require communication with other devices, making \textit{No Communication} necessary.
    \item \textit{Partition}: the tensor is divided into shards across devices. The head node gathers all shards from the devices and assembles the complete tensor through a process referred to as \textit{Gather}.
    \item \textit{PartialSum}: while similar to Partition, the aggregation sums the values across devices instead of concatenation. This operation is commonly known as \textit{Reduce}.
\end{itemize}
On the other hand, for \textit{Push}, the process of distributing the tensor to devices is described as follows:
\begin{itemize}
    \item \textit{Duplicated}: the tensor is replicated from the virtual head node to all devices using a \textit{Broadcast} operation.
    \item \textit{Partition}: each device receives its corresponding shard of the tensor through an operation called \textit{Scatter}.
    \item \textit{PartialSum}: Generally not used, as distributing a full tensor as partial sums is uncommon in practice.
\end{itemize}

\begin{table}[t]
    \centering
    \caption{Examples of matched collective communications. The symbolic tensor notations are defined in  \autoref{subsec:communication_matcher}.
    %\TK{font not visible. I suggest using letters like B, S, D for the Batch, Seq, Dmodel etc .. so that font is larger. And then expand these letters in the caption} \hw{Done - Edited the symbols across different parts in the design section to ensure the same symbols}
    }
    %\vspace{-10pt}
    \vspace{-5pt}
    % \resizebox{\linewidth}{!}{
 \begin{tabularx}{\linewidth}{>{\centering\arraybackslash}X|c|>{\centering\arraybackslash}X}
    \hline
         \textbf{Producer Tensor Distribution} & \textbf{Matched Coll-Comm} & \textbf{Consumer Tensor Distribution} \\
         \hline\hline
         $[B/\textcolor{red}{dp},S,H@1/\textcolor{blue}{tp}]$ & \textcolor{blue}{ReduceScatter} & $[B/\textcolor{red}{dp},S,H/\textcolor{blue}{tp}]$\\\hline
         $[B/\textcolor{red}{dp},S,H@1/\textcolor{blue}{tp}]$ & \textcolor{red}{AllToAll} & $[B,S/\textcolor{red}{dp},H@1/\textcolor{blue}{tp}]$\\\hline
         $[B/\textcolor{red}{dp},S,H@1/\textcolor{blue}{tp}]$ & \textcolor{red}{AllGather} & $[B,S,H@1/\textcolor{blue}{tp}]$\\\hline
         $[B/\textcolor{red}{dp},S,H@1/\textcolor{blue}{tp}]$ & \textcolor{blue}{AllReduce} & $[B/\textcolor{red}{dp},S,H]$\\\hline
         $[B/\textcolor{red}{dp},S,H@1/\textcolor{blue}{tp}]$ & \textcolor{blue}{ReduceScatter} + \textcolor{red}{AllToAll} & $[B/\textcolor{blue}{tp},S,H/\textcolor{red}{dp}]$\\\hline
         $[B/\textcolor{red}{dp},S,H@1/\textcolor{blue}{tp}]$ & \textcolor{blue}{AllReduce} + \textcolor{red}{AllGather} & $[B,S,H]$\\\hline
    \end{tabularx}
    \label{tab:coll_comm_example}
    \vspace{-7mm}
\end{table}

% Table \ref{tab:coll_comm_example} provides some examples of the results of collective matching, 

% To summarize the required communication patterns for tensor transformation, we define the collective communication requirements as shown in Table~\ref{tab:coll_comm_example}. By integrating a matching algorithm based on push-pull communication principles, our framework, \sys, identifies additional potential patterns that were previously overlooked but can arise from arbitrary tensor distribution schemes.

To summarize the required communication patterns for tensor transformation, we share examples how the collective communication matcher can be used as shown in ~\autoref{tab:coll_comm_example}. By integrating a matching algorithm based on push-pull communication principles, \sys identifies additional patterns that were previously overlooked but can arise from arbitrary tensor distribution schemes.

\subsubsection{\textbf{Graph-Level Distributor}}
% \ch{Seems very confusion, merge to the workload distributor?}
Graph-level distribution plays a critical role in modeling parallel strategies, particularly pipeline parallelism. Unlike tensor-level distribution, which distributes individual operators, graph-level distribution divides the compute graph into multiple subgraphs and assigns these subgraphs to different devices. 
%, as illustrated in Figure~\ref{fig:shadow_tensor}.
% \sr{fig 7?}\ju{Commented out the ref!}

% Two main challenges arise in graph-level distribution:
% 1) Determining how to partition the graph to reflect parallelization strategies effectively.
% 2) Managing edges that span across subgraphs to ensure correct execution.

% Graph-level partition is another important part of modeling parallel strategies, especially for pipeline parallel. Instead of distributing each operator like the tensor-level partition, a graph-level partition distributes workloads by dividing the compute graph into multiple subgraphs, and deploying subgraphs to devices, as shown in Figure \ref{}.

% Here are mainly two challenges about graph-level partition: 1) how to partition a graph, which reflects the parallel-strategies, and 2) after dividing the graph into subgraphs, how to handle the edges that will cross different subgraphs.

In \sys, a graph distribution can be defined with multiple lists of nodes, where each list contains the nodes within this subgraph. Furthermore, for specific parallel strategies like pipeline parallel, we predefine a rule-based script to partition the workload into multiple stages by evenly dividing models according to their layer.

By partitioning the graph into subgraphs, we create some cross-graph edges, which indicate where the tensor moves from one machine to another. \sys inserts send/recv pairs by identifying the rank of the source and destination nodes on each side of cross-graph edges.

\vspace{-2mm}
\subsection{Graph Instantiation: Symbolic to Numeric Conversion}
\vspace{-1mm}

At the final stage of the \sys pipeline, %symbolic \jy{\st{workloads} graphs (or even just STG)} 
the STG is transformed into fully instantiated execution graphs.
%through \textit{trace instantiation}. 
In this step, \sys replaces symbolic tensor shapes, operations, and communication patterns with concrete numeric values (such as batch size, sequence length, or hidden size), producing a detailed per-node representation of tensor sizes, communication volumes, and operator types.
%The primary mode of instantiation in \sys involves assigning numeric values to symbolic tensor dimensions, such as batch size, sequence length, or hidden size.
Once specified, these values are automatically propagated through the STG, resulting in a complete and consistent execution graph.
%This enables scalable generation of workloads for various model and system configurations without requiring access to real execution traces.

For advanced use cases, \sys also supports plugging in real-system values collected from profiling tools such as PyTorch or Kineto~\cite{pytorchkineto}. These real values can be selectively injected into the symbolic graph to guide the instantiation process, enabling hybrid scenarios where partial traces are extended or scaled. This feature allows users to maintain high fidelity to real system behaviors while still benefiting from the scalability of symbolic modeling.

By separating graph construction from value instantiation, \sys enables scalable, adaptable simulations across a wide design space.

% For network simulators, \sys encodes compute times to enable end-to-end evaluation of system performance. In operator-level simulators, it generates execution graphs without predefined execution times, allowing for detailed analysis of operator behavior. For accelerator-specific simulations, \sys provides workloads that include memory access patterns and detailed tensor dimensions, making it ideal for evaluating accelerator architectures. Additionally, \sys supports memory offloading configurations for system evaluations, such as parameter servers (e.g., ZeRO~\cite{ZeRO}), to explore the potential of future disaggregated memory systems.

% Network Simulators: \sys encodes compute times to enable end-to-end evaluation.
% Operator-Level Simulators: \sys generates execution graphs without predefined execution times, providing flexibility for analyzing operator behavior.
% Accelerator-Specific Simulations: \sys produces workloads that include memory access patterns and detailed tensor dimensions, making it suitable for evaluating accelerator architectures.
% System Evaluations: \sys supports memory offloading configurations for setups like parameter servers (e.g., ZeRO~\cite{zero}) forDES for future disagreegated memory system.

%% file: sections/validation.tex
\vspace{-2mm}
\section{Validation}
\label{sec:validation}
% \vspace{-1mm}

%\TK{need some figure showing STG to Chakra trace to ASTRA-sim ..} \ju{Added in Fig 1.}
% \ch{Add stacked figure of operator type to show majority ops contribute to runtime}
% \ch{For comms, add a table of total comm size between measuerd/synthesized.}

To ensure the fidelity of STAGE-generated workloads, we conducted a comprehensive comparison with real ETs. 
\vspace{-2mm}
\subsection{Methodology}
\vspace{-1mm}
Execution traces were collected from a system equipped with 128 NVIDIA H100 GPUs (SMX5) across 16 servers, each hosting 8 GPUs. The system was configured with NVIDIA NeMo 24.07, CUDA 12.5, and PyTorch 2.5.0. Additionally, each server was powered by dual Intel Sapphire Rapids CPUs (32-core, 2.8 GHz) and DDR5 DRAM.
We modified the NeMo framework to integrate PyTorch’s profiling features and enable Chakra trace collection. This setup employed CUDA Profiling Tools Interface (CUPTI)~\cite{nvidia_cupti} to capture kernel execution timelines and operator-level activity, offering detailed insights into computational and communication operations as well as device-memory usage. 

For validation, we focused on three aspects: 
(1) the peak device-memory usage, 
(2) computation operators and volume, 
(3) communication operators and volume.

\vspace{-2mm}
\subsection{Memory Footprint Validation}
\vspace{-1mm}

For memory‑footprint validation, we fed \sys‑synthesized graphs to the Chakra trace parser provided by MLCommons Chakra~\cite{chakra-wg}. We replayed the graphs via  ASTRA‑Sim~\cite{astraSim}, extending it to track memory usage over the simulation lifetime. Our modifications enable ASTRA‑Sim to utilize tensor metadata (e.g., name, size) from \sys graphs when generating tensor read/write events. These events are then post‑processed to determine each tensor’s lifetime, from creation to last use, assuming garbage collection immediately thereafter.

\autoref{tab:mem_profile} compares per-device peak memory usage across different hardware configurations, models, and parallelization strategies, using both measured traces and \sys‑synthesized execution graphs.
On average, the simulated peak memory usage is about 2GB lower than the measured value. This discrepancy primarily arises from PyTorch’s CUDA initialization, which consumes roughly 1GB of VRAM, and from delays in actual tensor garbage collection. After excluding this initialization overhead, the memory footprint predicted by \sys accounts for approximately 97\% of the measured footprint on average. This inaccuracy is within acceptable bounds for our targeted large‑scale simulations, with the error rate decreasing as model size increases (\autoref{tab:mem_profile}).

\begin{table}[t]
\caption{Peak per-GPU Memory Analysis.}
\vspace{-3mm}
\centering
\resizebox{\columnwidth}{!}
{
\begin{tabular}{l l l l l l}
\hline
Model     & Hardware           & Parallelization & Measured & Synthesized & Error Rate* \\
\hline
GPT-3 5B   & 1 x 8-H200-HGX     & FSDP=8           & \textcolor{blue}{18.1 GB} & \textcolor{orange}{16.1 GB} & 5.5\% \\
GPT-3 5B   & 1 x 8-H200-HGX     & TP=8             & \textcolor{blue}{15.4 GB} & \textcolor{orange}{13.7 GB} & 4.5\% \\
GPT-3 5B   & 1 x 8-H200-HGX     & PP=8             & \textcolor{blue}{17.5 GB} & \textcolor{orange}{15.2 GB} & 7.4\% \\
GPT-3 175B & 4 x 8-H200-HGX    & TP=32            & \textcolor{blue}{118.9 GB} & \textcolor{orange}{115.2 GB} & 2.3\% \\
LLaMA-3 70B & 2 x 8-H200-HGX   & TP=16            & \textcolor{blue}{94.3 GB} & \textcolor{orange}{92.1 GB} & 1.3\% \\
Mixtral8x7B & 8 x 8-H200-HGX   & TP=4, EP=8, PP=4 & \textcolor{blue}{15.8 GB} & \textcolor{orange}{16.07 GB} & 1.7\% \\
Mixtral8x7B & 4 x 8-H200-HGX   & EP=8, PP=4    & \textcolor{blue}{56.8 GB} & \textcolor{orange}{58.55 GB} & 3.0\% \\
DeepSeek-8E & 1 x 8-H200-HGX   & EP=8  & \textcolor{blue}{52.31 GB} & \textcolor{orange}{55.08 GB} & 5.3\% \\
DeepSeek-144E & 4 x 8-H200-HGX   & EP=8, TP=2, DP=2 & \textcolor{blue}{26.6 GB} & \textcolor{orange}{27.4 GB} & 2.9\% \\
\hline
\end{tabular}
}
% };
% \end{tikzpicture}
% \vspace{-5mm}
\label{tab:mem_profile}
\parbox{\linewidth}{\textit{*We remove the CUDA initialization footprint for error estimate.}}
\vspace{-10mm}
\end{table}

\begin{table*}[t]
    \caption{Operator Total Compute Time [ms] (\textcolor{blue}{Measured} / \textcolor{orange}{Synthesized}) 
    }
    % \vspace{-4mm}
    \vspace{-1em}
    \centering
    % \begin{tikzpicture}
    %   % Draw blue rectangle outline around the table contents
    % \node[draw=red, thick, rounded corners=2pt, inner sep=2pt] (tbl) {
    \resizebox{\linewidth}{!}{
    \begin{tabular}{|c|l|l|c|cccc|c|}
    \hline
    \textbf{Model} & \textbf{GPUs} & \textbf{Parallelization} & \textbf{\makecell{Micro Batch\\ / Batch}} & \textbf{GeMM} & \textbf{Attn*} & \textbf{ElementWise} & \textbf{Others} & \textbf{\makecell{Total Error}} \\
    \hline
    \multirow{3}{*}{GPT-3-5B} 
    & 8 & TP=8, w/ SP & 1 / 128 & \textcolor{blue}{2187.0} / \textcolor{orange}{2060.4} & \textcolor{blue}{210.8} / \textcolor{orange}{197.4} & \textcolor{blue}{106.9} / \textcolor{orange}{96.9\ } & \textcolor{blue}{50.7} / \textcolor{orange}{44.6} & \textbf{6.7\%}\\
    \cline{2-9}
    & 8 & PP=8 & 1 / 128 & \textcolor{blue}{1307.9} / \textcolor{orange}{1413.6} & \textcolor{blue}{184.0} / \textcolor{orange}{197.4} & \textcolor{blue}{\ 97.0} / \textcolor{orange}{100.1} & \textcolor{blue}{88.0} / \textcolor{orange}{67.0} & \textbf{5.9\%}\\
    \cline{2-9}
    & 8 & FSDP=8 & 8 / 128 & \textcolor{blue}{1834.1} / \textcolor{orange}{1771.4} & \textcolor{blue}{432.1} / \textcolor{orange}{432.1} & \textcolor{blue}{182.1} / \textcolor{orange}{173.2} & \textcolor{blue}{144.9} / \textcolor{orange}{101.1} & \textbf{4.6\%}\\
    \cline{2-9}
    \hline
    \multirow{2}{*}{GPT-3-175B} 
    & 32 & TP=32 w/ SP & 1 / 128 & \textcolor{blue}{3719.4} / \textcolor{orange}{3690.6} & \textcolor{blue}{444.1} / \textcolor{orange}{444.1} & \textcolor{blue}{165.0} / \textcolor{orange}{173.1} & \textcolor{blue}{164.3} / \textcolor{orange}{109.2} & \textbf{1.7\%}\\
    \cline{2-9}
    & 64 & TP = 4, DP = 2, PP = 8, w/SP & 1 / 128 & \textcolor{blue}{6697.4} / \textcolor{orange}{6685.8} & \textcolor{blue}{266.7} / \textcolor{orange}{266.7} & \textcolor{blue}{\ 61.4} / \textcolor{orange}{116.9} & \textcolor{blue}{224.4} / \textcolor{orange}{155.9} & \textbf{0.3\%}\\
    \cline{2-9}
    \hline
    \multirow{3}{*}{LLaMA-3 70B} 
    & 8 & TP=4, PP=2 & 1 / 32 & \textcolor{blue}{8913.1} / \textcolor{orange}{8775.1} & \textcolor{blue}{4401.4} / \textcolor{orange}{4399.2} & \textcolor{blue}{524.8} / \textcolor{orange}{487.0} & \textcolor{blue}{344.0} / \textcolor{orange}{281.8} & \textbf{1.7\%} \\
    \cline{2-9}
    & 8 & TP=8 & 1 / 128 & \textcolor{blue}{12156.5} / \textcolor{orange}{10993.0} & \textcolor{blue}{5126.3} / \textcolor{orange}{5126.3} & \textcolor{blue}{1896.8} / \textcolor{orange}{1811.0} & \textcolor{blue}{599.8} / \textcolor{orange}{435.3} & \textbf{7.4\%}\\
    \cline{2-9}
    & 16 & TP=4, PP=2, DP=2 & 1 / 128 & \textcolor{blue}{4222.1} / \textcolor{orange}{3635.4} & \textcolor{blue}{2197.4} / \textcolor{orange}{1922.3} & \textcolor{blue}{540.7} / \textcolor{orange}{508.9} & \textcolor{blue}{172.7} / \textcolor{orange}{122.5} & \textbf{14.2\%}\\
    \cline{2-9}
    \hline
    \multirow{2}{*}{Mixtral 8x7} 
    & 128 & TP=4, EP=8, PP=4 & 1 / 128 & \textcolor{blue}{444.7} / \textcolor{orange}{508.8} & \textcolor{blue}{43.6} / \textcolor{orange}{43.6} & \textcolor{blue}{222.8} / \textcolor{orange}{197.0} & \textcolor{blue}{47.7} / \textcolor{orange}{32.5} & \textbf{3.0\%}\\
    \cline{2-9}
    & 32 & EP=8, PP=4 & 1 / 128 & \textcolor{blue}{1688.1} / \textcolor{orange}{1967.3} & \textcolor{blue}{266.4} / \textcolor{orange}{266.4} & \textcolor{blue}{182.1} / \textcolor{orange}{184.1} & \textcolor{blue}{165.1} / \textcolor{orange}{120.8} & \textbf{9.8\%}\\
    \cline{2-9}
    \hline
    \multirow{1}{*}{DeepSeek-MoE 8E} 
    & 8 & EP=8 & 1 / 128 & \textcolor{blue}{1015.3} / \textcolor{orange}{1213.3} & \textcolor{blue}{89.5} / \textcolor{orange}{89.5} & \textcolor{blue}{111.8} / \textcolor{orange}{152.7} & \textcolor{blue}{182.9} / \textcolor{orange}{171.6} & \textbf{15.0\%}\\
    \hline
    \multirow{1}{*}{DeepSeek-MoE 144E}
& 32 & EP=8, TP=2, DP=2 & 1 / 128 & \textcolor{blue}{136.4} / \textcolor{orange}{152.6} & \textcolor{blue}{13.2} / \textcolor{orange}{13.2} & \textcolor{blue}{19.5} / \textcolor{orange}{26.0} & \textcolor{blue}{38.6} / \textcolor{orange}{34.9} & \textbf{8.8\%}\\
    \hline
    % \multicolumn{4}{|r|}{\textbf{Error Per-Category}} & \textbf{0.034} & \textbf{0.021} & \textbf{0.025} & \textbf{0.283} & \textbf{0.038}\\
    % \hline
\end{tabular}
}
% };
% \end{tikzpicture}
\begin{flushleft}
\vspace{-2mm}
{\footnotesize \textit{*Attn here is the fused kernel like flash attention}
}
\vspace{-2mm}
\end{flushleft}
\vspace{-6mm}
\label{tab:op_timing_comparison}
\end{table*}

\begin{table*}[t]
\caption{Communication Breakdown per GPU for a Single Epoch (\textcolor{blue}{Measured} / \textcolor{orange}{Synthesized})
%\TK{TODO - need footnote for the Mixtral case}
}
\vspace{-1em}
% \vspace{-4mm}
\centering
% \begin{tikzpicture}
%     \node[draw=red, thick, rounded corners=2pt, inner sep=2pt] (tbl) {
\resizebox{\linewidth}{!}{
\begin{tabular}{|c|l|l|c|ccccc|c|}
\hline
\textbf{\multirow{2}{*}{Model}} & \textbf{\multirow{2}{*}{GPUs}} & \textbf{\multirow{2}{*}{Parallelization}} & \textbf{\multirow{2}{*}{\makecell{Micro Batch\\ / Batch}}} & \multicolumn{5}{c|}{\textbf{Communication Volume (MB)}} & \textbf{\multirow{2}{*}{Total Error}} \\
\cline{5-9}
& & & & \textbf{Send} & \textbf{Receive} & \textbf{AllReduce} & \textbf{AllGather} & \textbf{ReduceScatter} & \\
\hline
\multirow{3}{*}{GPT-3-5B} 
& 8 & TP=8, w/ SP & 1 / 128 & \textcolor{blue}{0.0} / \textcolor{orange}{0.0} & \textcolor{blue}{0.0} / \textcolor{orange}{0.0} & \textcolor{blue}{1075.1} / \textcolor{orange}{1073.7} & \textcolor{blue}{19730.0} / \textcolor{orange}{19327.3} & \textcolor{blue}{104153.0} / \textcolor{orange}{103079.2} & 0.237\% \\
\cline{2-10}
& 8 & PP=8 & 1 / 128 & \textcolor{blue}{1073.7} / \textcolor{orange}{1073.7} & \textcolor{blue}{1073.7} / \textcolor{orange}{1073.7} & \textcolor{blue}{206.0} / \textcolor{orange}{206.0} & \textcolor{blue}{0.0} / \textcolor{orange}{0.} & \textcolor{blue}{0.0} / \textcolor{orange}{0.0} & 0.000\% \\
\cline{2-10}
& 8 & FSDP=8 & 8 / 128 & \textcolor{blue}{0.0} / \textcolor{orange}{0.0} & \textcolor{blue}{0.0} / \textcolor{orange}{0.0} & \textcolor{blue}{0.0} / \textcolor{orange}{0.0} & \textcolor{blue}{19761.3} / \textcolor{orange}{20401.1} & \textcolor{blue}{80760.9} / \textcolor{orange}{78383.2} & 0.346\% \\

\hline
\multirow{2}{*}{GPT-3-175B} 
& 32 & TP=32, w/ SP & 1 / 128 & \textcolor{blue}{0.0} / \textcolor{orange}{0.0} & \textcolor{blue}{0.0} / \textcolor{orange}{0.0} & \textcolor{blue}{812.6} / \textcolor{orange}{805.3} & \textcolor{blue}{14571.0} / \textcolor{orange}{14495.5} & \textcolor{blue}{310043.0} / \textcolor{orange}{309237.6} & 0.055\% \\
\cline{2-10}
& 64 & TP=4, DP=2, PP=8, w/ SP& 1 / 128 & \textcolor{blue}{13287.6} / \textcolor{orange}{13287.6} & \textcolor{blue}{13287.6} / \textcolor{orange}{13287.6} & \textcolor{blue}{1767.2} / \textcolor{orange}{1384.1} & \textcolor{blue}{29393.7} / \textcolor{orange}{28991.0} & \textcolor{blue}{77309.4} / \textcolor{orange}{77309.4} & 0.043\% \\
\hline
\multirow{3}{*}{LLaMA-3 70B} 
& 8 & TP=4, PP=2 & 1 / 32 & \textcolor{blue}{1073.7} / \textcolor{orange}{1073.7} & \textcolor{blue}{1073.7} / \textcolor{orange}{1073.7} & \textcolor{blue}{0.0} / \textcolor{orange}{0.0} & \textcolor{blue}{104153.0} / \textcolor{orange}{103210.3} & \textcolor{blue}{279172.9} / \textcolor{orange}{275009.0} & 0.265\% \\
\cline{2-10}
& 8 & TP=8 & 1 / 128 & \textcolor{blue}{0.0} / \textcolor{orange}{0.0} & \textcolor{blue}{0.0} / \textcolor{orange}{0.0} & \textcolor{blue}{558315.3} / \textcolor{orange}{587068.3} & \textcolor{blue}{0.0} / \textcolor{orange}{0.0} & \textcolor{blue}{0.0} / \textcolor{orange}{0.0} & 0.985\% \\
\cline{2-10}
& 16 & TP=4, DP=2, PP=2 & 1 / 128 & \textcolor{blue}{2147.5} / \textcolor{orange}{2147.5} & \textcolor{blue}{2147.5} / \textcolor{orange}{2147.5} & \textcolor{blue}{139552.9} / \textcolor{orange}{164257.3} & \textcolor{blue}{0.0} / \textcolor{orange}{0.0} & \textcolor{blue}{0.0} / \textcolor{orange}{0.0} & 2.980\% \\
\hline
\multirow{2}{*}{Mixtral 8x7} 
& 128 & TP=4, EP=8, PP=4 & 1 / 128 & \textcolor{blue}{4496.3} / \textcolor{orange}{4362.1} & \textcolor{blue}{4496.3} / \textcolor{orange}{4362.1} & \textcolor{blue}{0.3} / \textcolor{orange}{16.4}* & \textcolor{blue}{3825.2} / \textcolor{orange}{3590.4} & \textcolor{blue}{13153.3} / \textcolor{orange}{17716.7} & 2.755\% \\
\cline{2-10}
& 32 & EP=8, PP=4 & 1 / 128 & \textcolor{blue}{17935.2} / \textcolor{orange}{19327.4} & \textcolor{blue}{17935.2} / \textcolor{orange}{19327.4} & \textcolor{blue}{0.0} / \textcolor{orange}{0.0} & \textcolor{blue}{0.0} / \textcolor{orange}{0.0} & \textcolor{blue}{0.0} / \textcolor{orange}{0.0} & 1.399\% \\
\hline
\multirow{1}{*}{DeepSeek-MoE 8E}
& 8 & EP=8 & 1 / 128 & \textcolor{blue}{44767.8} / \textcolor{orange}{45097.2} & \textcolor{blue}{43486.5} / \textcolor{orange}{45097.2} & \textcolor{blue}{0.0} / \textcolor{orange}{0.0} & \textcolor{blue}{142.4} / \textcolor{orange}{3758.8}* & \textcolor{blue}{1138.9} / \textcolor{orange}{1138.9} & 0.945\% \\
\hline
\multirow{1}{*}{DeepSeek-MoE 144E}
& 32 & EP=8, TP=2, DP=2 & 1 / 128 & \textcolor{blue}{1981.9} / \textcolor{orange}{1961.1} & \textcolor{blue}{1981.8} / \textcolor{orange}{1961.1} & \textcolor{blue}{8.4} / \textcolor{orange}{16.1} & \textcolor{blue}{1720.7} / \textcolor{orange}{1814.3} & \textcolor{blue}{2954.9} / \textcolor{orange}{3025.8} & 1.501\% \\
\hline
\end{tabular}
}
% };
% \end{tikzpicture}
\parbox{\linewidth}{\footnotesize \textit{*Our trace uses a micro-batch size of 1, so not all experts are activated, which differs from STAGE's default behavior assuming all experts activated and causes mismatches. In practice during real training, larger batches are more common and would activate all experts.}}

\label{tab:comm_breakdown_merged}
\vspace{-5mm}
\end{table*}

\vspace{-2mm}
\subsection{Compute and Communication Validation.}\label{subsec:comp_validation}
\vspace{-1mm}
% We categorize compute into four types: (1) \textit{GeMM}, (2) \textit{Attention}, (3) \textit{ElementWise}, and (4) \textit{Other}.

%\ju{
We validate both the compute and communication components of \sys, as well as the end-to-end runtime. 

\noindent
%\ju{
\textbf{Compute Time Accuracy.}
\label{sec:stage_calibrated_roofline_model}
We estimate operator runtime using a hybrid model that combines benchmark-derived lookup tables with a calibrated roofline model, prioritizing trace-based lookups for observed operators and falling back to a coefficient-calibrated roofline model otherwise. As shown in \autoref{tab:op_timing_comparison}, timing error across workloads ranges from $0.3\%$ to $15.0\%$, averaging $4.25\%$.

\noindent
%\ju{
\textbf{Communication Volume Accuracy.}
\autoref{tab:comm_breakdown_merged} compares the communication volume for each operator type. NCCL implements \textit{AllToAll} by decomposing it into multiple \textit{Send} and \textit{Recv} operations, and Kineto records volume only for these decomposed primitives. To ensure a fair comparison, we similarly decompose \sys’s \textit{AllToAll} volume in the table. The resulting breakdown shows a strong match between real-system traces and \sys-generated workloads, indicating that \sys captures the communication volume accurately enough to model distributed behavior.
%}
\begin{figure}
    \centering
    % \vspace{-1em}
    \includegraphics[width=\linewidth]{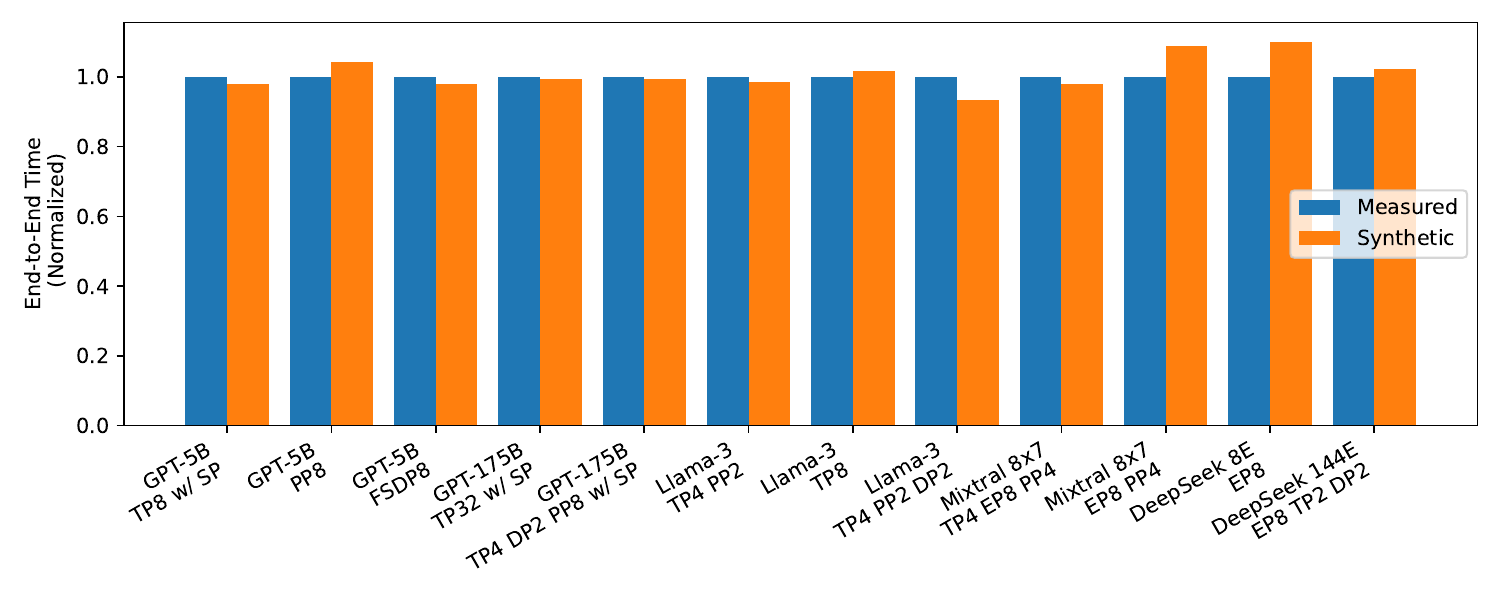}
    \vspace{-8mm}
    \caption{End-to-End Runtime Validation: Measured vs Synthetic}
    % \vspace{-8mm}
    \vspace{-2em}
    \label{fig:e1e_runtime_validation}
\end{figure}
\noindent
%\ch{
\textbf{End-to-End Runtime Accuracy.}
\label{sec:end-to-end-runtime-accuracy}
By combining the calibrated real-system compute model with ASTRA‑Sim \cite{astraSim} simulations for the communication operators and scheduling, we validate the end-to-end runtime of each model instance.
%Because the real-system implementations use suboptimal scheduling for overlapping compute and communication, we disable compute–communication overlap for all instances except the DeepSeek models.
As shown in \autoref{fig:e1e_runtime_validation}, our simulations closely match the real-system performance, achieving an average error of $3.53\%$ \footnote{Interestingly, we observed that many of the real-system ETs (collected using the PyTorch-Chakra stack) lacked overlapping compute and communication given contention for GPU cores by both kernels. For these models, we also disabled compute–communication within the simulator (which natively tries to overlap independent operators as much as possible). \sys achieves this across more models by relying on a graph-based representation with fine-grained operator modeling. Efficiently overlapping compute and communication remains an active research topic today.}. 
Thus, \sys outperforms prior SOTA modeling frameworks—Calculon \cite{calculon} (3.65\% error across only four dense Megatron models) and MADMAX \cite{madmax} (15.34\% error on LLaMA-70B)—in both modeling accuracy and model coverage. Notably, neither Calculon nor MADMAX validate MoE models properly, which introduce more dynamic communication behavior. \textit{This demonstrates that \sys accurately captures real-system workload characteristics and delivers reliable results \textbf{with high-quality performance models}.}
%}
% \vspace{-1em}
\subsection{\sys Modeling Assumptions}
% \vspace{-0.5em}
\label{sec:modeling_assumptions}
In this section, we clarify key modeling assumptions employed by \sys.

\noindent\textbf{MoE Expert Activation}: In Mixture-of-Experts models, each expert has a probability of being activated by a given token. \sys models this behavior using layer-wise expert activation histograms. By default, we assume a uniform distribution. However, users can override this default by specifying custom statistics derived from their own workloads.

\noindent\textbf{Element-wise Kernel Fusion}: \sys assumes that all element-wise kernels are fusible. This may occasionally result in performance estimates that exceed real-system performance, as actual fusion depends on the availability of specific kernel implementations. Because fusion implementation is highly hardware- and system-dependent—which stands in contrast to \sys's general design principles—we do not capture these specific constraints by default. However, \sys provides hook functions allowing users to model custom fusion behaviors if necessary.

\noindent\textbf{Data-Layout}: \sys assumes that all workload data remains on the same device unless offloading is explicitly specified. This assumption holds true for most real-world systems, as keeping data on-device is optimal for performance. While systems may occasionally offload data to the CPU or perform swapping when memory is constrained, \sys does not model this behavior by default because it significantly degrades performance. However, if necessary, users can specify data layout at the granularity of individual tensors.

\noindent\textbf{Memory Allocation/Deallocation}: \sys assumes ideal memory management: memory is allocated only when needed and freed immediately after its last use. Although real systems are not perfectly ideal, this approximation closely matches frameworks such as PyTorch, where allocation is typically lazy to support dynamic computation graphs and deallocation is handled by garbage collection or reference counting.

%% file: sections/evaluation.tex
\vspace{-2mm}
\section{Evaluation}
% \vspace{-2mm}
We present a suite of design space exploration (DSE) case studies to showcase the value of \sys for co-design. Unless specified otherwise, all experiments use the  ASTRA-Sim~\cite{astraSim} simulator to model diverse systems\footnote{ASTRA‑Sim natively supports the Chakra format enabling a proof-of-concept to run \sys-generated workloads. In addition, \sys is also being used by proprietary simulators.}.

\subsection{Impact of Parallelism Strategies}

We demonstrate how \sys can be utilized to explore the complex design space of various parallelization strategies and model optimization techniques and highlight some observations. 
These case studies are not intended to be comprehensive - and can be extended for deeper research enabled by \sys.  

\begin{figure*}[t]
    \centering
    % \begin{tikzpicture}
    %     \node[draw=red, line width=1pt, inner sep=10pt] (box) {
            \begin{minipage}{0.98\textwidth}
                \centering

                \begin{minipage}[t]{0.15\textwidth}
                    \centering
                    \includegraphics[width=\linewidth]{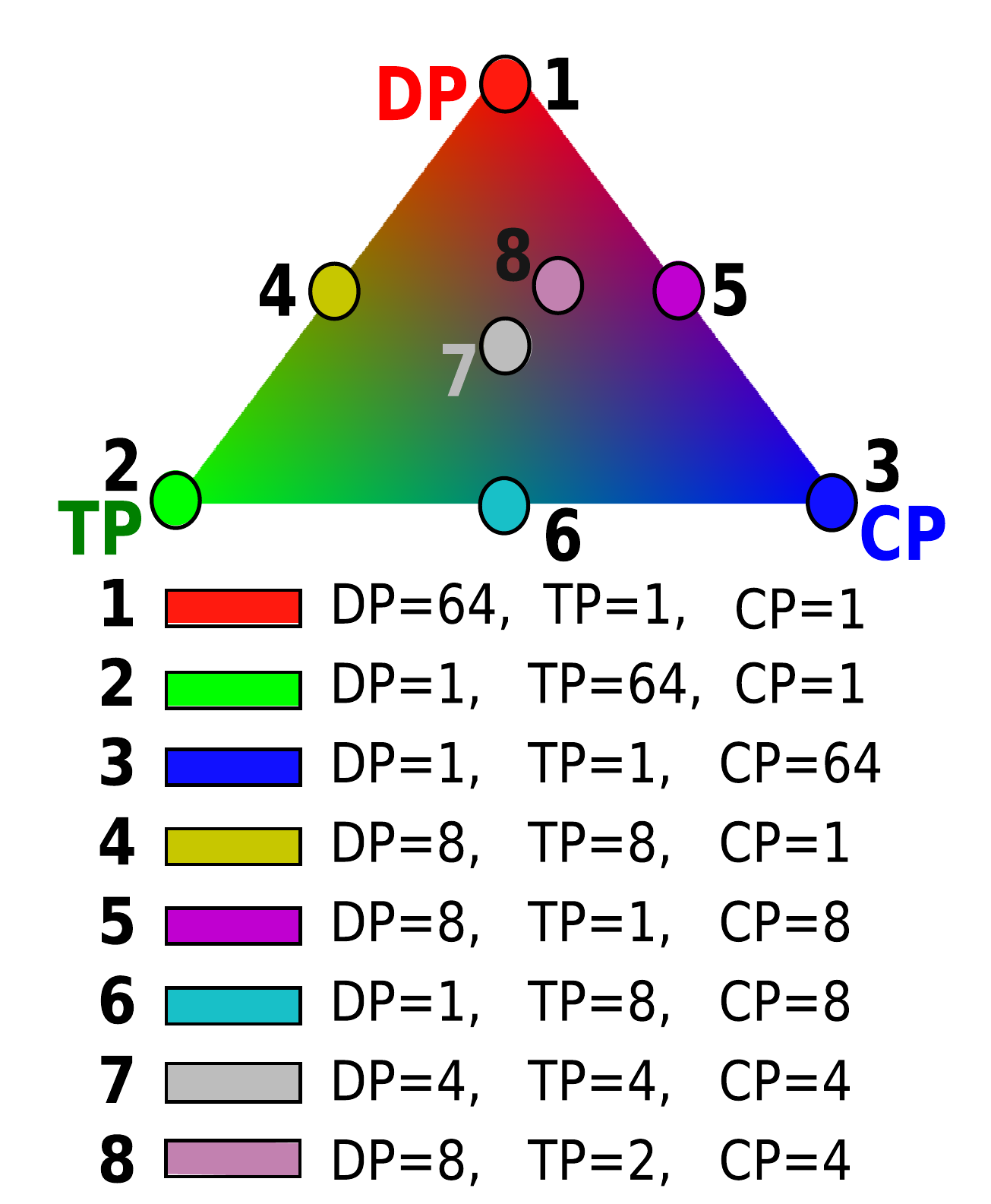}
                    
                    % \vspace{0.1cm}
                    \scriptsize Color-Coded Parallel Strategies
                \end{minipage}
                \hspace{0.01\textwidth}
                \begin{minipage}[t]{0.26\textwidth}
                    \centering
                    \includegraphics[width=\linewidth]{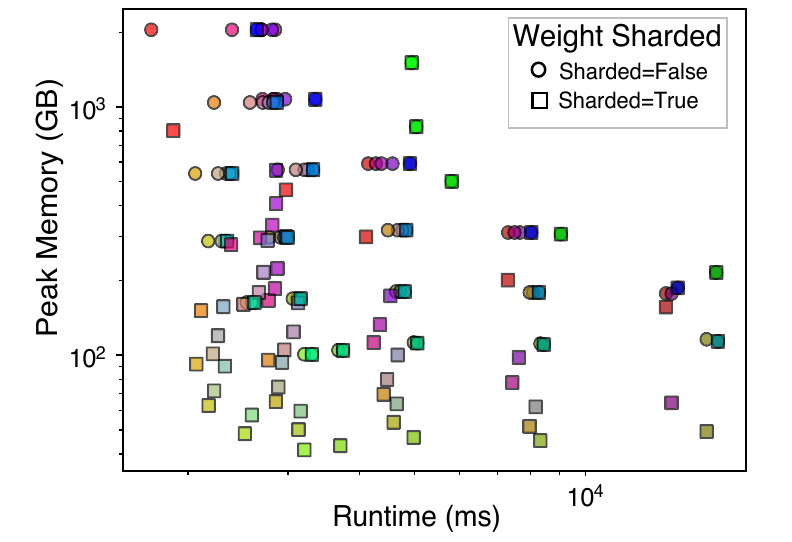}
                    
                    % \vspace{0.1cm}
                    \scriptsize (a) PaLM-540B, Batch=64 @ 64 H100
                \end{minipage}
                \hspace{0.01\textwidth}
                \begin{minipage}[t]{0.26\textwidth}
                    \centering
                    \includegraphics[width=\linewidth]{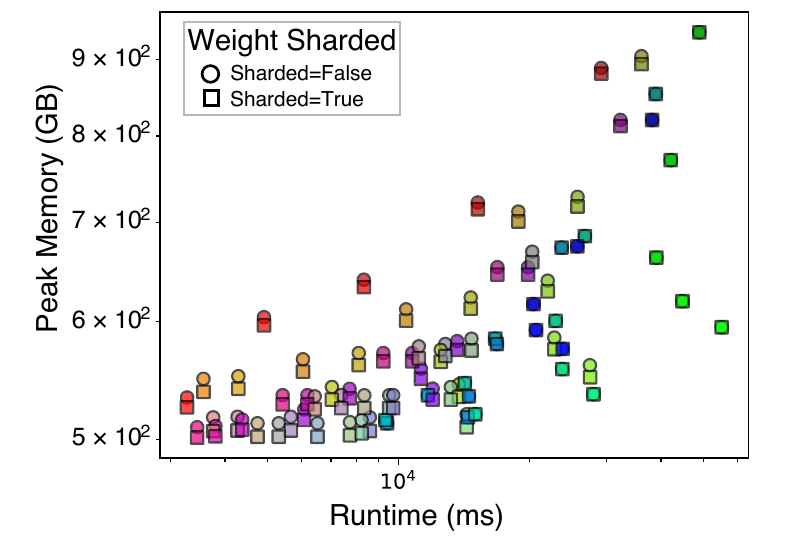}
                    
                    % \vspace{0.1cm}
                    \scriptsize (b) LLaMA-1B, Batch=2048 @ 64 H100
                \end{minipage}
                \hspace{0.01\textwidth}
                \begin{minipage}[t]{0.26\textwidth}
                    \centering
                    \includegraphics[width=\linewidth]{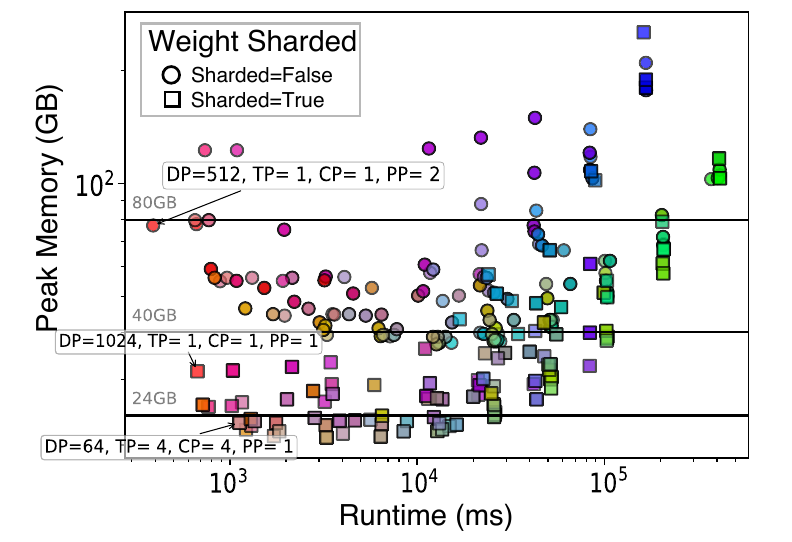}
                    
                    % \vspace{0.1cm}
                    \scriptsize (c) LLaMA-70B, Batch=1024 @ 1024 H100
                \end{minipage}

                \vspace{-0.5em}
                \caption{Peak Memory Usage vs Runtime across configurations.}
                \label{fig:merged_memory_runtime}
            \end{minipage}
    %     };
    % \end{tikzpicture}
    \vspace{-0.5cm}
\end{figure*}

\vspace{0.5em}
\noindent\fbox{%
    \parbox{\columnwidth}{%
        \textit{\textbf{Observation 1.} No single parallelism strategy fits all models; each model and system may prefer different strategies. }%
    }%
}
\vspace{0em}

%\ju{
This observation highlights the need for \sys to generate and evaluate a wide range of parallel strategies.
We simulate a system with 64 H100 GPUs connected in an 8$\times$8 NVLink+IB topology and run DSE on two setups:
(1) a large model with small batch size (PaLM-540B~\cite{palm}, batch = 64),
and (2) a small model with large batch size (LLaMA3.2-1B~\cite{llama3}, batch = 2048).
\autoref{fig:merged_memory_runtime}a and \autoref{fig:merged_memory_runtime}b show peak memory usage versus runtime for both settings.
%}

%\ju{
Data-point shapes indicate whether weight sharding is applied; colors denote DP/TP/CP configurations; and pipeline parallelism (PP) is computed as
$pp = \text{GPUs}/(dp \cdot tp \cdot cp)$, where larger PP values appear as darker points.
%}

%\ju{
For the small-batch, large-model case, two patterns emerge:
(i) higher data parallelism reduces runtime but increases memory usage, while higher tensor parallelism lowers memory but slows execution, reflecting a runtime-memory trade-off;
(ii) weight sharding significantly reduces memory footprint at the cost of a small runtime overhead.
%}

%\ju{
For the large-batch, small-model case, the behavior differs:
(i) memory and runtime no longer form a clear trade-off, as data-parallelism can achieve both low runtime and low memory usage;
(ii) weight sharding has smaller impact because the model contains fewer large parameters worth sharding.
%}

%\ju{
These results show that different models and training regimes favor different parallel strategies.
Real-world scenarios can be even more nuanced:
\autoref{fig:merged_memory_runtime}c shows results for LLaMA-70B (batch = 1024) on a 1024-GPU H100 system, combining characteristics of both earlier cases.
Weight sharding again lowers memory footprint.
The most memory-efficient configurations are mixed parallel strategies-visible as blended-color points near the bottom.
Data parallelism still yields the fastest runtime, but only when memory capacity is sufficient:
high-DP configurations are feasible on both 80\,GB and 40\,GB H100s.
Under a tighter 24\,GB constraint, however, the optimal configuration becomes a composite strategy such as
\textit{(dp = 64, tp = 4, cp = 4, with FSDP)}.

\vspace{0.5em}
\noindent\fbox{%
    \parbox{\columnwidth}{%
        \textit{\textbf{Observation 2.} Optimal parallelization strategies vary with hardware constraints, not just model architecture.}%
    }%
}
\vspace{0em}

\begin{figure}[t]
    \centering
            \begin{minipage}{0.95\linewidth} 
                \centering
                \vspace{-1mm}
                \includegraphics[width=\linewidth]{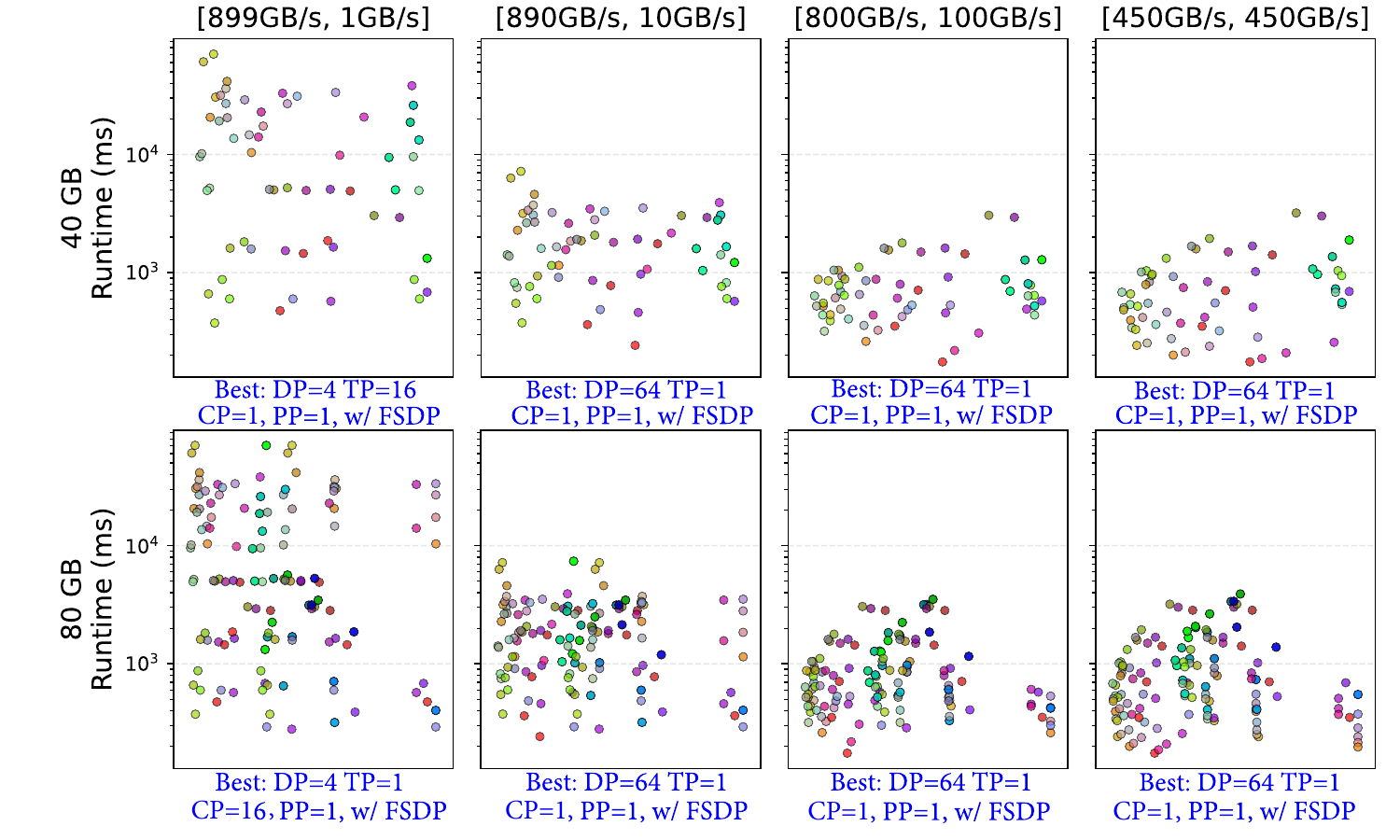}
                
                \vspace{-1EM}
                \caption{Runtime on different HBM capacity and Network Bandwidth. Llama70B @ 64 $\times$ H100}
                \vspace{-1em}
                \label{fig:dse_hw.pdf}
            \end{minipage}
\end{figure}

%\ju{
\autoref{fig:dse_hw.pdf}\footnote{The x-axis represents peak memory usage; however, we omit the specific labels for simplification to focus on demonstrating the runtime.} presents DSE results for various parallel strategies under different hardware configurations. 
We fix the network topology to an 8$\times$8 2D torus and vary both the per-dimension bandwidth distribution and the available HBM capacity, while keeping the total bandwidth per GPU constant across all setups. 
The figure shows that, under certain hardware constraints, the optimal parallel strategy shifts from pure data parallelism to hybrid configurations. 
This underscores the importance of DSE, enabled by \sys, for selecting strategies that best match a given system’s hardware characteristics.
%}

\vspace{0.5em}
\noindent\fbox{%
    \parbox{\columnwidth}{%
        \textit{\textbf{Observation 3.} More communication might not mean more runtime. Communication and compute overlap also matters. }%
    }%
}
\vspace{0em}

\begin{figure}[t]
    \centering
    % \begin{tikzpicture}
    %     % draw=red: Red border
    %     % inner sep=5pt: Padding
    %     \node[draw=red, line width=1.5pt, inner sep=5pt] (box) {
            
            \begin{minipage}{0.95\linewidth}
                \centering
                % Image width is relative to the minipage now
                \includegraphics[width=0.8\linewidth]{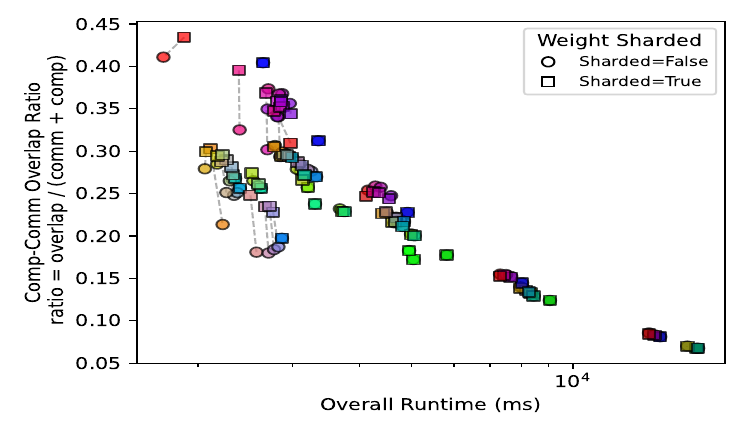}
                
                \vspace{-5mm}
                \caption{Compute-Comms Overlap vs Runtime, PaLM-540B @ 64 H100}
                \vspace{-1.5em}
                \label{fig:comp_comm_overlap}
            \end{minipage}

\end{figure}

%\ju{
From the previous DSE experiments, we observe that FSDP can substantially reduce memory footprint in many cases while having minimal impact on runtime. 
At first glance, this is counterintuitive: FSDP reconstructs weights every time they are used, which should introduce additional communication and increase runtime.
%}

%\ju{
To understand this behavior, \autoref{fig:comp_comm_overlap} visualizes the ratio of compute overlapping with communication versus the overall runtime. 
Dashed lines pair configurations with the same parallel degree, comparing setups with and without weight sharding. 
The figure shows that, in most situations where FSDP has an effect, the amount of overlap increases. 
This suggests that the additional communication introduced by FSDP is largely hidden behind ongoing computation. 
Furthermore, runtime often improves slightly, likely because optimizer states are sharded across nodes, reducing per-node computation.

\begin{figure}[t]
    \centering
    \includegraphics[width=8cm]{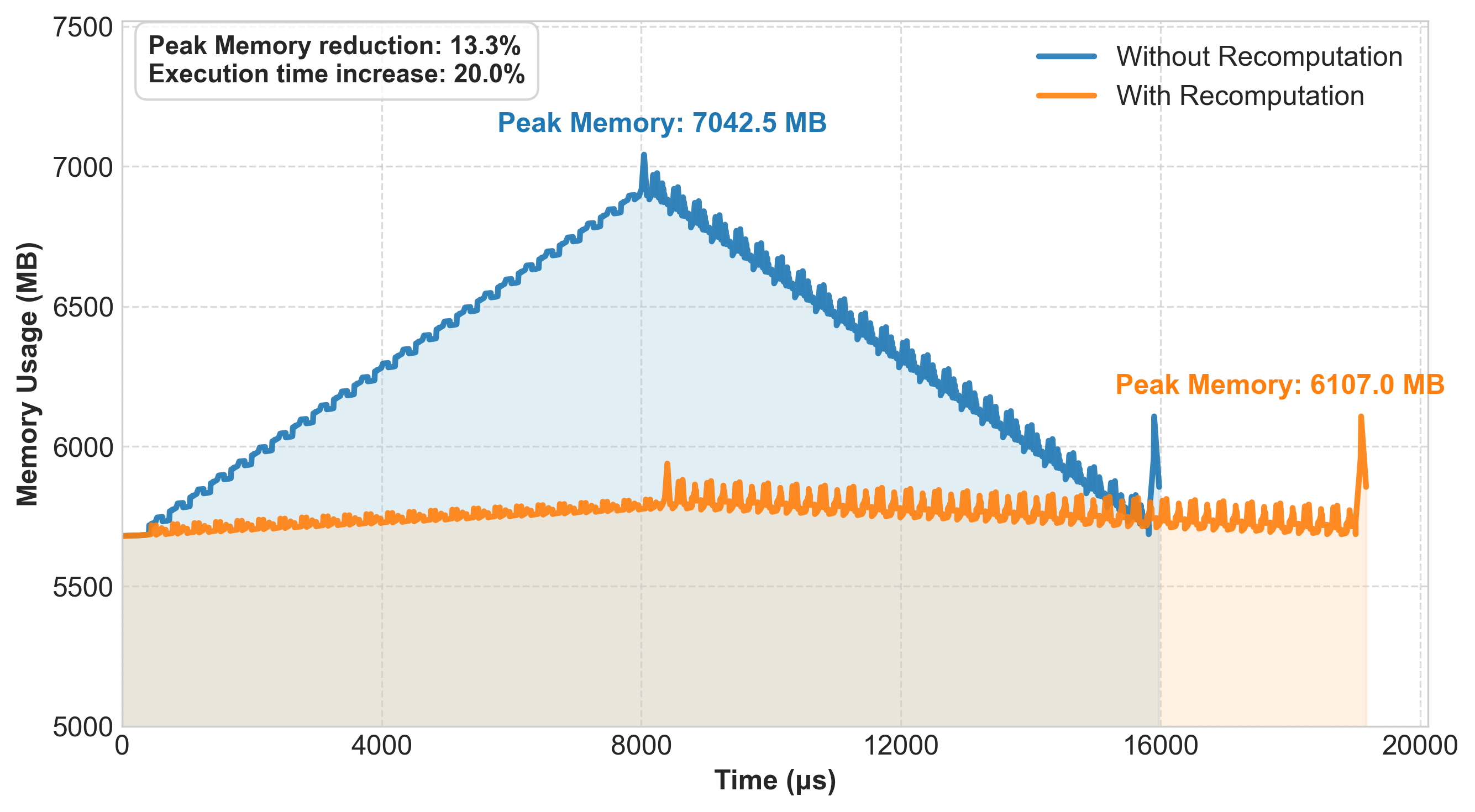}
    % \vspace{-5mm}
    \caption{Memory w/ and w/o Activation Recomputation}
    \label{fig:recompute_comparison}
        \vspace{-5mm}
\end{figure}

\vspace{0.5em}
\noindent\fbox{%
    \parbox{\columnwidth}{%
        \textit{\textbf{Observation 4.} Activation Recompute is a promising trade-off.}%
    }%
}
\vspace{0.2em}
For a given model and parallel strategy, STAGE can generate workloads both with and without activation recomputation~\cite{llama3, korthikanti2022reduceact}.
For LLaMA-7B with batch = 1, TP = 8, and SP, \autoref{fig:recompute_comparison} shows that activation recomputation lowers peak memory usage while increasing runtime. This reduction in memory footprint can enable larger data-parallel degrees, which may be beneficial based on the earlier analysis.
%}

\noindent\fbox{%
    \parbox{\columnwidth}{%
        \textit{\textbf{Guideline. } Choosing parallelism strategies in practice.}%
    }%
}
\vspace{0.2em}
In the workloads and system settings studied in \autoref{fig:merged_memory_runtime}, higher DP often delivers the lowest runtime among feasible configurations. However, different models might lead to different behavior in memory usage. 
For large models, this exposes a clear runtime-memory trade-off, where higher DP improves runtime but can exceed memory capacity, requiring hybrid DP/TP configurations.
For small models, high DP often provides both low runtime and low memory usage, so there is little trade-off. 
Most real deployments lie between these extremes, where a practical default is therefore,  starting from the largest memory-feasible DP, then increasing TP/CP only as needed to satisfy per-device memory limits and utilization constraints. 
Weight sharding and activation recomputation usually improve memory feasibility, possibly enabling faster configurations. However, their runtime impact depends on the available communication and compute resources of the target system. Therefore simulation-based exploration is still needed to choose the best parallel strategies.
% }

\vspace{-2mm}
\subsection{Workload Scalability Studies with \sys}
\label{sec:evaluation_workload_scalability}
\vspace{-2mm}
In this section, we demonstrate how \sys supports workload-level scalability analysis. We study how communication behavior changes as parallelization strategies vary under a fixed system configuration. This complements the next section, which examines \emph{system-level scalability} by scaling the system configuration to support larger models.

\begin{figure}
    \centering
    \includegraphics[width=0.95\linewidth]{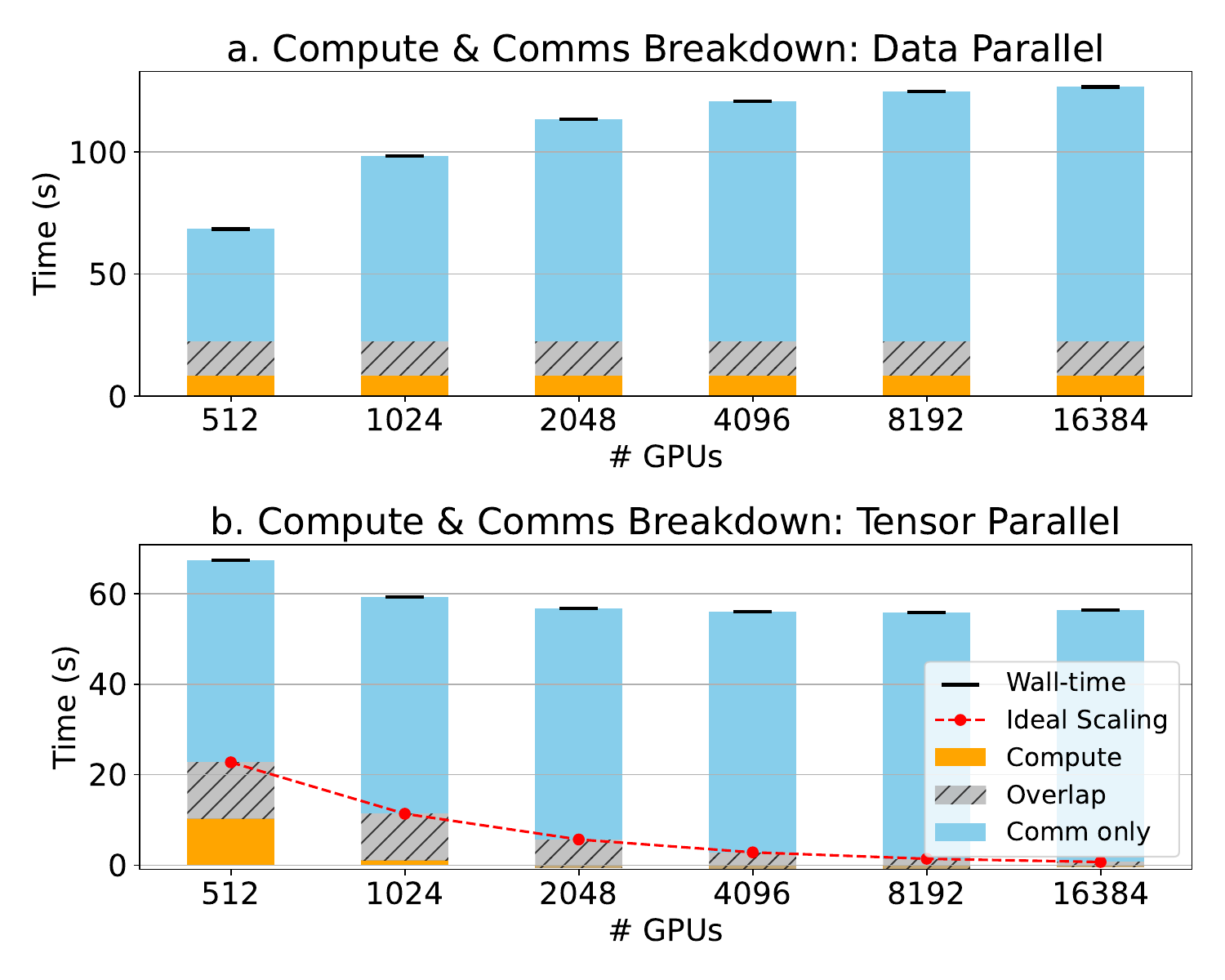}
    \vspace{-5mm}
    \caption{Time Breakdown for Scalability Studies}
    \label{fig:system_scaling}
        \vspace{-6mm}
\end{figure}

\noindent
\textbf{Target System Setup:}
We simulate a large-scale system built from NVIDIA DGX nodes, each with 8 H100 GPUs connected via NVLink. Sixteen nodes form a pod connected by a local ring, and multiple pods are linked through a global ring. Our experiments cover system sizes from 512 to 16K GPUs.\footnote{To support these scales without running out of memory, we extended the ASTRA-sim workload feeder with disk-backed trace processing and caching.}

% \TK{@Changhai - pls make sure the text regarding assumption of weak / strong scaling etc for both is correct}

\noindent
\textbf{Scaling Data Parallelism.} We analyze how data parallelism impacts performance with a fixed 
microbatch
% \jy{"microbatch" in p.3}
size per GPU (i.e., \textit{weak scaling}), simulating scenarios where batch size is scaled out for more stable convergence and improved training. Using LLaMA-70B with PP=4, we keep the per-GPU batch size at 8 and scale DP. ~\autoref{fig:system_scaling}a presents the breakdown of computation and communication times. As expected, compute time stays constant due to fixed per-device batch size and minor contribution to overall runtime. With scaling, communication overhead increases and finally converges, matching the behavior of data-parallel ring all-reduce.

\noindent
\textbf{Fixed Model, Scaling Tensor Parallelism.} We evaluate tensor parallelism’s impact on training on same PaLM-540B~\cite{palm} (DP=4, CP=4, micro-batch=256), scaling TP w/ SP from 4 to 1024 GPUs to simulate faster training (i.e., \textit{strong scaling}). As shown in \autoref{fig:system_scaling}b, compute time decreases with more GPUs, while communication time remains nearly constant. This is because tensor parallelism with sequence parallelism mainly uses ring reduce-scatter. As the TP degree grows, group size and communication steps increase, but per-device communication volume decreases, keeping total communication time stable. Furthermore, compute time reductions taper off at scale, causing scalability to plateau—especially beyond 2048 GPUs.
% \vspace{-8mm}
\vspace{-1.5em}
\subsection{System Scalability Studies with \sys}
% \vspace{-2mm}
\begin{figure}
    \centering
    \includegraphics[width=\linewidth]{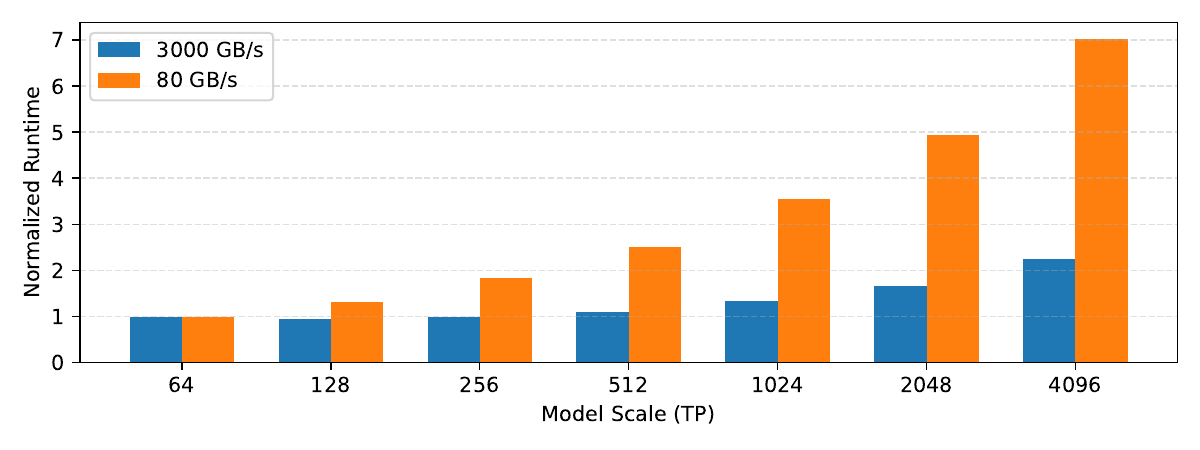}
    \vspace{-8mm}
    \caption{Normalized runtime vs. system scale for high and low bandwidth system}
    \vspace{-8mm}
    \label{fig:hi-lo-bw-compare}
\end{figure}
\begin{figure}
\vspace{-2em}
    \centering
    \includegraphics[width=\linewidth]{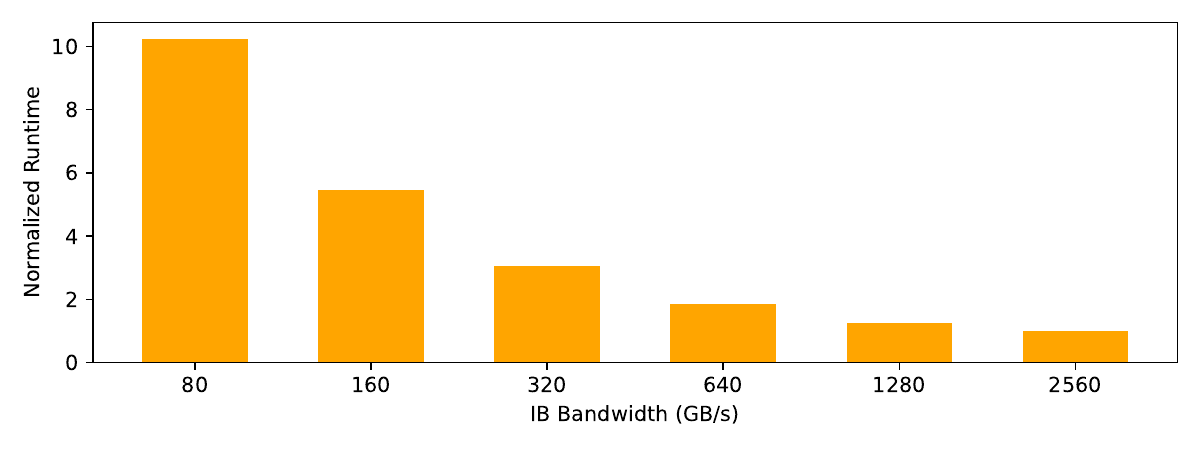}
    \vspace{-8mm}
    \caption{Normalized runtime vs. system bandwidth}
    \vspace{-6mm}
    \label{fig:bw_scan}
\end{figure}
Similar to \autoref{sec:evaluation_workload_scalability}, we evaluate the effect of system properties on model scaling. We keep per-GPU compute and model size constant while scaling up the system. Starting from a LLaMA-8B model on 64 GPUs, we increase the model size by proportionally expanding TP. We then investigate how network bandwidth influences scaling, using the same H100-DGX nodes (8 H100 per node connected via NVLink) linked through Infiniband~\cite{infiniband} switches with varying bandwidths.

\autoref{fig:hi-lo-bw-compare} shows the normalized runtime on systems with high (3000 GB/s) and low (80 GB/s) Infiniband bandwidths. For the high-bandwidth system, runtime remains largely unaffected because scaling out mainly increases communication while compute and I/O per GPU stay constant, making it more suitable for very large-scale systems. In contrast, for the low-bandwidth system, communication overhead grows rapidly with scale, which limits the training of very large models.

% \TK{@Changhai - are you wokring on this?}
Furthermore, \autoref{fig:bw_scan} shows the impact of bandwidth on model performance under the same TP configuration. As bandwidth increases, runtime decreases, but the benefit tapers off once bandwidth becomes sufficiently large.

In conclusion, larger network bandwidth helps accelerate large-model training. However, when the model scale is limited, there exists a bandwidth sweet spot that offers near-optimal performance while maintaining a good cost–performance trade-off.

\subsection{Real-world Application Study: DeepSeek-R1 Inference System}
\vspace{-2mm}
\label{sec:dsr1_inference}
In this section, we demonstrate that \sys can model real-world LLM workloads using the DeepSeek-R1 inference architecture\cite{deepseekV3R1_inference_system_2025}, which separates prefilling and decoding. These two phases exhibit distinct performance characteristics and require different parallelism configurations.

We evaluate a system with 144 GPUs, partitioned into either 4 clusters of 36 GPUs, 2 clusters of 72 GPUs, or a single 144-GPU cluster. Within each cluster, we use expert parallelism for MoE layers and data parallelism for the remaining layers. The total batch size across clusters is fixed at 2048. The resulting decoding and prefilling performance under different EP degrees is shown in ~\autoref{tab:decode_prefill_perf}.

Prefilling generally prefers lower EP degrees because it operates on long sequences and large batches, making it compute-bound while reducing all-to-all overhead. Conversely, decoding handles short sequences per step and benefits from larger effective batch sizes, thus achieving higher throughput with larger clusters and higher EP degrees.

% \shephered{
\subsection{Architectural-Oriented Case Study: HBM/Communication Bandwidth Distribution under Fixed Budget}
We demonstrate how \sys supports architectural design exploration by studying bandwidth partitioning under a fixed off-chip bandwidth budget per accelerator. The total budget is divided between HBM and scale-up interconnect bandwidth. Using \sys-generated workloads, we sweep HBM bandwidth shares and assign the remaining budget to interconnects.

\begin{figure}[t]
    \centering
    % \fcolorbox{blue}{white}{
        \includegraphics[width=0.96\linewidth]{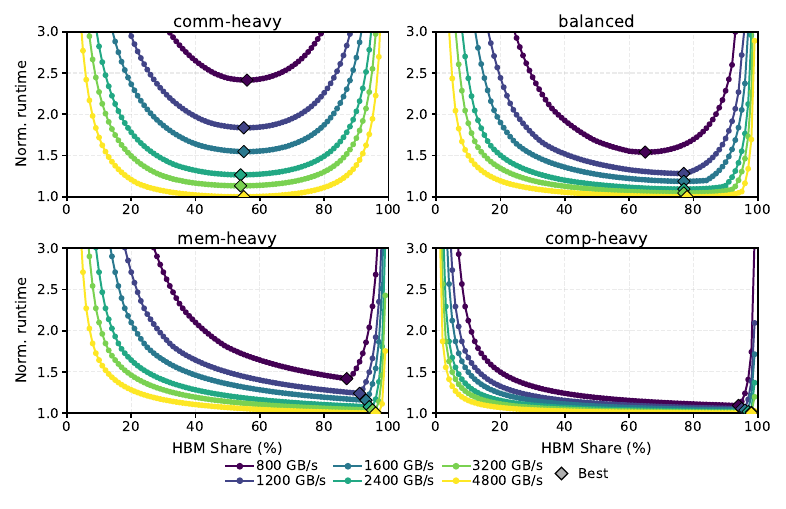}
    % }
    \vspace{-1.5em}
    \caption{Optimal HBM bandwidth share under different workloads and total bandwidth budget 
    % \hw{Would it be better to put the unit on the plot (I think it should be GB/s in here?)? Also is this interconnect bandwidth for scale-up or scale-out?}
    }
    \vspace{-1em}
    \label{fig:fixed_sum_hbm_comm_split}
\end{figure}

\autoref{fig:fixed_sum_hbm_comm_split} reports normalized runtime across multiple total bandwidth budgets for four workload classes: communication-heavy, balanced, memory-heavy, and compute-heavy.

The results highlight three key insights. First, bandwidth provisioning should be workload-aware, as different workloads may prefer different bandwidth distributions. Second, the preferred split is primarily determined by workload characteristics and is relatively insensitive to the total bandwidth budget: while changing the total budget affects overall runtime, the preferred split remains stable. Third, the optimal HBM share consistently exceeds 50\%, as most interconnect traffic originates from HBM, while direct communication from on-chip memory is rare due to limited on-chip capacity and ML workload compute patterns.
% }

\begin{table}[t]
\centering
\caption{Decode and Prefilling performance across different EP configurations for DeepSeek-R1.
% \TK{this is for deepseek right? caption should say that}
}
\vspace{-1em}
\label{tab:decode_prefill_perf}
%\begin{tikzpicture}
      % Draw blue rectangle outline around the table contents
%    \node[draw=red, thick, rounded corners=2pt, inner sep=2pt] (tbl) {
\resizebox{0.5\textwidth}{!}{%
\begin{tabular}{|l|ccc|ccc|}
\Xhline{1pt}
\textbf{Phase} & \multicolumn{3}{c|}{\textbf{Decode}} & \multicolumn{3}{c|}{\textbf{Prefilling}} \\
\Xhline{1pt}
\textbf{Cluster Size} & 36 & 72 & 144 & 36 & 72 & 144 \\
\Xhline{1pt}
\textbf{Batch Size} & 512 & 1024 & 2048 & 512 & 1024 & 2048 \\
\hline
\textbf{\#\ Tokens} & 512 & 1024 & 2048 & 524{,}288 & 1{,}048{,}576 & 2{,}097{,}152 \\
\hline
\textbf{Step Time (ms)} & 227.483 & 187.483 & 163.681 & 2051.994 & 2866.145 & 3723.360 \\
\hline
\textbf{Throughput*} & 62.520 & 75.859 & 86.890 & 7097.270 & 5081.235 & 3911.401 \\
\hline
\end{tabular}%
}
%};
%\end{tikzpicture}
% \vspace{1mm}
{\footnotesize \textit{*Throughput here is number of tokens processed per second, per GPU.}}
\vspace{-2.5em}
\end{table}
%}
% \vspace{-0.8em}
\subsection{\sys for Different Simulators and Architectures}
\vspace{-0.2em}
% \ch{We need to show proof that we run multiple simulators, but their results are not really meaningful here, so how to demonstrate?}

While our primary evaluation leverages AstraSim with the Chakra format, focusing primarily on H100/200 systems, \sys is architecturally decoupled from any particular simulator or workload schema. The generated execution graphs serve as simulator-agnostic artifacts that can be consumed by diverse performance modeling frameworks.

To validate this portability, we integrate \sys with multiple simulators, including SimAI~\cite{simai} from Alibaba, ScaleSim~\cite{scalesim} from Georgia Tech, and Genie~\cite{genie} from HPE, using lightweight translation layers without modifying workload semantics.
Each simulator models different aspects of AI systems at high fidelity: SimAI captures NVIDIA NCCL and NVLink semantics, ScaleSim models TPU-like compute arrays, and Genie emulates AI traffic over real physical network fabrics such as RDMA.

In \autoref{tab:stage_multiple_simulation}, we present the results obtained across the three different simulators and setups\footnote{Note that due to differences in modeling scope, target systems, and execution environments across backends, the reported runtimes by each simulator do not encompass all workload components, and so comparisons across the different simulators is not the focus of this experiment.}.
For SimAI, we compare $8\times$H100 and $8\times$H200 systems with NVLink interconnects; for ScaleSim, we %evaluate  GEMM 
contrast compute times across TPUv5e and TPUv4 configurations; for Genie, an RDMA traffic emulator we study the runtime for a 100Gbps versus 400Gbps InfiniBand network with a single-layer switch. 

These experiments highlight that \sys-generated workloads can be instantiated and executed across heterogeneous simulation environments without redesigning workload logic, underscoring the value of decoupling workload generation from simulation (\autoref{sec:design_principle1}. Furthermore, we report the Lines-of-Code (LoC) required to adapt \sys to each simulator backend. For most simulators, fewer than one hundred lines of translation code are required, demonstrating that \sys maintains a shared workload generation pipeline while isolating simulator-specific graph instantiation logic.

\begin{table}[t]
    \centering
    % \begin{tikzpicture}
    %     % draw=red: Red border
    %     % inner sep=5pt: Padding inside the box
    %     \node[] () {
            
            % Create a container slightly smaller than the column width
            \begin{minipage}{0.9\linewidth}
                \centering
                
                \caption{\sys with Different Simulation/Emulation Backends\\
                Llama3.1-70B, Training, DP=2, TP=4, 32 Micro-Batches
                }
                % \vspace{-1em}
                \label{tab:stage_multiple_simulation}
                % Resize the table to fit the width of the minipage (the red box)
                \resizebox{0.8\linewidth}{!}{%
                    \begin{tabular}{|c|c|c|c|}
                    \hline
                    Simulator   &   Target System   & \makecell[c]{Runtime [ms]}  & \makecell[c]{LoC for\\ Adaption}\\
                    \hline
                    \multirow{2}{*}{SimAI}       &   8xH100          & 3,909.5 & \multirow{2}{*}{73} \\\cline{2-3}
                           &   8xH200         & 3,791.5 &  \\\hline
                    \multirow{2}{*}{ScaleSim} & 8xTPUv5e & 1,843.8 & \multirow{2}{*}{34}\\\cline{2-3}
                        & 8xTPUv4 & 1,452.9 & \\\hline
                    \multirow{2}{*}{Genie}       &   8x100Gbps IB     &  33,128.5                 & \multirow{2}{*}{46}\\\cline{2-3}
                                                 &   8x400Gbps IB      &  11,441.7         & \\\hline
                    \end{tabular}%
                }
            \end{minipage}
    %     };
    % \end{tikzpicture}
\end{table}

\subsection{\sys performance}
% \vspace{-2mm}

% \hw{Separate \sys performance as a section with two parts: 1. Runtime performance discussion, using STG as DSE to find the best parallelization strategy. 2. Memory tracking, using performance to }

We evaluate \sys in terms of runtime and memory footprint across scales. Results show that \sys significantly reduces the time required to collect graph workloads for simulation. Experiments are conducted on a Linux server with four Intel Xeon E7-8880 v4 processors (2.2 GHz) and 354 GiB of DDR3-1333 memory.

\begin{figure}[t]
    \centering
    % \begin{tikzpicture}
    %     % draw=red: Red border
    %     % inner sep=5pt: Padding
    %     \node[] () {
            \vspace{-1em}
            \begin{minipage}{0.95\linewidth}
                \centering
                \includegraphics[width=\linewidth]{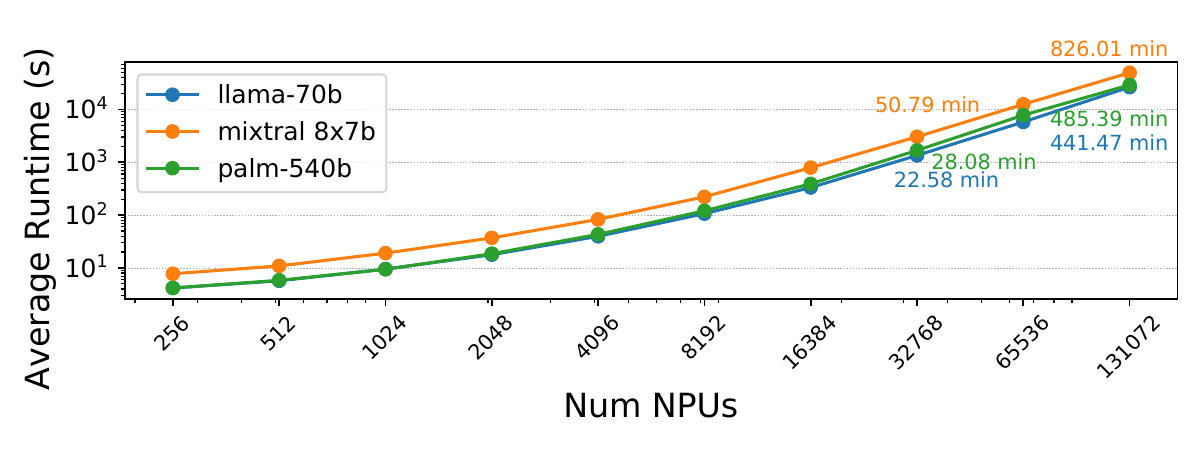}
                \vspace{-5mm}
                
                \caption{\sys Runtime Scaling with Number of GPUs 
                % \ju{Can we make this graph and figure 10 more flattened?}
                % \jy{Try to use dashed lines or square/triangle points if possible}
                }
                \label{fig:runtime_memory_vs_npus}
            \end{minipage}
    %     };
    % \end{tikzpicture}
    % Place the negative vspace outside to pull the main text up
    \vspace{-6mm}
\end{figure}

We also evaluated \sys across a wide range of GPU scales to assess how generation time grows with model and system size. As shown in~\autoref{fig:runtime_memory_vs_npus}, runtime increases non-linearly due to the expanding parallel configuration space, yet \sys remains highly efficient. At 32K GPUs, it generates graphs for a 540B dense LLM in just 28 minutes. For more complex models like Mixtral-8x7B, with added expert parallelism, generation remains practical at around 50 minutes. For a larger scale of 128K devices, for which to the best of our knowledge, no publicly accessible real-world system currently exists, \sys still generates the workload within hours while keeping memory usage below 400 MB in all cases. In contrast, real-system trace generation is expensive and slow. For instance, collecting an execution trace for training LLaMA-3.1-70B with 128 micro-batches on 32 H100 GPUs takes approximately \textbf{47 GPU-minutes}, whereas \sys synthesizes the corresponding workload in only \textbf{37 CPU-seconds}. Moreover, real-system traces require re-collection when the target system changes, as traces are inherently system-specific. In contrast, \sys provides a more generalized solution. Furthermore, access to the physical system required for trace collection may not always be feasible.

%% file: sections/discussion.tex
\vspace{-2mm}
\subsection{Discussion: Synthesizing advanced models and parallelization with \sys} 
\vspace{-1mm}
% \subsection{Modeling more LLM architecture and Communication Pattern with \sys} 
\label{sec:discussion}
% \vspace{-1mm}

% \TK{BTW this should be a sub-section instead of a full-section}
% \hw{do we need this still?}
% \TK{depends on space .. can be shortened to just keep the unconventional parallleism instead of Tensor Train}

% \section{\sys can handle more than LLMs}

% \ch{Move after evaluation? That should be a supplement, not a points. We shouldn't break other's attention.. Just show there are more opportunity for a follow-up work}
% \ju{We might need to tone down little bit for this section. Maybe replace this section with Memory Tracker}
While the evaluation in this work focuses on conventional LLMs (including MoEs), the symbolic representation employed by \sys is not inherently limited to LLMs. The framework's design allows it to generalize to any tensor computation workloads from ML or other fields. 
%This flexibility positions \sys as a versatile tool for analyzing and simulating diverse workloads across a wide range of architectures, extending its applicability beyond the immediate scope of LLMs. 
Here we show the flexibility of \sys through some application cases.

% \subsection{Other ML model: ResNet-50}
% Before vision transformers, people used Convolution Neural Network (CNN) to handle vision applications. ResNet-50\cite{} is one of those CNN models, which is made up mainly of convolutions. Here we will show that \sys is capable of modeling ResNet-50.

% One main challenge of modeling ResNet-50 is that there is no conventional operator that can be used to represent a convolution. One solution is to add another operator of the convolution in the operator set of \sys, however, It is still possible to model it with conventional \sys. Fundamentally convolution is also tensor multiplication with complex access patterns, thus it can be possibly represented in the form of einsum in some way. 

% First we represent a

% % y[B, N, X, Y] = Conv[](w[N, C, W, H], x[B, C, X, Y])
% % y[B, N, X, Y] = Conv[](w[N, C, W, H], x[B, C, X, Y, W, H])
% % y[B, N, X, Y] = einsum[ncwh,bcxywh->bnxy](w, x) reduce c, w, h

\begin{table}[t]
    \centering
    \caption{State-Space Model}
    %\vspace{-10pt}
    \vspace{-8pt}
    \resizebox{\linewidth}{!}{
    \begin{tabular}{p{0.48\linewidth} p{0.48\linewidth}}
        \toprule
        \multicolumn{1}{l}{\textbf{Inputs}} & \multicolumn{1}{l}{\textbf{Output}} \\
        \hline
        \texttt{x[B/\textcolor{red}{p1},S,D/\textcolor{blue}{p2}]} & \multirow{6}{*}{\texttt{y[B/\textcolor{red}{p1},S,D]}} \\
        \texttt{wdt1[D/\textcolor{blue}{p2},R]}, \texttt{wdt2[R,D/\textcolor{blue}{p2}]} & \\
        \texttt{A[D/\textcolor{blue}{p2},P]}, \texttt{B[B/\textcolor{red}{p1},S,P]} & \\
        \texttt{C[B/\textcolor{red}{p1},S,P]}, \texttt{D[D/\textcolor{blue}{p2}]} & \\
        \hline
        \multicolumn{2}{l}{\textbf{Compute}:} \\
        \hline
        \multicolumn{2}{l}{\texttt{dt1[B/\textcolor{red}{p1},S,R] = \textcolor{blue}{AllReduce}(einsum[bsd,de->bse](x, wdt1))}} \\
        \multicolumn{2}{l}{\texttt{dt[B/\textcolor{red}{p1},S,D/\textcolor{blue}{p2}] = einsum[bse,ed->bsd](dt1, wdt2)}} \\
        \multicolumn{2}{l}{\texttt{dA[B/\textcolor{red}{p1},S,D/\textcolor{blue}{p2},P] = einsum[dp,bsd->bsdp](A, dt)}} \\
        \multicolumn{2}{l}{\texttt{dB[B/\textcolor{red}{p1},S,D/\textcolor{blue}{p2},P] = einsum[bsp,bsd->bsdp](B, dt)}} \\
        \multicolumn{2}{l}{\texttt{deltaB[B/\textcolor{red}{p1},S,D/\textcolor{blue}{p2},P] = einsum[bsdp,bsd->bsdp](dB, x)}} \\
        \multicolumn{2}{l}{\texttt{hs[B/\textcolor{red}{p1},S,D/\textcolor{blue}{p2},P] = pscan[dim=1](dA, deltaB)}} \\
        \multicolumn{2}{l}{\texttt{y0[B/\textcolor{red}{p1},S,D/\textcolor{blue}{p2}] = einsum[bsdp,bsp->bsd](hs, C)}} \\
        \multicolumn{2}{l}{\texttt{y[B/\textcolor{red}{p1},S,D/\textcolor{blue}{p2}] = einsum[bsd,d](y0, D)}} \\
        \bottomrule
    \end{tabular}
    }
    \label{tab:state-space}
    \vspace{-0.2cm}
\end{table}

\noindent
\textbf{Emerging Model Architecture: State Space Model (SSM) \cite{mamba}: } SSMs are emerging as a compelling alternative to traditional transformer architectures in LLMs, primarily due to their linear computational and memory complexity, which allows for efficient handling of long sequences. Therefore, to showcase the flexibility of \sys, ~\autoref{tab:state-space} shows how users can model a state-space model using \sys, where we denote data-parallel and tensor-parallel as \textcolor{red}{p1} and \textcolor{blue}{p2}. 

\begin{table}[t]
    \centering
    \caption{Fully-Sharded Tensor Parallel}
    \vspace{-0.2cm}
    \label{tab:fstp}
%    \begin{tikzpicture}
      % Draw blue rectangle outline around the table contents
%    \node[draw=red, thick, rounded corners=2pt, inner sep=2pt] (tbl) {
    \resizebox{0.9\linewidth}{!}{
    \begin{tabular}{p{0.45\linewidth} p{0.45\linewidth}}
        \toprule
        \multicolumn{1}{l}{\textbf{Inputs}} & \multicolumn{1}{l}{\textbf{Output}} \\
        \hline
        \texttt{X[Batch/\textcolor{red}{dp}, D1/\textcolor{blue}{tp}]} & \multirow{2}{*}{\texttt{Y[Batch/\textcolor{red}{dp}, D2/\textcolor{blue}{tp}]}} \\
        \texttt{W[D1/\textcolor{blue}{tp}, D2]} & \\
        \hline
        \multicolumn{2}{l}{\textbf{Compute}:} \\
        \hline
        \multicolumn{2}{l}{\texttt{X*[Batch/\textcolor{red}{dp},D1]=AllGather[\textcolor{blue}{tp}](X)}} \\
        \multicolumn{2}{l}{\texttt{Y*[Batch/\textcolor{red}{dp}, D2@1/\textcolor{blue}{tp}]=einsum[bm,mn->bn](X*, W)}} \\
        \multicolumn{2}{l}{\texttt{Y[Batch/\textcolor{red}{dp}, D2/\textcolor{blue}{tp}]=ReduceScatter[\textcolor{blue}{tp}](Y*)}} \\
        \bottomrule
    \end{tabular}
    }
    \vspace{-1em}
%    };
%\end{tikzpicture}
\end{table}

% \subsection{\textbf{Unconventional Parallel Strategies: Interleaved DP/TP}} 
\noindent
%\revise{
\textbf{Emerging Parallel Strategies: }
%\sys supports modeling emgering parallel strategies, as long as it can be represent in a tensor symbolic way. 
\autoref{tab:fstp} illustrates a hypothetical symmetric parallel strategy for FSDP that we call Fully-Sharded Tensor Parallel (FSTP), based on tensor parallelism (TP) with activation sharding. Although this strategy does not currently exist in ML frameworks, \sys can model it, enabling rapid prototyping and exploration in real or hypothetical systems before investing engineering effort to implement it in a framework.

\label{subsec:unconvential}

%% file: sections/relatedwork.tex
\vspace{-0.2cm}
\section{Related Works}
\label{sec:related}
\vspace{-1mm}

%\subsection{Benchmarking for Distributed Training}

\noindent
\textbf{Benchmarking for Distributed Training.}
DeepBench~\cite{bai2019deepbench} and MLPerf~\cite{mlperf} offer standardized metrics for evaluating the performance of training and inference tasks. While these tools excel in providing reproducible benchmarks, they do not support detailed profiling data.
PyTorch Execution Observer~\cite{pytorch-profiler} and NVIDIA CUPTI~\cite{cupti} provide performance profiling results for training systems. 
However, they require actual runs to collect traces. Moreover, the generated execution traces lack annotations for optimizations and dependencies, which are essential for profiling system architectures. 
PyTorch FX~\cite{pytorch-fx} can capture static model behaviors with dependency graph during compile time but it lacks post-execution information and requires optimized code for analysis.
% , limiting its flexibility in exploring configurations. 
In contrast, \sys automatically partitions the operators, generating an updated computational graph that incorporates the appropriate parallelization annotations and dependencies.

%\subsection{Performance Modeling for Distributed Training} %\label{sec:background_perf_modeling}

\noindent
\textbf{Performance Modeling for Distributed Training.}
Recent efforts on performance modeling such as vTrain~\cite{vtrain}, MADMAX~\cite{madmax}, and Calculon~\cite{calculon} have significantly advanced the community’s understanding of distributed LLM workloads through detailed analytical modeling or trace-driven simulation. However, these frameworks share a common limitation in terms of flexibility and configurability, making it difficult to systematically explore emerging models such as MoE and state-space model in detail, as identified in \autoref{tab:comparison_mlsys_perf_simulations}.
% \begin{table}[t]
% \centering
% \caption{Comparison between \sys and State-of-the-Art Performance Modeling for Distributed Training}
% \label{tab:comparison_mlsys_perf_simulations}
% \resizebox{\linewidth}{!}{
% \begin{tabular}{lcccc}
% \toprule
% \textbf{Method} & \makecell[c]{\textbf{Supported}\\\textbf{Workloads}} & \makecell[c]{\textbf{Supported}\\\textbf{Accelerators}} & \makecell[c]{\textbf{Workload Extension}\\\textbf{Mechanism}} & \makecell[c]{\textbf{Performance}\\\textbf{Model}}\\
% \midrule
% % vTrain
% vTrain~\cite{vtrain} & 
% \makecell[c]{Dense} & 
% \makecell[c]{A100/V100} & 
% \makecell[c]{Code Change + \\ Re-profiling} & 
% \makecell[c]{Trace-Driven} \\
% \addlinespace

% % MADMAX
% MADMAX~\cite{madmax} & 
% \makecell[c]{Dense, MoE \\ DLRM} & 
% \makecell[c]{Roofline \\ Parametrics} & 
% \makecell[c]{Code Change} & 
% \makecell[c]{Analytical} \\
% \addlinespace

% % Calculon
% Calculon~\cite{calculon} & 
% \makecell[c]{Dense} & 
% \makecell[c]{A100/H100} & 
% \makecell[c]{Code Change + \\ Perf. Model Update} & 
% \makecell[c]{Analytical}\\
% \addlinespace

% % SimAI
% SimAI~\cite{simai} &
% \makecell[c]{Dense, MoE} &
% \makecell[c]{NVIDIA GPUs} &
% \makecell[c]{Code Change} &
% \makecell[c]{Cycle-Accurate}\\
% \midrule
% % \sys (Ours)
% % \rowcolor{gray!10} 
% \textbf{\sys (Ours)} & 
% \makecell[c]{Arbitrary \\Tensor Graphs\\(Dense, MoE, \\DLRM, SSM, etc)} & 
% \makecell[c]{Hardware\\Agnostic} & 
% \makecell[c]{Input Change \\ (No Code Mod.)} & 
% \makecell[c]{Plug-and-Play \\ (Analytical, \\ Trace-Driven, \\Emulated)} \\
% \bottomrule
% \end{tabular}
% }
% \end{table}
\begin{table}[t]
    \centering
    % \begin{tikzpicture}
    %     % draw=red: Red border
    %     % inner sep=5pt: Padding
    %     \node[] () {
            
            \begin{minipage}{0.95\linewidth}
                \centering
                \caption{Comparison between \sys and State-of-the-Art Performance Modeling for Distributed Training}
                \vspace{-1em}
                \label{tab:comparison_mlsys_perf_simulations}
                \resizebox{\linewidth}{!}{
                    \begin{tabular}{lcccc}
                    \toprule
                    \textbf{Method} & \makecell[c]{\textbf{Supported}\\\textbf{Workloads}} & \makecell[c]{\textbf{Supported}\\\textbf{Accelerators}} & \makecell[c]{\textbf{Workload Extension}\\\textbf{Mechanism}} & \makecell[c]{\textbf{Performance}\\\textbf{Model}}\\
                    \midrule
                    % vTrain
                    vTrain~\cite{vtrain} & 
                    \makecell[c]{Dense} & 
                    \makecell[c]{A100/V100} & 
                    \makecell[c]{Code Change + \\ Re-profiling} & 
                    \makecell[c]{Trace-Driven} \\
                    \addlinespace
                    
                    % MADMAX
                    MADMAX~\cite{madmax} & 
                    \makecell[c]{Dense, MoE \\ DLRM} & 
                    \makecell[c]{Roofline \\ Parametrics} & 
                    \makecell[c]{Code Change} & 
                    \makecell[c]{Analytical} \\
                    \addlinespace
                    
                    % Calculon
                    Calculon~\cite{calculon} & 
                    \makecell[c]{Dense} & 
                    \makecell[c]{A100/H100} & 
                    \makecell[c]{Code Change + \\ Perf. Model Update} & 
                    \makecell[c]{Analytical}\\
                    \addlinespace
                    
                    % SimAI
                    SimAI~\cite{simai} &
                    \makecell[c]{Dense, MoE} &
                    \makecell[c]{NVIDIA GPUs} &
                    \makecell[c]{Code Change} &
                    \makecell[c]{Cycle-Accurate}\\
                    \midrule
                    % \sys (Ours)
                    % \rowcolor{gray!10} 
                    \textbf{\makecell[l]{\sys\\(Ours)}} & 
                    \makecell[c]{Arbitrary \\Tensor Graphs\\(Dense, MoE, \\DLRM, SSM, etc)} & 
                    \makecell[c]{Hardware\\Agnostic} & 
                    \makecell[c]{Input Change \\ (No Code Mod.)} & 
                    \makecell[c]{Plug-and-Play \\ (Analytical, \\ Trace-Driven, \\Emulated)} \\
                    \bottomrule
                    \end{tabular}
                }
            \end{minipage}
    %     };
    % \end{tikzpicture}
    \vspace{-1.5em}
\end{table}

In this context, our work, \sys, aims not to compete but rather to complement these existing frameworks. By providing a flexible and configurable workload generation mechanism, \sys can interface with diverse backends in a plug-and-play manner, adapting to the specific performance modeling requirements of different tools.

%\subsection{Tensor Representation for System-level Optimization} 
% \hw{updated} \ju{Shrinked}

\noindent
\textbf{Tensor Representation for System-level Optimizations.}
Tensor representation is commonly utilized for system-level optimization of deep learning models~\cite{Souffle, TVM, TensorComprehend}, enabling computational graph optimizations for frameworks including PyTorch~\cite{pytorch} and TensorFlow~\cite{tensorflow}. Techniques such as operator fusion leverage tensor representations to enhance parallel processing and memory efficiency~\cite{Apollo, DNNFusion}. 
FlexFlow~\cite{FlexFlow} and Unity~\cite{Unity} employ system-level compilation to determine effective parallelization strategies in distributed settings, while Mist~\cite{Zhu2025Mist} recently proposed symbolic tensor representations specifically for memory parallelism. 
In contrast, we propose a symbolic tensor graph that systematically annotates key operators with parallelization dimensions to guide runtime optimization for large-scale LLM training.

%% file: sections/conclusion.tex
\vspace{-0.15cm}
\section{Conclusion}
\vspace{-0.1cm}

We introduce \sys, 
a framework for generating high-fidelity workload graphs for distributed LLM training. It provides practitioners with a  robust tool for system-level design exploration and scalable benchmarking in future AI infrastructure research.
The symbolic tensor graph allows for a structured representation of parallelization strategies, moving beyond ad-hoc methods and enabling the exploration of previously unattainable system configurations. Our validation against real-world traces and scalability up to 128K GPUs demonstrate its effectiveness and practicality.

% In this work, we introduce \sys, an open-source framework designed to generate synthetic traces tailored to the unique demands of Large Language Model (LLM) training. 
% Through comprehensive analysis of real-world traces, \sys provides detailed configuration files encompassing various model and system parameters, allowing for highly accurate and customizable workload simulations. By offering pre-defined configurations and multiple parallelization strategies, \sys addresses the challenge of trace accessibility, facilitating high-fidelity simulations that support both model optimization and system-level performance tuning. 
% By disseminating \sys to the research community, we aim to advance the efficiency and scalability of LLM training, thereby enhancing access to simulation tools for future research and development.

%% file: reference.bib
@misc{korthikanti2022reduceact,
      title={Reducing Activation Recomputation in Large Transformer Models}, 
      author={Vijay Korthikanti and Jared Casper and Sangkug Lym and Lawrence McAfee and Michael Andersch and Mohammad Shoeybi and Bryan Catanzaro},
      year={2022},
      eprint={2205.05198},
      archivePrefix={arXiv},
      primaryClass={cs.LG},
      url={https://arxiv.org/abs/2205.05198}, 
}

@misc{cerebrasCS3,
  author       = {{Cerebras Systems, Inc.}},
  title        = {{CS‑3 System}},
  howpublished = {\url{https://www.cerebras.ai/system}},
  note         = {Accessed: 2025‑08‑01},
  year         = {n.d.}
}

@misc{liang2025lumos,
      title={Lumos: Efficient Performance Modeling and Estimation for Large-scale LLM Training}, 
      author={Mingyu Liang and Hiwot Tadese Kassa and Wenyin Fu and Brian Coutinho and Louis Feng and Christina Delimitrou},
      year={2025},
      eprint={2504.09307},
      archivePrefix={arXiv},
      primaryClass={cs.DC},
      url={https://arxiv.org/abs/2504.09307}, 
}

@misc{duan2023proteus,
      title={Proteus: Simulating the Performance of Distributed DNN Training}, 
      author={Jiangfei Duan and Xiuhong Li and Ping Xu and Xingcheng Zhang and Shengen Yan and Yun Liang and Dahua Lin},
      year={2023},
      eprint={2306.02267},
      archivePrefix={arXiv},
      primaryClass={cs.DC},
      url={https://arxiv.org/abs/2306.02267}, 
}

@online{deepseekV3R1_inference_system_2025,
  author       = {deepseek-ai},
  title        = {DeepSeek V3/R1 Inference System Overview},
  year         = {2025},
  month        = feb,
  url          = {https://github.com/deepseek-ai/open-infra-index/blob/main/202502OpenSourceWeek/day_6_one_more_thing_deepseekV3R1_inference_system_overview.md},
  note         = {GitHub: Open Infra Index, Day 6 of 2025 Open Source Week. Accessed: 2025-10-20}
}

@inproceedings{AWSTrainium,
author = {Bshara, Nafea},
title = {AWS Trainium: The Journey for Designing and Optimization Full Stack ML Hardware},
year = {2024},
isbn = {9798400703867},
publisher = {Association for Computing Machinery},
address = {New York, NY, USA},
url = {https://doi.org/10.1145/3620666.3655592},
doi = {10.1145/3620666.3655592},
booktitle = {Proceedings of the 29th ACM International Conference on Architectural Support for Programming Languages and Operating Systems, Volume 3},
pages = {4},
numpages = {1},
location = {La Jolla, CA, USA},
series = {ASPLOS '24}
}

@inproceedings {TPUv4Cluster,
author = {Yazhou Zu and Alireza Ghaffarkhah and Hoang-Vu Dang and Brian Towles and Steven Hand and Safeen Huda and Adekunle Bello and Alexander Kolbasov and Arash Rezaei and Dayou Du and Steve Lacy and Hang Wang and Aaron Wisner and Chris Lewis and Henri Bahini},
title = {Resiliency at Scale: Managing {Google{\textquoteright}s} {TPUv4} Machine Learning Supercomputer},
booktitle = {21st USENIX Symposium on Networked Systems Design and Implementation (NSDI 24)},
year = {2024},
isbn = {978-1-939133-39-7},
address = {Santa Clara, CA},
pages = {761--774},
url = {https://www.usenix.org/conference/nsdi24/presentation/zu},
publisher = {USENIX Association},
month = apr
}

@misc{nvidiaHGX,
  author       = {{NVIDIA Corporation}},
  title        = {{NVIDIA HGX Platform}},
  howpublished = {\url{https://www.nvidia.com/en-us/data-center/hgx/}},
  note         = {Accessed: 2025‑08‑01},
  year         = {n.d.}
}

@misc{pytorchkineto,
  author       = {{PyTorch Team}},
  title        = {Kineto: Performance Profiling Library for PyTorch},
  howpublished = {\url{https://github.com/pytorch/kineto}},
  note         = {Accessed: 2025-07-31},
  year         = {2025}
}

@article{infiniband,
  title={An introduction to the infiniband architecture},
  author={Pfister, Gregory F},
  journal={High performance mass storage and parallel I/O},
  volume={42},
  number={617-632},
  pages={10},
  year={2001}
}

@misc{nvidia_cupti,
  title        = {NVIDIA CUPTI - CUDA Profiling Tools Interface},
  author       = {{NVIDIA}},
  year         = {2024},
  url          = {https://developer.nvidia.com/cupti},
  note         = {Accessed: 2024-11-23}
}

@article{chollmservingsim,
  title={LLMServingSim: A Simulation Infrastructure for LLM Inference Serving Systems},
  author={Cho, Jaehong and Kim, Minsu and Choi, Hyunmin and Park, Jongse}
}

@inproceedings{madmax,
  title={MAD-Max Beyond Single-Node: Enabling Large Machine Learning Model Acceleration on Distributed Systems},
  author={Hsia, Samuel and Golden, Alicia and Acun, Bilge and Ardalani, Newsha and DeVito, Zachary and Wei, Gu-Yeon and Brooks, David and Wu, Carole-Jean},
  booktitle={2024 ACM/IEEE 51st Annual International Symposium on Computer Architecture (ISCA)},
  pages={818--833},
  year={2024},
  organization={IEEE}
}

@inproceedings {simai,
author = {Xizheng Wang and Qingxu Li and Yichi Xu and Gang Lu and Dan Li and Li Chen and Heyang Zhou and Linkang Zheng and Sen Zhang and Yikai Zhu and Yang Liu and Pengcheng Zhang and Kun Qian and Kunling He and Jiaqi Gao and Ennan Zhai and Dennis Cai and Binzhang Fu},
title = {{SimAI}: Unifying Architecture Design and Performance Tuning for {Large-Scale} Large Language Model Training with Scalability and Precision},
booktitle = {22nd USENIX Symposium on Networked Systems Design and Implementation (NSDI 25)},
year = {2025},
isbn = {978-1-939133-46-5},
address = {Philadelphia, PA},
pages = {541--558},
url = {https://www.usenix.org/conference/nsdi25/presentation/wang-xizheng-simai},
publisher = {USENIX Association},
month = apr
}

@article{vtrain,
  title={vtrain: A simulation framework for evaluating cost-effective and compute-optimal large language model training},
  author={Bang, Jehyeon and Choi, Yujeong and Kim, Myeongwoo and Kim, Yongdeok and Rhu, Minsoo},
  journal={arXiv preprint arXiv:2312.12391},
  year={2023}
}

@online{meta2025llama4,
  author       = {{Meta AI}},
  title        = {The Llama 4 Herd: The Beginning of a New Era of Natively Multimodal AI Innovation},
  year         = {2025},
  month        = apr,
  url          = {https://ai.meta.com/blog/llama-4-multimodal-intelligence/},
  note         = {Accessed: 2025-04-22}
}

@misc{mystique,
      title={Mystique: Enabling Accurate and Scalable Generation of Production AI Benchmarks}, 
      author={Mingyu Liang and Wenyin Fu and Louis Feng and Zhongyi Lin and Pavani Panakanti and Shengbao Zheng and Srinivas Sridharan and Christina Delimitrou},
      year={2023},
      eprint={2301.04122},
      archivePrefix={arXiv},
      primaryClass={cs.DC},
      url={https://arxiv.org/abs/2301.04122}, 
}

@inproceedings{astraSim,
  title={Astra-sim2. 0: Modeling hierarchical networks and disaggregated systems for large-model training at scale},
  author={Won, William and Heo, Taekyung and Rashidi, Saeed and Sridharan, Srinivas and Srinivasan, Sudarshan and Krishna, Tushar},
  booktitle={2023 IEEE International Symposium on Performance Analysis of Systems and Software (ISPASS)},
  pages={283--294},
  year={2023},
  organization={IEEE}
}

@inproceedings{calculon,
  title={Calculon: a methodology and tool for high-level co-design of systems and large language models},
  author={Isaev, Mikhail and McDonald, Nic and Dennison, Larry and Vuduc, Richard},
  booktitle={Proceedings of the International Conference for High Performance Computing, Networking, Storage and Analysis},
  pages={1--14},
  year={2023}
}

@misc{palm,
      title={PaLM: Scaling Language Modeling with Pathways}, 
      author={Aakanksha Chowdhery and Sharan Narang and Jacob Devlin and Maarten Bosma and Gaurav Mishra and Adam Roberts and Paul Barham and Hyung Won Chung and Charles Sutton and Sebastian Gehrmann and Parker Schuh and Kensen Shi and Sasha Tsvyashchenko and Joshua Maynez and Abhishek Rao and Parker Barnes and Yi Tay and Noam Shazeer and Vinodkumar Prabhakaran and Emily Reif and Nan Du and Ben Hutchinson and Reiner Pope and James Bradbury and Jacob Austin and Michael Isard and Guy Gur-Ari and Pengcheng Yin and Toju Duke and Anselm Levskaya and Sanjay Ghemawat and Sunipa Dev and Henryk Michalewski and Xavier Garcia and Vedant Misra and Kevin Robinson and Liam Fedus and Denny Zhou and Daphne Ippolito and David Luan and Hyeontaek Lim and Barret Zoph and Alexander Spiridonov and Ryan Sepassi and David Dohan and Shivani Agrawal and Mark Omernick and Andrew M. Dai and Thanumalayan Sankaranarayana Pillai and Marie Pellat and Aitor Lewkowycz and Erica Moreira and Rewon Child and Oleksandr Polozov and Katherine Lee and Zongwei Zhou and Xuezhi Wang and Brennan Saeta and Mark Diaz and Orhan Firat and Michele Catasta and Jason Wei and Kathy Meier-Hellstern and Douglas Eck and Jeff Dean and Slav Petrov and Noah Fiedel},
      year={2022},
      eprint={2204.02311},
      archivePrefix={arXiv},
      primaryClass={cs.CL},
      url={https://arxiv.org/abs/2204.02311}, 
}

@misc{mistral,
      title={Mistral 7B}, 
      author={Albert Q. Jiang and Alexandre Sablayrolles and Arthur Mensch and Chris Bamford and Devendra Singh Chaplot and Diego de las Casas and Florian Bressand and Gianna Lengyel and Guillaume Lample and Lucile Saulnier and Lélio Renard Lavaud and Marie-Anne Lachaux and Pierre Stock and Teven Le Scao and Thibaut Lavril and Thomas Wang and Timothée Lacroix and William El Sayed},
      year={2023},
      eprint={2310.06825},
      archivePrefix={arXiv},
      primaryClass={cs.CL},
      url={https://arxiv.org/abs/2310.06825}, 
}

@misc{mixtral,
      title={Mixtral of Experts}, 
      author={Albert Q. Jiang and Alexandre Sablayrolles and Antoine Roux and Arthur Mensch and Blanche Savary and Chris Bamford and Devendra Singh Chaplot and Diego de las Casas and Emma Bou Hanna and Florian Bressand and Gianna Lengyel and Guillaume Bour and Guillaume Lample and Lélio Renard Lavaud and Lucile Saulnier and Marie-Anne Lachaux and Pierre Stock and Sandeep Subramanian and Sophia Yang and Szymon Antoniak and Teven Le Scao and Théophile Gervet and Thibaut Lavril and Thomas Wang and Timothée Lacroix and William El Sayed},
      year={2024},
      eprint={2401.04088},
      archivePrefix={arXiv},
      primaryClass={cs.LG},
      url={https://arxiv.org/abs/2401.04088}, 
}

@misc{gpt-3,
      title={Language Models are Few-Shot Learners}, 
      author={Tom B. Brown and Benjamin Mann and Nick Ryder and Melanie Subbiah and Jared Kaplan and Prafulla Dhariwal and Arvind Neelakantan and Pranav Shyam and Girish Sastry and Amanda Askell and Sandhini Agarwal and Ariel Herbert-Voss and Gretchen Krueger and Tom Henighan and Rewon Child and Aditya Ramesh and Daniel M. Ziegler and Jeffrey Wu and Clemens Winter and Christopher Hesse and Mark Chen and Eric Sigler and Mateusz Litwin and Scott Gray and Benjamin Chess and Jack Clark and Christopher Berner and Sam McCandlish and Alec Radford and Ilya Sutskever and Dario Amodei},
      year={2020},
      eprint={2005.14165},
      archivePrefix={arXiv},
      primaryClass={cs.CL},
      url={https://arxiv.org/abs/2005.14165}, 
}

@misc{deepseek-moe,
      title={DeepSeekMoE: Towards Ultimate Expert Specialization in Mixture-of-Experts Language Models}, 
      author={Damai Dai and Chengqi Deng and Chenggang Zhao and R. X. Xu and Huazuo Gao and Deli Chen and Jiashi Li and Wangding Zeng and Xingkai Yu and Y. Wu and Zhenda Xie and Y. K. Li and Panpan Huang and Fuli Luo and Chong Ruan and Zhifang Sui and Wenfeng Liang},
      year={2024},
      eprint={2401.06066},
      archivePrefix={arXiv},
      primaryClass={cs.CL},
      url={https://arxiv.org/abs/2401.06066}, 
}

@misc{MQA,
      title={GQA: Training Generalized Multi-Query Transformer Models from Multi-Head Checkpoints}, 
      author={Joshua Ainslie and James Lee-Thorp and Michiel de Jong and Yury Zemlyanskiy and Federico Lebrón and Sumit Sanghai},
      year={2023},
      eprint={2305.13245},
      archivePrefix={arXiv},
      primaryClass={cs.CL},
      url={https://arxiv.org/abs/2305.13245}, 
}

@misc{megatron-lm,
      title={Megatron-LM: Training Multi-Billion Parameter Language Models Using Model Parallelism}, 
      author={Mohammad Shoeybi and Mostofa Patwary and Raul Puri and Patrick LeGresley and Jared Casper and Bryan Catanzaro},
      year={2020},
      eprint={1909.08053},
      archivePrefix={arXiv},
      primaryClass={cs.CL},
      url={https://arxiv.org/abs/1909.08053}, 
}

@misc{zhao2023pytorchfsdp,
      title={PyTorch FSDP: Experiences on Scaling Fully Sharded Data Parallel}, 
      author={Yanli Zhao and Andrew Gu and Rohan Varma and Liang Luo and Chien-Chin Huang and Min Xu and Less Wright and Hamid Shojanazeri and Myle Ott and Sam Shleifer and Alban Desmaison and Can Balioglu and Pritam Damania and Bernard Nguyen and Geeta Chauhan and Yuchen Hao and Ajit Mathews and Shen Li},
      year={2023},
      eprint={2304.11277},
      archivePrefix={arXiv},
      primaryClass={cs.DC},
      url={https://arxiv.org/abs/2304.11277}, 
}

@inproceedings{Zhu2025Mist, series={EuroSys ’25},
   title={Mist: Efficient Distributed Training of Large Language Models via Memory-Parallelism Co-Optimization},
   url={http://dx.doi.org/10.1145/3689031.3717461},
   DOI={10.1145/3689031.3717461},
   booktitle={Proceedings of the Twentieth European Conference on Computer Systems},
   publisher={ACM},
   author={Zhu, Zhanda and Giannoula, Christina and Andoorveedu, Muralidhar and Su, Qidong and Mangalam, Karttikeya and Zheng, Bojian and Pekhimenko, Gennady},
   year={2025},
   month=mar, pages={1298–1316},
   collection={EuroSys ’25} }

@misc{mamba,
      title={Mamba: Linear-Time Sequence Modeling with Selective State Spaces}, 
      author={Albert Gu and Tri Dao},
      year={2024},
      eprint={2312.00752},
      archivePrefix={arXiv},
      primaryClass={cs.LG},
      url={https://arxiv.org/abs/2312.00752}, 
}

@misc{huang2019gpipe,
      title={GPipe: Efficient Training of Giant Neural Networks using Pipeline Parallelism}, 
      author={Yanping Huang and Youlong Cheng and Ankur Bapna and Orhan Firat and Mia Xu Chen and Dehao Chen and HyoukJoong Lee and Jiquan Ngiam and Quoc V. Le and Yonghui Wu and Zhifeng Chen},
      year={2019},
      eprint={1811.06965},
      archivePrefix={arXiv},
      primaryClass={cs.CV},
      url={https://arxiv.org/abs/1811.06965}, 
}

@misc{harlap2018pipedream,
      title={PipeDream: Fast and Efficient Pipeline Parallel DNN Training}, 
      author={Aaron Harlap and Deepak Narayanan and Amar Phanishayee and Vivek Seshadri and Nikhil Devanur and Greg Ganger and Phil Gibbons},
      year={2018},
      eprint={1806.03377},
      archivePrefix={arXiv},
      primaryClass={cs.DC},
      url={https://arxiv.org/abs/1806.03377}, 
}

@misc{lepikhin2020gshard,
      title={GShard: Scaling Giant Models with Conditional Computation and Automatic Sharding}, 
      author={Dmitry Lepikhin and HyoukJoong Lee and Yuanzhong Xu and Dehao Chen and Orhan Firat and Yanping Huang and Maxim Krikun and Noam Shazeer and Zhifeng Chen},
      year={2020},
      eprint={2006.16668},
      archivePrefix={arXiv},
      primaryClass={cs.CL},
      url={https://arxiv.org/abs/2006.16668}, 
}

@misc{li2020pytorchdistributed,
      title={PyTorch Distributed: Experiences on Accelerating Data Parallel Training}, 
      author={Shen Li and Yanli Zhao and Rohan Varma and Omkar Salpekar and Pieter Noordhuis and Teng Li and Adam Paszke and Jeff Smith and Brian Vaughan and Pritam Damania and Soumith Chintala},
      year={2020},
      eprint={2006.15704},
      archivePrefix={arXiv},
      primaryClass={cs.DC},
      url={https://arxiv.org/abs/2006.15704}, 
}

@misc{switch-transformer,
      title={Switch Transformers: Scaling to Trillion Parameter Models with Simple and Efficient Sparsity}, 
      author={William Fedus and Barret Zoph and Noam Shazeer},
      year={2022},
      eprint={2101.03961},
      archivePrefix={arXiv},
      primaryClass={cs.LG},
      url={https://arxiv.org/abs/2101.03961}, 
}

@misc{bert,
      title={BERT: Pre-training of Deep Bidirectional Transformers for Language Understanding}, 
      author={Jacob Devlin and Ming-Wei Chang and Kenton Lee and Kristina Toutanova},
      year={2019},
      eprint={1810.04805},
      archivePrefix={arXiv},
      primaryClass={cs.CL},
      url={https://arxiv.org/abs/1810.04805}, 
}

@misc{deepseek-v2,
      title={DeepSeek-V2: A Strong, Economical, and Efficient Mixture-of-Experts Language Model}, 
      eprint={2405.04434},
      archivePrefix={arXiv},
      primaryClass={cs.CL},
      url={https://arxiv.org/abs/2405.04434}, 
}

@misc{vaswani2023attention,
      title={Attention Is All You Need}, 
      author={Ashish Vaswani and Noam Shazeer and Niki Parmar and Jakob Uszkoreit and Llion Jones and Aidan N. Gomez and Lukasz Kaiser and Illia Polosukhin},
      year={2023},
      eprint={1706.03762},
      archivePrefix={arXiv},
      primaryClass={cs.CL},
      url={https://arxiv.org/abs/1706.03762}, 
}

@misc{jamba,
      title={Jamba: A Hybrid Transformer-Mamba Language Model}, 
      author={Opher Lieber and Barak Lenz and Hofit Bata and Gal Cohen and Jhonathan Osin and Itay Dalmedigos and Erez Safahi and Shaked Meirom and Yonatan Belinkov and Shai Shalev-Shwartz and Omri Abend and Raz Alon and Tomer Asida and Amir Bergman and Roman Glozman and Michael Gokhman and Avashalom Manevich and Nir Ratner and Noam Rozen and Erez Shwartz and Mor Zusman and Yoav Shoham},
      year={2024},
      eprint={2403.19887},
      archivePrefix={arXiv},
      primaryClass={cs.CL},
      url={https://arxiv.org/abs/2403.19887}, 
}

@misc{zamba,
      title={Zamba: A Compact 7B SSM Hybrid Model}, 
      author={Paolo Glorioso and Quentin Anthony and Yury Tokpanov and James Whittington and Jonathan Pilault and Adam Ibrahim and Beren Millidge},
      year={2024},
      eprint={2405.16712},
      archivePrefix={arXiv},
      primaryClass={cs.LG},
      url={https://arxiv.org/abs/2405.16712}, 
}

@misc{deepseekr1,
      title={DeepSeek-R1: Incentivizing Reasoning Capability in LLMs via Reinforcement Learning}, 
      author={DeepSeek-AI and Daya Guo and Dejian Yang and Haowei Zhang and Junxiao Song and Ruoyu Zhang and Runxin Xu and Qihao Zhu and Shirong Ma and Peiyi Wang and Xiao Bi and Xiaokang Zhang and Xingkai Yu and Yu Wu and Z. F. Wu and Zhibin Gou and Zhihong Shao and Zhuoshu Li and Ziyi Gao and Aixin Liu and Bing Xue and Bingxuan Wang and Bochao Wu and Bei Feng and Chengda Lu and Chenggang Zhao and Chengqi Deng and Chenyu Zhang and Chong Ruan and Damai Dai and Deli Chen and Dongjie Ji and Erhang Li and Fangyun Lin and Fucong Dai and Fuli Luo and Guangbo Hao and Guanting Chen and Guowei Li and H. Zhang and Han Bao and Hanwei Xu and Haocheng Wang and Honghui Ding and Huajian Xin and Huazuo Gao and Hui Qu and Hui Li and Jianzhong Guo and Jiashi Li and Jiawei Wang and Jingchang Chen and Jingyang Yuan and Junjie Qiu and Junlong Li and J. L. Cai and Jiaqi Ni and Jian Liang and Jin Chen and Kai Dong and Kai Hu and Kaige Gao and Kang Guan and Kexin Huang and Kuai Yu and Lean Wang and Lecong Zhang and Liang Zhao and Litong Wang and Liyue Zhang and Lei Xu and Leyi Xia and Mingchuan Zhang and Minghua Zhang and Minghui Tang and Meng Li and Miaojun Wang and Mingming Li and Ning Tian and Panpan Huang and Peng Zhang and Qiancheng Wang and Qinyu Chen and Qiushi Du and Ruiqi Ge and Ruisong Zhang and Ruizhe Pan and Runji Wang and R. J. Chen and R. L. Jin and Ruyi Chen and Shanghao Lu and Shangyan Zhou and Shanhuang Chen and Shengfeng Ye and Shiyu Wang and Shuiping Yu and Shunfeng Zhou and Shuting Pan and S. S. Li and Shuang Zhou and Shaoqing Wu and Shengfeng Ye and Tao Yun and Tian Pei and Tianyu Sun and T. Wang and Wangding Zeng and Wanjia Zhao and Wen Liu and Wenfeng Liang and Wenjun Gao and Wenqin Yu and Wentao Zhang and W. L. Xiao and Wei An and Xiaodong Liu and Xiaohan Wang and Xiaokang Chen and Xiaotao Nie and Xin Cheng and Xin Liu and Xin Xie and Xingchao Liu and Xinyu Yang and Xinyuan Li and Xuecheng Su and Xuheng Lin and X. Q. Li and Xiangyue Jin and Xiaojin Shen and Xiaosha Chen and Xiaowen Sun and Xiaoxiang Wang and Xinnan Song and Xinyi Zhou and Xianzu Wang and Xinxia Shan and Y. K. Li and Y. Q. Wang and Y. X. Wei and Yang Zhang and Yanhong Xu and Yao Li and Yao Zhao and Yaofeng Sun and Yaohui Wang and Yi Yu and Yichao Zhang and Yifan Shi and Yiliang Xiong and Ying He and Yishi Piao and Yisong Wang and Yixuan Tan and Yiyang Ma and Yiyuan Liu and Yongqiang Guo and Yuan Ou and Yuduan Wang and Yue Gong and Yuheng Zou and Yujia He and Yunfan Xiong and Yuxiang Luo and Yuxiang You and Yuxuan Liu and Yuyang Zhou and Y. X. Zhu and Yanhong Xu and Yanping Huang and Yaohui Li and Yi Zheng and Yuchen Zhu and Yunxian Ma and Ying Tang and Yukun Zha and Yuting Yan and Z. Z. Ren and Zehui Ren and Zhangli Sha and Zhe Fu and Zhean Xu and Zhenda Xie and Zhengyan Zhang and Zhewen Hao and Zhicheng Ma and Zhigang Yan and Zhiyu Wu and Zihui Gu and Zijia Zhu and Zijun Liu and Zilin Li and Ziwei Xie and Ziyang Song and Zizheng Pan and Zhen Huang and Zhipeng Xu and Zhongyu Zhang and Zhen Zhang},
      year={2025},
      eprint={2501.12948},
      archivePrefix={arXiv},
      primaryClass={cs.CL},
      url={https://arxiv.org/abs/2501.12948}, 
}

@misc{deepseekv3,
      title={DeepSeek-V3 Technical Report}, 
      author={DeepSeek-AI and Aixin Liu and Bei Feng and Bing Xue and Bingxuan Wang and Bochao Wu and Chengda Lu and Chenggang Zhao and Chengqi Deng and Chenyu Zhang and Chong Ruan and Damai Dai and Daya Guo and Dejian Yang and Deli Chen and Dongjie Ji and Erhang Li and Fangyun Lin and Fucong Dai and Fuli Luo and Guangbo Hao and Guanting Chen and Guowei Li and H. Zhang and Han Bao and Hanwei Xu and Haocheng Wang and Haowei Zhang and Honghui Ding and Huajian Xin and Huazuo Gao and Hui Li and Hui Qu and J. L. Cai and Jian Liang and Jianzhong Guo and Jiaqi Ni and Jiashi Li and Jiawei Wang and Jin Chen and Jingchang Chen and Jingyang Yuan and Junjie Qiu and Junlong Li and Junxiao Song and Kai Dong and Kai Hu and Kaige Gao and Kang Guan and Kexin Huang and Kuai Yu and Lean Wang and Lecong Zhang and Lei Xu and Leyi Xia and Liang Zhao and Litong Wang and Liyue Zhang and Meng Li and Miaojun Wang and Mingchuan Zhang and Minghua Zhang and Minghui Tang and Mingming Li and Ning Tian and Panpan Huang and Peiyi Wang and Peng Zhang and Qiancheng Wang and Qihao Zhu and Qinyu Chen and Qiushi Du and R. J. Chen and R. L. Jin and Ruiqi Ge and Ruisong Zhang and Ruizhe Pan and Runji Wang and Runxin Xu and Ruoyu Zhang and Ruyi Chen and S. S. Li and Shanghao Lu and Shangyan Zhou and Shanhuang Chen and Shaoqing Wu and Shengfeng Ye and Shengfeng Ye and Shirong Ma and Shiyu Wang and Shuang Zhou and Shuiping Yu and Shunfeng Zhou and Shuting Pan and T. Wang and Tao Yun and Tian Pei and Tianyu Sun and W. L. Xiao and Wangding Zeng and Wanjia Zhao and Wei An and Wen Liu and Wenfeng Liang and Wenjun Gao and Wenqin Yu and Wentao Zhang and X. Q. Li and Xiangyue Jin and Xianzu Wang and Xiao Bi and Xiaodong Liu and Xiaohan Wang and Xiaojin Shen and Xiaokang Chen and Xiaokang Zhang and Xiaosha Chen and Xiaotao Nie and Xiaowen Sun and Xiaoxiang Wang and Xin Cheng and Xin Liu and Xin Xie and Xingchao Liu and Xingkai Yu and Xinnan Song and Xinxia Shan and Xinyi Zhou and Xinyu Yang and Xinyuan Li and Xuecheng Su and Xuheng Lin and Y. K. Li and Y. Q. Wang and Y. X. Wei and Y. X. Zhu and Yang Zhang and Yanhong Xu and Yanhong Xu and Yanping Huang and Yao Li and Yao Zhao and Yaofeng Sun and Yaohui Li and Yaohui Wang and Yi Yu and Yi Zheng and Yichao Zhang and Yifan Shi and Yiliang Xiong and Ying He and Ying Tang and Yishi Piao and Yisong Wang and Yixuan Tan and Yiyang Ma and Yiyuan Liu and Yongqiang Guo and Yu Wu and Yuan Ou and Yuchen Zhu and Yuduan Wang and Yue Gong and Yuheng Zou and Yujia He and Yukun Zha and Yunfan Xiong and Yunxian Ma and Yuting Yan and Yuxiang Luo and Yuxiang You and Yuxuan Liu and Yuyang Zhou and Z. F. Wu and Z. Z. Ren and Zehui Ren and Zhangli Sha and Zhe Fu and Zhean Xu and Zhen Huang and Zhen Zhang and Zhenda Xie and Zhengyan Zhang and Zhewen Hao and Zhibin Gou and Zhicheng Ma and Zhigang Yan and Zhihong Shao and Zhipeng Xu and Zhiyu Wu and Zhongyu Zhang and Zhuoshu Li and Zihui Gu and Zijia Zhu and Zijun Liu and Zilin Li and Ziwei Xie and Ziyang Song and Ziyi Gao and Zizheng Pan},
      year={2025},
      eprint={2412.19437},
      archivePrefix={arXiv},
      primaryClass={cs.CL},
      url={https://arxiv.org/abs/2412.19437}, 
}

@inproceedings{Souffle,
author = {Xia, Chunwei and Zhao, Jiacheng and Sun, Qianqi and Wang, Zheng and Wen, Yuan and Yu, Teng and Feng, Xiaobing and Cui, Huimin},
title = {Optimizing Deep Learning Inference via Global Analysis and Tensor Expressions},
year = {2024},
isbn = {9798400703720},
publisher = {Association for Computing Machinery},
address = {New York, NY, USA},
url = {https://doi.org/10.1145/3617232.3624858},
doi = {10.1145/3617232.3624858},
booktitle = {Proceedings of the 29th ACM International Conference on Architectural Support for Programming Languages and Operating Systems, Volume 1},
pages = {286–301},
numpages = {16},
location = {La Jolla, CA, USA},
series = {ASPLOS '24}
}

@misc{llama,
      title={LLaMA: Open and Efficient Foundation Language Models}, 
      author={Hugo Touvron and Thibaut Lavril and Gautier Izacard and Xavier Martinet and Marie-Anne Lachaux and Timothée Lacroix and Baptiste Rozière and Naman Goyal and Eric Hambro and Faisal Azhar and Aurelien Rodriguez and Armand Joulin and Edouard Grave and Guillaume Lample},
      year={2023},
      eprint={2302.13971},
      archivePrefix={arXiv},
      primaryClass={cs.CL},
      url={https://arxiv.org/abs/2302.13971}, 
}

@misc{llama3,
      title={The Llama 3 Herd of Models}, 
      author={Aaron Grattafiori and Abhimanyu Dubey and Abhinav Jauhri and Abhinav Pandey and Abhishek Kadian and Ahmad Al-Dahle and Aiesha Letman and Akhil Mathur and Alan Schelten and Alex Vaughan and Amy Yang and Angela Fan and Anirudh Goyal and Anthony Hartshorn and Aobo Yang and Archi Mitra and Archie Sravankumar and Artem Korenev and Arthur Hinsvark and Arun Rao and Aston Zhang and Aurelien Rodriguez and Austen Gregerson and Ava Spataru and Baptiste Roziere and Bethany Biron and Binh Tang and Bobbie Chern and Charlotte Caucheteux and Chaya Nayak and Chloe Bi and Chris Marra and Chris McConnell and Christian Keller and Christophe Touret and Chunyang Wu and Corinne Wong and Cristian Canton Ferrer and Cyrus Nikolaidis and Damien Allonsius and Daniel Song and Danielle Pintz and Danny Livshits and Danny Wyatt and David Esiobu and Dhruv Choudhary and Dhruv Mahajan and Diego Garcia-Olano and Diego Perino and Dieuwke Hupkes and Egor Lakomkin and Ehab AlBadawy and Elina Lobanova and Emily Dinan and Eric Michael Smith and Filip Radenovic and Francisco Guzmán and Frank Zhang and Gabriel Synnaeve and Gabrielle Lee and Georgia Lewis Anderson and Govind Thattai and Graeme Nail and Gregoire Mialon and Guan Pang and Guillem Cucurell and Hailey Nguyen and Hannah Korevaar and Hu Xu and Hugo Touvron and Iliyan Zarov and Imanol Arrieta Ibarra and Isabel Kloumann and Ishan Misra and Ivan Evtimov and Jack Zhang and Jade Copet and Jaewon Lee and Jan Geffert and Jana Vranes and Jason Park and Jay Mahadeokar and Jeet Shah and Jelmer van der Linde and Jennifer Billock and Jenny Hong and Jenya Lee and Jeremy Fu and Jianfeng Chi and Jianyu Huang and Jiawen Liu and Jie Wang and Jiecao Yu and Joanna Bitton and Joe Spisak and Jongsoo Park and Joseph Rocca and Joshua Johnstun and Joshua Saxe and Junteng Jia and Kalyan Vasuden Alwala and Karthik Prasad and Kartikeya Upasani and Kate Plawiak and Ke Li and Kenneth Heafield and Kevin Stone and Khalid El-Arini and Krithika Iyer and Kshitiz Malik and Kuenley Chiu and Kunal Bhalla and Kushal Lakhotia and Lauren Rantala-Yeary and Laurens van der Maaten and Lawrence Chen and Liang Tan and Liz Jenkins and Louis Martin and Lovish Madaan and Lubo Malo and Lukas Blecher and Lukas Landzaat and Luke de Oliveira and Madeline Muzzi and Mahesh Pasupuleti and Mannat Singh and Manohar Paluri and Marcin Kardas and Maria Tsimpoukelli and Mathew Oldham and Mathieu Rita and Maya Pavlova and Melanie Kambadur and Mike Lewis and Min Si and Mitesh Kumar Singh and Mona Hassan and Naman Goyal and Narjes Torabi and Nikolay Bashlykov and Nikolay Bogoychev and Niladri Chatterji and Ning Zhang and Olivier Duchenne and Onur Çelebi and Patrick Alrassy and Pengchuan Zhang and Pengwei Li and Petar Vasic and Peter Weng and Prajjwal Bhargava and Pratik Dubal and Praveen Krishnan and Punit Singh Koura and Puxin Xu and Qing He and Qingxiao Dong and Ragavan Srinivasan and Raj Ganapathy and Ramon Calderer and Ricardo Silveira Cabral and Robert Stojnic and Roberta Raileanu and Rohan Maheswari and Rohit Girdhar and Rohit Patel and Romain Sauvestre and Ronnie Polidoro and Roshan Sumbaly and Ross Taylor and Ruan Silva and Rui Hou and Rui Wang and Saghar Hosseini and Sahana Chennabasappa and Sanjay Singh and Sean Bell and Seohyun Sonia Kim and Sergey Edunov and Shaoliang Nie and Sharan Narang and Sharath Raparthy and Sheng Shen and Shengye Wan and Shruti Bhosale and Shun Zhang and Simon Vandenhende and Soumya Batra and Spencer Whitman and Sten Sootla and Stephane Collot and Suchin Gururangan and Sydney Borodinsky and Tamar Herman and Tara Fowler and Tarek Sheasha and Thomas Georgiou and Thomas Scialom and Tobias Speckbacher and Todor Mihaylov and Tong Xiao and Ujjwal Karn and Vedanuj Goswami and Vibhor Gupta and Vignesh Ramanathan and Viktor Kerkez and Vincent Gonguet and Virginie Do and Vish Vogeti and Vítor Albiero and Vladan Petrovic and Weiwei Chu and Wenhan Xiong and Wenyin Fu and Whitney Meers and Xavier Martinet and Xiaodong Wang and Xiaofang Wang and Xiaoqing Ellen Tan and Xide Xia and Xinfeng Xie and Xuchao Jia and Xuewei Wang and Yaelle Goldschlag and Yashesh Gaur and Yasmine Babaei and Yi Wen and Yiwen Song and Yuchen Zhang and Yue Li and Yuning Mao and Zacharie Delpierre Coudert and Zheng Yan and Zhengxing Chen and Zoe Papakipos and Aaditya Singh and Aayushi Srivastava and Abha Jain and Adam Kelsey and Adam Shajnfeld and Adithya Gangidi and Adolfo Victoria and Ahuva Goldstand and Ajay Menon and Ajay Sharma and Alex Boesenberg and Alexei Baevski and Allie Feinstein and Amanda Kallet and Amit Sangani and Amos Teo and Anam Yunus and Andrei Lupu and Andres Alvarado and Andrew Caples and Andrew Gu and Andrew Ho and Andrew Poulton and Andrew Ryan and Ankit Ramchandani and Annie Dong and Annie Franco and Anuj Goyal and Aparajita Saraf and Arkabandhu Chowdhury and Ashley Gabriel and Ashwin Bharambe and Assaf Eisenman and Azadeh Yazdan and Beau James and Ben Maurer and Benjamin Leonhardi and Bernie Huang and Beth Loyd and Beto De Paola and Bhargavi Paranjape and Bing Liu and Bo Wu and Boyu Ni and Braden Hancock and Bram Wasti and Brandon Spence and Brani Stojkovic and Brian Gamido and Britt Montalvo and Carl Parker and Carly Burton and Catalina Mejia and Ce Liu and Changhan Wang and Changkyu Kim and Chao Zhou and Chester Hu and Ching-Hsiang Chu and Chris Cai and Chris Tindal and Christoph Feichtenhofer and Cynthia Gao and Damon Civin and Dana Beaty and Daniel Kreymer and Daniel Li and David Adkins and David Xu and Davide Testuggine and Delia David and Devi Parikh and Diana Liskovich and Didem Foss and Dingkang Wang and Duc Le and Dustin Holland and Edward Dowling and Eissa Jamil and Elaine Montgomery and Eleonora Presani and Emily Hahn and Emily Wood and Eric-Tuan Le and Erik Brinkman and Esteban Arcaute and Evan Dunbar and Evan Smothers and Fei Sun and Felix Kreuk and Feng Tian and Filippos Kokkinos and Firat Ozgenel and Francesco Caggioni and Frank Kanayet and Frank Seide and Gabriela Medina Florez and Gabriella Schwarz and Gada Badeer and Georgia Swee and Gil Halpern and Grant Herman and Grigory Sizov and Guangyi and Zhang and Guna Lakshminarayanan and Hakan Inan and Hamid Shojanazeri and Han Zou and Hannah Wang and Hanwen Zha and Haroun Habeeb and Harrison Rudolph and Helen Suk and Henry Aspegren and Hunter Goldman and Hongyuan Zhan and Ibrahim Damlaj and Igor Molybog and Igor Tufanov and Ilias Leontiadis and Irina-Elena Veliche and Itai Gat and Jake Weissman and James Geboski and James Kohli and Janice Lam and Japhet Asher and Jean-Baptiste Gaya and Jeff Marcus and Jeff Tang and Jennifer Chan and Jenny Zhen and Jeremy Reizenstein and Jeremy Teboul and Jessica Zhong and Jian Jin and Jingyi Yang and Joe Cummings and Jon Carvill and Jon Shepard and Jonathan McPhie and Jonathan Torres and Josh Ginsburg and Junjie Wang and Kai Wu and Kam Hou U and Karan Saxena and Kartikay Khandelwal and Katayoun Zand and Kathy Matosich and Kaushik Veeraraghavan and Kelly Michelena and Keqian Li and Kiran Jagadeesh and Kun Huang and Kunal Chawla and Kyle Huang and Lailin Chen and Lakshya Garg and Lavender A and Leandro Silva and Lee Bell and Lei Zhang and Liangpeng Guo and Licheng Yu and Liron Moshkovich and Luca Wehrstedt and Madian Khabsa and Manav Avalani and Manish Bhatt and Martynas Mankus and Matan Hasson and Matthew Lennie and Matthias Reso and Maxim Groshev and Maxim Naumov and Maya Lathi and Meghan Keneally and Miao Liu and Michael L. Seltzer and Michal Valko and Michelle Restrepo and Mihir Patel and Mik Vyatskov and Mikayel Samvelyan and Mike Clark and Mike Macey and Mike Wang and Miquel Jubert Hermoso and Mo Metanat and Mohammad Rastegari and Munish Bansal and Nandhini Santhanam and Natascha Parks and Natasha White and Navyata Bawa and Nayan Singhal and Nick Egebo and Nicolas Usunier and Nikhil Mehta and Nikolay Pavlovich Laptev and Ning Dong and Norman Cheng and Oleg Chernoguz and Olivia Hart and Omkar Salpekar and Ozlem Kalinli and Parkin Kent and Parth Parekh and Paul Saab and Pavan Balaji and Pedro Rittner and Philip Bontrager and Pierre Roux and Piotr Dollar and Polina Zvyagina and Prashant Ratanchandani and Pritish Yuvraj and Qian Liang and Rachad Alao and Rachel Rodriguez and Rafi Ayub and Raghotham Murthy and Raghu Nayani and Rahul Mitra and Rangaprabhu Parthasarathy and Raymond Li and Rebekkah Hogan and Robin Battey and Rocky Wang and Russ Howes and Ruty Rinott and Sachin Mehta and Sachin Siby and Sai Jayesh Bondu and Samyak Datta and Sara Chugh and Sara Hunt and Sargun Dhillon and Sasha Sidorov and Satadru Pan and Saurabh Mahajan and Saurabh Verma and Seiji Yamamoto and Sharadh Ramaswamy and Shaun Lindsay and Shaun Lindsay and Sheng Feng and Shenghao Lin and Shengxin Cindy Zha and Shishir Patil and Shiva Shankar and Shuqiang Zhang and Shuqiang Zhang and Sinong Wang and Sneha Agarwal and Soji Sajuyigbe and Soumith Chintala and Stephanie Max and Stephen Chen and Steve Kehoe and Steve Satterfield and Sudarshan Govindaprasad and Sumit Gupta and Summer Deng and Sungmin Cho and Sunny Virk and Suraj Subramanian and Sy Choudhury and Sydney Goldman and Tal Remez and Tamar Glaser and Tamara Best and Thilo Koehler and Thomas Robinson and Tianhe Li and Tianjun Zhang and Tim Matthews and Timothy Chou and Tzook Shaked and Varun Vontimitta and Victoria Ajayi and Victoria Montanez and Vijai Mohan and Vinay Satish Kumar and Vishal Mangla and Vlad Ionescu and Vlad Poenaru and Vlad Tiberiu Mihailescu and Vladimir Ivanov and Wei Li and Wenchen Wang and Wenwen Jiang and Wes Bouaziz and Will Constable and Xiaocheng Tang and Xiaojian Wu and Xiaolan Wang and Xilun Wu and Xinbo Gao and Yaniv Kleinman and Yanjun Chen and Ye Hu and Ye Jia and Ye Qi and Yenda Li and Yilin Zhang and Ying Zhang and Yossi Adi and Youngjin Nam and Yu and Wang and Yu Zhao and Yuchen Hao and Yundi Qian and Yunlu Li and Yuzi He and Zach Rait and Zachary DeVito and Zef Rosnbrick and Zhaoduo Wen and Zhenyu Yang and Zhiwei Zhao and Zhiyu Ma},
      year={2024},
      eprint={2407.21783},
      archivePrefix={arXiv},
      primaryClass={cs.AI},
      url={https://arxiv.org/abs/2407.21783}, 
}

@misc{pytorch,
      title={PyTorch: An Imperative Style, High-Performance Deep Learning Library}, 
      author={Adam Paszke and Sam Gross and Francisco Massa and Adam Lerer and James Bradbury and Gregory Chanan and Trevor Killeen and Zeming Lin and Natalia Gimelshein and Luca Antiga and Alban Desmaison and Andreas Köpf and Edward Yang and Zach DeVito and Martin Raison and Alykhan Tejani and Sasank Chilamkurthy and Benoit Steiner and Lu Fang and Junjie Bai and Soumith Chintala},
      year={2019},
      eprint={1912.01703},
      archivePrefix={arXiv},
      primaryClass={cs.LG},
      url={https://arxiv.org/abs/1912.01703}, 
}

@misc{tensorflow,
title={ {TensorFlow}: Large-Scale Machine Learning on Heterogeneous Systems},
url={https://www.tensorflow.org/},
note={Software available from tensorflow.org},
author={
    Mart\'{i}n~Abadi and
    Ashish~Agarwal and
    Paul~Barham and
    Eugene~Brevdo and
    Zhifeng~Chen and
    Craig~Citro and
    Greg~S.~Corrado and
    Andy~Davis and
    Jeffrey~Dean and
    Matthieu~Devin and
    Sanjay~Ghemawat and
    Ian~Goodfellow and
    Andrew~Harp and
    Geoffrey~Irving and
    Michael~Isard and
    Yangqing Jia and
    Rafal~Jozefowicz and
    Lukasz~Kaiser and
    Manjunath~Kudlur and
    Josh~Levenberg and
    Dandelion~Man\'{e} and
    Rajat~Monga and
    Sherry~Moore and
    Derek~Murray and
    Chris~Olah and
    Mike~Schuster and
    Jonathon~Shlens and
    Benoit~Steiner and
    Ilya~Sutskever and
    Kunal~Talwar and
    Paul~Tucker and
    Vincent~Vanhoucke and
    Vijay~Vasudevan and
    Fernanda~Vi\'{e}gas and
    Oriol~Vinyals and
    Pete~Warden and
    Martin~Wattenberg and
    Martin~Wicke and
    Yuan~Yu and
    Xiaoqiang~Zheng},
  year={2015},
}

@inproceedings{DNNFusion,
author = {Niu, Wei and Guan, Jiexiong and Wang, Yanzhi and Agrawal, Gagan and Ren, Bin},
title = {DNNFusion: accelerating deep neural networks execution with advanced operator fusion},
year = {2021},
isbn = {9781450383912},
publisher = {Association for Computing Machinery},
address = {New York, NY, USA},
url = {https://doi.org/10.1145/3453483.3454083},
doi = {10.1145/3453483.3454083},
booktitle = {Proceedings of the 42nd ACM SIGPLAN International Conference on Programming Language Design and Implementation},
pages = {883–898},
numpages = {16},
location = {Virtual, Canada},
series = {PLDI 2021}
}

@misc{TVM,
      title={TVM: An Automated End-to-End Optimizing Compiler for Deep Learning}, 
      author={Tianqi Chen and Thierry Moreau and Ziheng Jiang and Lianmin Zheng and Eddie Yan and Meghan Cowan and Haichen Shen and Leyuan Wang and Yuwei Hu and Luis Ceze and Carlos Guestrin and Arvind Krishnamurthy},
      year={2018},
      eprint={1802.04799},
      archivePrefix={arXiv},
      primaryClass={cs.LG},
      url={https://arxiv.org/abs/1802.04799}, 
}

@misc{TensorComprehend,
      title={Tensor Comprehensions: Framework-Agnostic High-Performance Machine Learning Abstractions}, 
      author={Nicolas Vasilache and Oleksandr Zinenko and Theodoros Theodoridis and Priya Goyal and Zachary DeVito and William S. Moses and Sven Verdoolaege and Andrew Adams and Albert Cohen},
      year={2018},
      eprint={1802.04730},
      archivePrefix={arXiv},
      primaryClass={cs.PL},
      url={https://arxiv.org/abs/1802.04730}, 
}

@inproceedings{Apollo,
 author = {Zhao, Jie and Gao, Xiong and Xia, Ruijie and Zhang, Zhaochuang and Chen, Deshi and Chen, Lei  and Zhang, Renwei and Geng, Zhen and Cheng, Bin and Jin, Xuefeng},
 booktitle = {Proceedings of Machine Learning and Systems},
 editor = {D. Marculescu and Y. Chi and C. Wu},
 pages = {1--19},
 title = {Apollo: Automatic Partition-based Operator Fusion through Layer by Layer Optimization},
 url = {https://proceedings.mlsys.org/paper_files/paper/2022/file/e175e8a86d28d935be4f43719651f86d-Paper.pdf},
 volume = {4},
 year = {2022}
}

@misc{FlexFlow,
      title={Beyond Data and Model Parallelism for Deep Neural Networks}, 
      author={Zhihao Jia and Matei Zaharia and Alex Aiken},
      year={2018},
      eprint={1807.05358},
      archivePrefix={arXiv},
      primaryClass={cs.DC},
      url={https://arxiv.org/abs/1807.05358}, 
}

@inproceedings {Unity,
author = {Colin Unger and Zhihao Jia and Wei Wu and Sina Lin and Mandeep Baines and Carlos Efrain Quintero Narvaez and Vinay Ramakrishnaiah and Nirmal Prajapati and Pat McCormick and Jamaludin Mohd-Yusof and Xi Luo and Dheevatsa Mudigere and Jongsoo Park and Misha Smelyanskiy and Alex Aiken},
title = {Unity: Accelerating {DNN} Training Through Joint Optimization of Algebraic Transformations and Parallelization},
booktitle = {16th USENIX Symposium on Operating Systems Design and Implementation (OSDI 22)},
year = {2022},
isbn = {978-1-939133-28-1},
address = {Carlsbad, CA},
pages = {267--284},
url = {https://www.usenix.org/conference/osdi22/presentation/unger},
publisher = {USENIX Association},
month = jul
}

@inproceedings{bai2019deepbench,
author = {Belloni, Stefano and Ritter, Daniel and Schr\"{o}der, Marco and R\"{o}rup, Nils},
title = {DeepBench: Benchmarking JSON Document Stores},
year = {2022},
isbn = {9781450393539},
publisher = {Association for Computing Machinery},
address = {New York, NY, USA},
url = {https://doi.org/10.1145/3531348.3532176},
doi = {10.1145/3531348.3532176},
pages = {1–9},
numpages = {9},
location = {Philadelphia, PA, USA},
series = {DBTest '22}
}

@misc{cupti,
  author       = {{NVIDIA Corporation}},
  title        = {CUDA Profiling Tools Interface (CUPTI)},
  year         = {2024},
  howpublished = {\url{https://developer.nvidia.com/cupti}},
  note         = {Accessed: 2024-11-21}
}

@misc{pytorch-fx,
      title={Torch.fx: Practical Program Capture and Transformation for Deep Learning in Python}, 
      author={James K. Reed and Zachary DeVito and Horace He and Ansley Ussery and Jason Ansel},
      year={2022},
      eprint={2112.08429},
      archivePrefix={arXiv},
      primaryClass={cs.LG},
      url={https://arxiv.org/abs/2112.08429}, 
}

@misc{pytorch-profiler,
  author       = {{PyTorch Contributors}},
  title        = {PyTorch Profiler Recipe},
  year         = {2024},
  howpublished = {\url{https://pytorch.org/tutorials/recipes/recipes/profiler_recipe.html}},
  note         = {Accessed: 2024-11-21}
}

@misc{astrasim2,
      title={ASTRA-sim2.0: Modeling Hierarchical Networks and Disaggregated Systems for Large-model Training at Scale}, 
      author={William Won and Taekyung Heo and Saeed Rashidi and Srinivas Sridharan and Sudarshan Srinivasan and Tushar Krishna},
      year={2023},
      eprint={2303.14006},
      archivePrefix={arXiv},
      primaryClass={cs.DC},
      url={https://arxiv.org/abs/2303.14006}, 
}

@inproceedings{sridharan2023chakra,
  title     = {MLCommons Chakra: Advancing Performance Benchmarking and Co-design using Standardized Execution Traces},
  author    = {Sridharan, Srinivas and Balogh, Andy and Beckmann, Bradford M. and Coutinho, Brian and Feng, Louis and Fu, Sheng and Gao, Sanshan and Garakani, Mehryar and Heo, Taekyung and Kanter, David and Ladd, Josh and Li, Ziwei and Liu, Winston and Man, Changhai and Mihailescu, Dan and More, Spandan and Park, Joongun and Ramachandran, Ashwin and Ramakrishnaiah, Vinay and Rashidi, Saeed and Reddi, Vijay Janapa and Sharma, Puneet and Tian, Phio and Won, William and Wu, Hanjiang and Xu, Huan and Yoo, Jinsun and Krishna, Tushar},
  booktitle = {Proceedings of the Ninth Annual Conference on Machine Learning and Systems (MLSys 2026), Industry Track},
  year      = {2026},
  address   = {Bellevue, WA, USA}
}

@misc{chakra-wg,
  title = {Chakra Working Group},
  author = {{MLCommons}},
  howpublished = {\url{https://mlcommons.org/working-groups/research/chakra/}},
  year = 2023
}

@misc{chakra-schema,
  title = {Chakra Schema},
  author = {{MLCommons}},
  howpublished = {\url{https://github.com/mlcommons/chakra/wiki/Chakra-Schema}},
  year = 2024
}

@misc{scaling-law,
      title={Scaling Laws for Neural Language Models}, 
      author={Jared Kaplan and Sam McCandlish and Tom Henighan and Tom B. Brown and Benjamin Chess and Rewon Child and Scott Gray and Alec Radford and Jeffrey Wu and Dario Amodei},
      year={2020},
      eprint={2001.08361},
      archivePrefix={arXiv},
      primaryClass={cs.LG},
      url={https://arxiv.org/abs/2001.08361}, 
}

@inproceedings{themis,
    title = {{Themis: A Network Bandwidth-Aware Collective Scheduling Policy for Distributed Training of DL Models}},
    author = {Rashidi, Saeed and Won, William and Srinivasan, Sudarshan and Sridharan, Srinivas and Krishna, Tushar},
    year = 2022,
    booktitle = {Proceedings of the 49th Annual International Symposium on Computer Architecture (ISCA '22)},
    pages = {581–596},
    doi = {10.1145/3470496.3527382},
    isbn = 9781450386104
}

@INPROCEEDINGS{wang_hoti2024,
author = { Wang, Weiyang and Ghobadi, Manya and Shakeri, Kayvon and Zhang, Ying and Hasani, Naader },
booktitle = { Proceedings of the 2024 IEEE Symposium on High-Performance Interconnects (HOTI) },
title = {{ Rail-only: A Low-Cost High-Performance Network for Training LLMs with Trillion Parameters }},
year = {2024},
doi = {10.1109/HOTI63208.2024.00013}}

@inproceedings{llm_memory,
      title={Mini-batch Coresets for Memory-efficient Training of Large Language Models}, 
      author={Dang Nguyen and Wenhan Yang and Rathul Anand and Yu Yang and Baharan Mirzasoleiman},
      year={2024},
      booktitle={arXiv:2407.19580 [cs.LG]}
}

@misc{mlperf,
      title={MLPerf Inference Benchmark}, 
      author={Vijay Janapa Reddi and Christine Cheng and David Kanter and Peter Mattson and Guenther Schmuelling and Carole-Jean Wu and Brian Anderson and Maximilien Breughe and Mark Charlebois and William Chou and Ramesh Chukka and Cody Coleman and Sam Davis and Pan Deng and Greg Diamos and Jared Duke and Dave Fick and J. Scott Gardner and Itay Hubara and Sachin Idgunji and Thomas B. Jablin and Jeff Jiao and Tom St. John and Pankaj Kanwar and David Lee and Jeffery Liao and Anton Lokhmotov and Francisco Massa and Peng Meng and Paulius Micikevicius and Colin Osborne and Gennady Pekhimenko and Arun Tejusve Raghunath Rajan and Dilip Sequeira and Ashish Sirasao and Fei Sun and Hanlin Tang and Michael Thomson and Frank Wei and Ephrem Wu and Lingjie Xu and Koichi Yamada and Bing Yu and George Yuan and Aaron Zhong and Peizhao Zhang and Yuchen Zhou},
      year={2020},
      eprint={1911.02549},
      archivePrefix={arXiv},
      primaryClass={cs.LG},
      url={https://arxiv.org/abs/1911.02549}, 
}

@INPROCEEDINGS{ZeRO,
  author={Rajbhandari, Samyam and Rasley, Jeff and Ruwase, Olatunji and He, Yuxiong},
  booktitle={SC20: International Conference for High Performance Computing, Networking, Storage and Analysis}, 
  title={ZeRO: Memory optimizations Toward Training Trillion Parameter Models}, 
  year={2020},
  volume={},
  number={},
  pages={1-16},
  doi={10.1109/SC41405.2020.00024}}

@misc{scalesim,
      title={SCALE-Sim: Systolic CNN Accelerator Simulator}, 
      author={Ananda Samajdar and Yuhao Zhu and Paul Whatmough and Matthew Mattina and Tushar Krishna},
      year={2019},
      eprint={1811.02883},
      archivePrefix={arXiv},
      primaryClass={cs.DC},
      url={https://arxiv.org/abs/1811.02883}, 
}

@misc{param,
  author       = {Facebook Research},
  title        = {Param: A Trace Abstraction for ML Workloads},
  year         = {2023},
  howpublished = {\url{https://github.com/facebookresearch/param}},
  note         = {Accessed: 2025-04-14}
}

@misc{kineto,
  author       = {PyTorch Team},
  title        = {Kineto: A CPU+GPU Profiling Library for PyTorch},
  year         = {2023},
  howpublished = {\url{https://github.com/pytorch/kineto}},
  note         = {Accessed: 2025-04-14}
}

@article{genie,
  author  = {Jinsun Yoo},
  title   = {Towards Easy and Realistic Network Infrastructure Testing for Large-scale Machine Learning},
  journal = {arXiv preprint arXiv:2504.20854},
  year    = {2026},
  url     = {https://arxiv.org/abs/2504.20854},
  note    = {Accessed: 2026-03-06}
}
